\documentclass[acmsmall,nonacm]{acmart}
\usepackage{microtype}
\usepackage{subfigure}
\usepackage{booktabs} 
\usepackage{enumitem}
\usepackage{titlesec}
\usepackage[most]{tcolorbox}
\usepackage{hyperref}
\usepackage{cleveref}
\crefformat{section}{\S#2#1#3}
\crefname{section}{\S}{\S\S}

\usepackage{algorithm}
\usepackage{amsmath}
\usepackage{xcolor}
\usepackage{colortbl}
\usepackage{multirow}
\usepackage{rotating}
\usepackage{adjustbox}
\usepackage{caption}

\usepackage{listings}

\usepackage{color}
\definecolor{codegray}{rgb}{0.5,0.5,0.5}
\lstdefinestyle{mystyle}{
    keywordstyle=\color{blue}\textbf,
    numberstyle=\scriptsize\color{codegray},
    basicstyle=\scriptsize\fontfamily{txtt}\selectfont,
    breakatwhitespace=false,
    breaklines=true,
    captionpos=b,
    keepspaces=true,
    numbers=left,
    numbersep=5pt,
    showspaces=false,
    showstringspaces=false,
    showtabs=false,
    tabsize=2,
    xleftmargin=20pt,
}
\newcommand{\myName}{PyGim }
\newcommand{\myNameN}{PyGim}
\newcommand{\spmm}{SpMM }

\newcommand{\spmmN}{SpMM}

\newcommand{\gnn}{GNN }

\newcommand{\gnns}{GNNs }
\newcommand{\gnnsN}{GNNs}
\newcommand{\CPUName}{PyTorch }
\newcommand{\CPUNameN}{PyTorch}
\newcommand{\aggravgcpuspeedup}{3.09}
\newcommand{\aggrmaxcpuspeedup}{4.00}
\newcommand{\aggravgpimspeedup}{4.10}
\newcommand{\aggrmaxpimspeedup}{7.70}
\newcommand{\totalavgcpuspeedup}{3.04}
\newcommand{\totalmaxcpuspeedup}{3.44}
\newcommand{\totalavgpimspeedup}{4.38}
\newcommand{\totalmaxpimspeedup}{7.20}
\newcommand{\totalavgcpuenergyspeedup}{1.55}
\newcommand{\totalmaxcpuenergyspeedup}{1.75}
\newcommand{\totalavgpimenergyspeedup}{2.86}
\newcommand{\totalmaxpimenergyspeedup}{3.68}
\newcommand{\aggravggpuspeeduputilization}{11.6}

\newcommand\shep[1]{\noindent{\color{black}{#1}}} 

\tcbset{
    left=8pt,
    right=8pt,
    enhanced,
    halign=center,
    colback=gray!16,
    colframe=black,
    boxrule=0.8pt,
    width=\dimexpr\textwidth\relax,
}
\AtBeginDocument{%
  }

\setcopyright{acmcopyright}
\copyrightyear{2022}
\acmYear{2022}
\acmDOI{}

\acmJournal{POMACS}
\acmVolume{6}
\acmNumber{1}
\acmArticle{}
\acmMonth{2}

\begin{document}

\title{\myNameN: An Efficient Graph Neural Network Library for Real Processing-In-Memory Architectures  }

\author{Christina Giannoula}
\email{christina.giann@gmail.com}
\affiliation{%
  \institution{University of Toronto}
  \country{Canada}
}  
\affiliation{%
  \institution{ETH Zürich}
  \country{Switzerland}
}
\affiliation{%
  \institution{Vector Institute}
  \country{Canada}
}
\affiliation{%
  \institution{CentML}
  \country{Canada}
}  

\author{Peiming Yang}
\affiliation{%
  \institution{University of Toronto}
  \country{Canada}
}

\author{Ivan Fernandez}
\affiliation{%
  \institution{Barcelona Supercomputing Center}
  \country{Spain}
}
\affiliation{%
  \institution{Universitat Politècnica de Catalunya}
  \country{Spain}
}
\affiliation{%
  \institution{ETH Z{\"u}rich}
  \country{Switzerland}
}

\author{Jiacheng Yang}
\affiliation{%
  \institution{University of Toronto}
  \country{Canada}
}
\affiliation{%
  \institution{Vector Institute}
  \country{Canada}
}

\author{Sankeerth Durvasula}
\affiliation{%
  \institution{University of Toronto}
  \country{Canada}
}
\affiliation{%
  \institution{Vector Institute}
  \country{Canada}
}

\author{Yu Xin Li}
\affiliation{%
  \institution{University of Toronto}
  \country{Canada}
}

\author{Mohammad Sadrosadati}
\affiliation{%
  \institution{ETH Z{\"u}rich}
  \country{Switzerland}
}

\author{Juan Gomez Luna}
\affiliation{%
  \institution{NVIDIA}
  \country{Switzerland}
}

\author{Onur Mutlu}
\affiliation{%
  \institution{ETH Z{\"u}rich}
  \country{Switzerland}
}

\author{Gennady Pekhimenko}
\affiliation{%
  \institution{University of Toronto}
  \country{Canada}
}
\affiliation{%
  \institution{Vector Institute}
  \country{Canada}
}
\affiliation{%
  \institution{CentML}
  \country{Canada}
}

\renewcommand{\shortauthors}{Christina Giannoula et al.}


\begin{abstract}
Graph Neural Networks (\gnnsN) are emerging models to analyze graph-structure data. 
The \gnn execution involves both compute-intensive and memory-intensive kernels. 
The memory-intensive kernels dominate execution time, because they are significantly bottlenecked by data movement between memory and processors. 
Processing-In-Memory (PIM) systems can alleviate this data movement bottleneck by placing simple processors near or inside memory arrays. 
To this end, we investigate the potential of PIM systems to alleviate the data movement bottleneck in \gnnsN, and introduce \myNameN, an efficient and easy-to-use \gnn library for real PIM systems. We propose intelligent parallelization techniques for memory-intensive kernels of \gnns tailored for real PIM systems, and develop an easy-to-use Python API for them. \myName employs a cooperative \gnn execution, in which the compute- and memory-intensive kernels are executed in processor-centric and memory-centric computing systems, respectively, to fully exploit the hardware capabilities. 
\myName integrates a lightweight tuner that configures the parallelization strategy of the memory-intensive kernel of \gnns to provide high system performance, while also enabling high programming ease. We extensively evaluate \myName on a real-world PIM system that has 16 PIM DIMMs with 1992 PIM cores connected to a Host CPU.
In \gnn inference, we demonstrate that it outperforms prior state-of-the-art PIM works by on average  \totalavgpimspeedup $\times$ (up to \totalmaxpimspeedup$\times$), and the state-of-the-art \CPUName implementation running on Host (on Intel Xeon CPU)
by on average  \totalavgcpuspeedup $\times$ (up to \totalmaxcpuspeedup$\times$). \myName improves energy efficiency by \totalavgpimenergyspeedup $\times$ (up to \totalmaxpimenergyspeedup$\times$) and \totalavgcpuenergyspeedup $\times$ (up to \totalmaxcpuenergyspeedup$\times$)  over prior PIM and \CPUName Host schemes, respectively. In memory-intensive kernel of \gnnsN, \myName provides \aggravggpuspeeduputilization $\times$ higher resource utilization in PIM system than that of PyTorch library (optimized CUDA implementation) in GPU systems.
 Our work provides useful recommendations for software, system and hardware designers. \myName  is publicly and freely available at \url{https://github.com/CMU-SAFARI/PyGim} to facilitate the widespread use of PIM systems in \gnnsN.
\end{abstract}



\maketitle

\section{Introduction}

Graph Neural Networks (\gnnsN)~\cite{kipf2016semi,xu2018powerful,hamilton2017inductive,Zheng2020DistDGL} have emerged as state-of-the-art Machine Learning (ML) models to depict dependent relations in graph-structure data, providing high accuracy in vertex classification  and  link (edge) prediction tasks~\cite{shang2021discrete,Thorpe2021Dorylus,mohamed2020discovering,ying2018graph}. 
Thus, they have been adopted to many real-world applications, including point-cloud analysis~\cite{qian2021pu}, recommendation systems~\cite{wu2019session}, social network analysis~\cite{fan2019graph}, and drug discovery~\cite{stokes2020deep}.  
\gnns comprise a few layers, and each layer consists of two steps: \emph{aggregation} and \emph{combination}. The former aggregates the input feature vectors of the neighboring vertices for each vertex in the graph via a permutation-invariant operator (e.g., average). The latter processes the aggregated vectors of all vertices through a small neural network (e.g., a multilayer perceptron~\cite{bishop1995neural}) to produce the output feature vectors, which will be fed as input feature vectors to the next layer.

The key operators of combination are dense matrix matrix multiplications (GEMMs), while aggregation degenerates to a Sparse Matrix Matrix Multiplication (\spmmN) kernel, processing the graph data that is represented as a sparse matrix~\cite{Gong2022Graphite,yun2023grande,zhou2022gnnear}.
In ~\cref{sec:backround_gnn}, we profile  the \gnn execution in a high-end GPU system, and find that aggregation dominates execution time and exhibits high memory intensity, while combination is compute-intensive.
The compute-intensive combination
fits 
to be executed in processor-centric systems (CPUs or GPUs). 
However, aggregation is significantly bottlenecked by data movement between memory and processors in such systems, since \spmm is typically memory-bandwidth-bound in CPUs and GPUs~\cite{yun2023grande,Gu2020Bandwidth,Gong2022Graphite} (See also \cref{sec:backround_gnn}).


A promising way to alleviate the data movement cost is Processing-In-Memory (PIM)~\cite{lee20221ynm,devaux2019,mutlu2020modern,Gomez2022Benchmarking,giannoula2022sparsep,Diab2023Framework,Gomez2023Evaluating,Lim2023Design,Item2023TransPimLib,Das2022Implementation,Jibril2024Aggregation,chen2023simplepim,Shin2023PIMFlow,rhyner2024analysis,giannoula2021syncron,fernandez2024Matsa,Kwon2021Function,Cho2020Near,tensordimm,Gao2015Practical,fernandez2020natsa,ke2019recnmp,Hadi2016Chameleon,Nair2015Active,ahn2015scalable,Hsieh2016accelerating,zhuo2019graphq,Boroumand2018Google,Gomez2021Benchmarking,Drumond2017mondrian,Liu2018Processing,liu2017concurrent,choe2019concurrent,Gokhale2015Near,Zhang2018GraphP,Mutlu2020AMP,Ghose2019Workload,Mutlu2019Processing} computing paradigm.
PIM enables computation to be performed close to the application data by equipping memory chips with processing capabilities (in-memory processors).
To provide significantly higher memory bandwidth for the in-memory processors than standard DRAM modules, manufacturers have commercialized \emph{near-bank} PIM designs~\cite{devaux2019}. Near-bank PIM memory modules tightly couple a PIM core with one (or a few) DRAM bank, exploiting bank-level parallelism to expose the high aggregated on-chip memory bandwidth of standard DRAM to processors. A real PIM system supports multiple near-bank PIM memory modules, which are connected to a CPU or GPU, henceforth referred to as \emph{Host}. The UPMEM PIM architecture~\cite{devaux2019} is the first PIM system to become commercially available. HBM-PIM~\cite{Lee2021Hardware} and AiM~\cite{lee20221ynm} are near-bank PIM systems that have been prototyped and evaluated in real systems.

A few works~\cite{zhou2022gnnear,tian2022GNMP,yun2023grande} propose hybrid Host-PIM accelerators for \gnnsN. However, none of them considers real-world PIM systems. These works design new microarchitectures for \emph{near-rank} PIM systems, i.e., accelerator cores are placed at each rank of memory modules. Near-rank PIM designs have not been commercialized yet, and are not always able to provide significantly higher memory bandwidth for processors than standard DRAM~\cite{Lee2021Hardware,Hadi2016Chameleon}. In the software level, these works have simple \emph{fixed} parallelization strategies for \gnn aggregation in PIM cores, which would cause out-of-memory errors for medium-/large-size graphs (See \cref{sec:background_prior_work}) or achieve very low performance in \emph{real} near-bank PIM systems, as we show in Figs.~\ref{fig:eval-aggregation} and ~\ref{fig:eval-full-infer} of ~\cref{sec:evaluation}. Moreover, these works use software emulators for their evaluations (not a real PIM system), and do not describe the engineering efforts needed to deploy \gnns in their accelerators.

Our \textbf{goal} in this work is to efficiently map \gnns on near-bank PIM systems and quantify the potential of \emph{real} PIM architectures in \gnn executions. Efficiently executing \gnns in real PIM systems encounters three key challenges. 
1) \gnn execution has repeated compute-intensive (combination)  and memory-intensive (aggregation) kernels. On the one hand, executing both types of kernels in PIM cores would incur high performance overheads in combination, since PIM cores are low-area and low-power cores with relatively low computation capabilities~\cite{Gomez2023Evaluating,Lee2021Hardware,Hadi2016Chameleon}. On the other hand, executing combination on Host cores and aggregation on PIM cores, respectively, necessitates minimizing the overheads of passing the output result of one kernel as input to the next kernel.
2) Real-world graphs exhibit diverse characteristics, e.g., the average, min or max vertex neighboring degrees vary across different graphs. Therefore, as discussed in prior works~\cite{kanellopoulos2019smash,giannoula2022sparsep,lenadora2024sensei,Ye2023SparseTIR,Strati2019Adaptive}, the 
execution behavior of sparse kernels, such as the SpMV/\spmmN, depends on the particular characteristics of the input given, and there is no typically one-size-fits-all parallelization solution that performs best across various inputs~\cite{giannoula2022sparsep}.
3) Programming a real near-bank PIM system for a high-level application is a hard task~\cite{chen2023simplepim,hyun2024pathfinding,Gilbert2024Scalabity}, since software stacks for PIM systems are still in an early stage.  Thus, ML programmers may need to  distribute data of \gnns across thousands of DRAM banks in a fine-grained and  careful way, have expertise of  the PIM hardware~\cite{Gilbert2024Scalabity,chen2023simplepim} and/or program the PIM cores  using low-level APIs~\cite{Gomez2023Evaluating,chen2023simplepim}.

To address the aforementioned challenges, we design \myName~\cite{PyGimLibrary}\footnote{\myName  is publicly available at \url{https://github.com/CMU-SAFARI/PyGim}.}, a high-level ML library to efficiently execute \gnns in real PIM systems. \myName provides high system performance in \emph{real} Host-PIM executions of \gnnsN, and bridges the gap between ML engineers, who prefer high-level programming interfaces (e.g., Python), and real PIM systems, that typically provide complex and low-level APIs and may need deep knowledge of PIM hardware.

\myName co-designs a \textbf{Cooperative Acceleration (CoA)} model with a novel \textbf{Parallelism Fusion (PaF)} method. CoA runs heterogeneous kernels to the best-fit underlying hardware: the processor-centric Host (CPU/GPU) system executes the compute-intensive \gnn combination, and the memory-centric PIM system executes the memory-intensive aggregation. PaF serves a dual purpose: it (i) strives a balance between computation and data transfer costs in \gnn aggregation executed in PIM cores, minimizing the overheads of passing the output result of combination as input to aggregation, and vice versa, and (ii) provides various parallelization techniques to cover many real-world graphs with diverse characteristics. 
Specifically, in \gnn aggregation, we enable three parallelism levels on the hardware PIM side and, at each level,  we provide different parallelization techniques on the software side. 1) We group the available PIM cores of the system in clusters, and design \textit{edge-} and \textit{feature-level} parallelism \textit{across PIM clusters}. 2) We enable \textit{vertex-} or \textit{edge-level} parallelism across cores \textit{within PIM cluster}. 3) We employ either \textit{vertex-} or \textit{edge-level} parallelism across threads \textit{within a PIM core}. 
The technique of the first parallelism level reduces data transfer overheads to/from PIM memory modules, thus reducing costs when passing the output of one \gnn kernel as input to the next one. The techniques of the second and third parallelism levels enable load balancing schemes that provide high compute balance across low-power PIM cores and across threads of a PIM core. 
PaF enables various \gnn aggregation configurations and load balancing strategies, by configuring the number of PIM cores per cluster, vertex- or edge-level parallelism within a PIM cluster or within a PIM core, such that to efficiently support diverse real-world graphs.


We design \myName to adapt to the graph's characteristics with minimal programmer intervention. 
We integrate in \myName  a \textbf{lightweight tuner} that predicts the best-performing PaF aggregation configuration based on the particular characteristics of the input graph. \myNameN's tuner employs effective performance models to estimate performance of different \gnn aggregation configurations in PIM systems at low cost. This way, we automate the selection of the \myName PaF configuration and eliminate the need for manual programmer intervention, while also providing high system performance. We develop a PIM backend for our optimized implementations and expose them with a \textbf{handy Python API} (See Alg.~\ref{alg:api}), so that programmers can easily use them.  We integrate our API with PyTorch~\cite{paszke2019pytorch} (it can be integrated to other frameworks~\cite{abadi2016tensorflow,gulli2017deep,chen2015mxnet,jia2014caffe}) to support either CPU or GPU as the Host (GPU-PIM systems are expected to be commercialized) in \gnn  PIM-based executions. \myName supports two widely-used compression formats for real-world graphs.
To our knowledge, \myName is \emph{the first easy-to-use and high-level ML library to deploy \gnn models in real PIM systems}, and is available as open-source to 
enable the wide use of PIM systems in \gnnsN.

We comprehensively characterize \gnn execution on the UPMEM PIM system, the first real-world PIM architecture, which has 16 PIM DIMMs with 1992 PIM cores connected to Host CPU. We evaluate our techniques in terms of scalability, data transfer costs, aggregation kernel and inference performance and energy efficincy using various real-world graphs and various \gnn models. We compare \myName over prior state-of-the-art PIM-based works for \gnns and show that it achieves significantly higher performance by on average \totalavgpimspeedup $\times$ (up to \totalmaxpimspeedup $\times$) 
and higher energy efficiency by on average \totalavgpimenergyspeedup $\times$ (up to \totalmaxpimenergyspeedup$\times$)  in \gnn inference. \myName improves \gnn inference performance and energy efficiency over the state-of-the-art \CPUName scheme running on Host by on average \totalavgcpuspeedup $\times$ (up to \totalmaxcpuspeedup $\times$) 
and \totalavgcpuenergyspeedup $\times$ (up to \totalmaxcpuenergyspeedup$\times$), respectively. Moreover, \myName achieves on average \aggravggpuspeeduputilization $\times$ higher resource utilization on PIM system than that of PyTorch’s backend on GPU systems, which is an optimized CUDA implementation from pytorch\_sparse library~\cite{Fey2019Fast}. This means that \myName uses the PIM system more effectively than PyTorch’s backend library uses the GPU system. Our extensive study provides  recommendations to improve multiple aspects of future  PIM hardware, systems and software. We hope that our ML library encourages further research and deployment of \gnns and sparse ML models in real PIM systems.

Overall, we make the following contributions:
\begin{itemize}[noitemsep,topsep=0pt,leftmargin=8pt]
    \item We investigate the challenges of efficiently implementing 
    \gnns on real-world PIM architectures, and propose an easy-to-use high-level \gnn library, named \myNameN, for such systems. \myName is open-source to enable further research.
    \item We combine the execution of heterogeneous kernels running on Host and PIM cores with a multi-level parallelization model. We enable three levels of parallelism and reduce the data transfer overheads from/to PIM memory modules in a heterogeneous \gnn execution. We provide various parallelization strategies and load balancing schemes to cover diverse real-world graphs. 
    \item We design a fast tuning mechanism to adapt the parallelization configuration of \gnns in PIM systems to the particular characteristics of the input graph, eliminating the need for manual programmer intervention. We expose our optimized PIM backend implementations as a handy Python API that can be integrated with state-of-the-art ML frameworks such as PyTorch.
    \item We extensively study the potential of real-world PIM architectures in \gnns  using various real-world graphs and  models. We show that \myName significantly outperforms prior approaches both in performance and energy efficiency,  and provides high resource utilization on real PIM systems.
\end{itemize}

\section{Background \& Motivation}
\label{sec:background_motivation}

\subsection{\gnns in Commodity Systems}\label{sec:backround_gnn}

\gnns are emerging ML models that analyze graph-structured data (\shep{e.g.,} knowledge graphs, social and road networks). A \gnn has a few layers. Each layer takes as input (i) the graph $G = (V, E)$, where $V$ and $E$ represent the graph's vertices and edges (connections between vertices), respectively, which is stored as a matrix, 
referred to as \textbf{adjacency matrix} $A$, and (ii) the \emph{feature} matrix $F$, that has one feature vector per vertex in the graph, each vector encodes the vertex's characteristics. Most real-world graphs are typically sparse (less than 1\% density)
~\cite{yun2023grande,giannoula2022sparsep}, i.e., they have relatively few connections between vertices compared to the total number of possible connections. Thus, the adjacency matrix is stored in memory in a compressed format, e.g., Compress Sparse Row (CSR)~\cite{bjorck1996numerical}. 
The feature matrix is dense with size $N \times K$, where $N$ is the number of vertices and $K$ is the number of features per vertex (henceforth referred to as \textbf{hidden size}).

\begin{figure}[t]
    \vspace{2pt}
    \centering
    \includegraphics[width=0.8\linewidth]{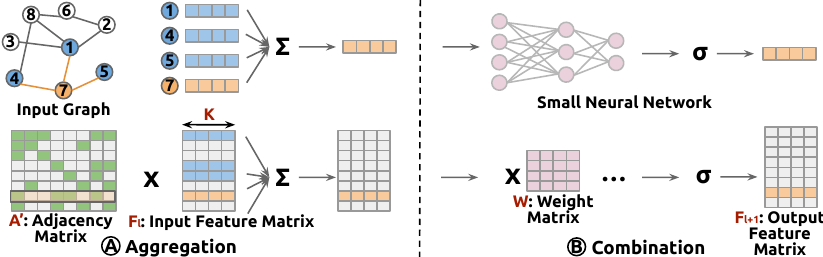}
    \vspace{4pt}
    \caption{Overview of the \gnn layer execution workflow.}
    \label{fig:gnn-overview}
    \vspace{-10pt}
\end{figure}

Fig.~\ref{fig:gnn-overview} shows the \gnn layer execution that has two steps: the \emph{aggregation} and \emph{combination}. In aggregation, each vertex gathers the feature vectors of its neighbors, and produces an aggregated vector though an operator (e.g., average). In combination, the aggregated vectors of all vertices are processed through a small neural network, that typically has dense operators (e.g., GEMMs) and finishes with an activation. The output feature vectors of all vertices serve as an input feature matrix of the next layer in \gnn model.


The aggregation and combination operators vary slightly across \gnn models. For example, GCN~\cite{kipf2016semi} uses a \emph{weighted sum} function for aggregation, while GIN~\cite{xu2018powerful} uses a \emph{sum} function. GIN uses an MLP for combination, while SAGE~\cite{hamilton2017inductive} uses a fully-connected operator. Assuming that $A'$ is a normalized adjacency matrix based on the aggregation function of each particular model, $F^l$ is the input feature matrix of a layer $l$ with hidden size $K$, and $W_{i}^l$ matrices are weight matrices used in the small neural network of combination, the \gnn computation can be expressed as:

\vspace{5pt}
\hspace{100pt} $ F^{l+1} = \sigma(\sigma((A' * F^l) * W_1^l) ... * W_w^l) $ 
\vspace{5pt}

$F^{l+1}$ will be the input feature matrix of the next layer. The aggregation step corresponds to the computation $A' * F^l$, which is a Sparse Matrix Matrix Multiplication (\spmmN).





\begin{figure}[h] 
\vspace{4pt}
    \centering
    \includegraphics[width=0.7\linewidth]{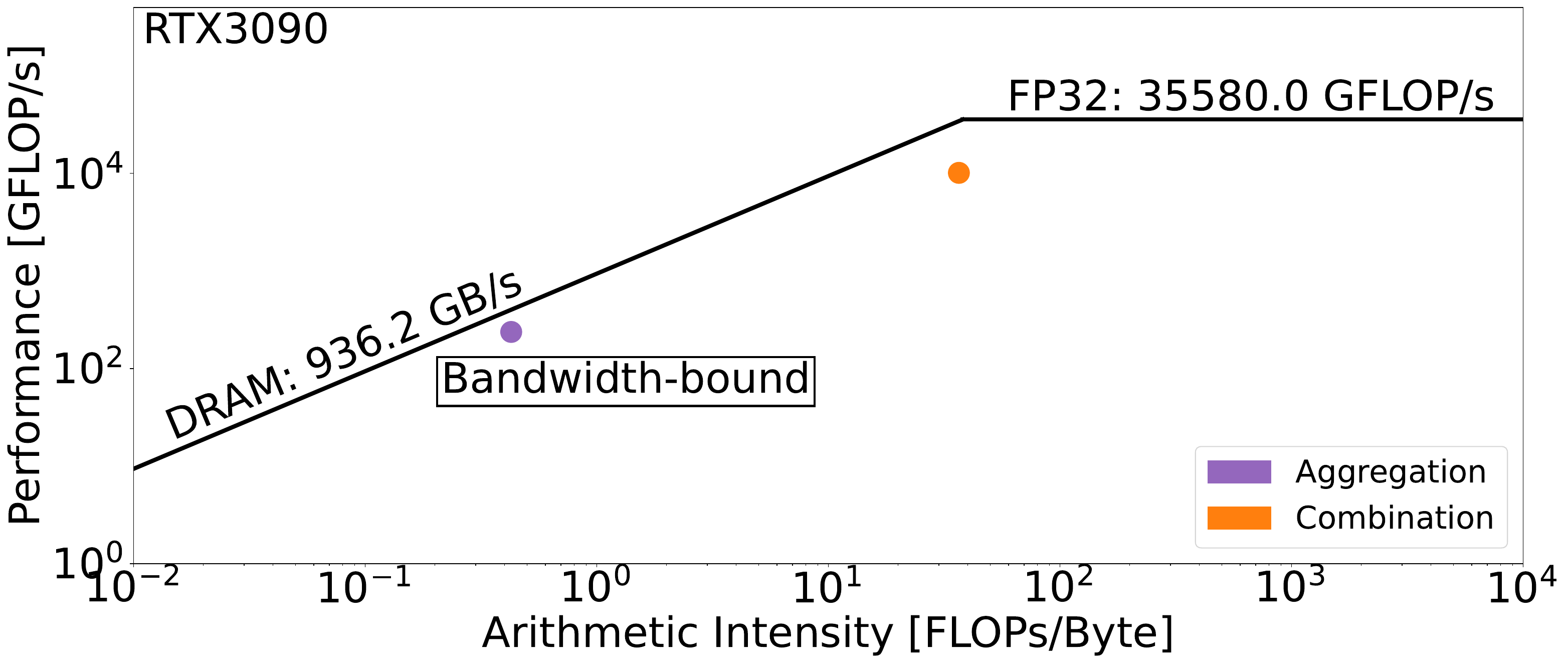}
    \vspace{3pt}
    \caption{Roofline model in the NVIDIA RTX 3090 GPU for aggregation and  combination kernels.}
    \label{fig:roofline}
\end{figure}

Recent works~\cite{zhou2022gnnear,tian2022GNMP,yun2023grande} show that \gnn aggregation takes the largest portion of the execution time, 
because it is bottlenecked by memory bandwidth in processor-centric systems (CPUs/GPUs). We evaluate a 3-layer \gnn in a RTX 3090 GPU and observe that aggregation takes $\sim$91\% of the inference time.
Fig.~\ref{fig:roofline} shows the roofline model, when executing a \gnn layer in the GPU. Even in a high-end GPU with more than 900GB/s bandwidth,  aggregation is highly limited by memory bandwidth.
Moreover, as we show in Table~\ref{tab:utilization-aggregation}, the resource utilization in aggregation is very low, i.e., on average 0.44\% 
and 1.19\% in CPU and GPU systems, respectively, due to  the bottleneck of moving data from memory to processors. Therefore, we conclude that \gnn aggregation is significantly limited by data movement in processor-centric systems like CPUs and GPUs.



\subsection{Processing-In-Memory (PIM) Systems}\label{sec:background_pim} 
PIM computing 
paradigm~\cite{mutlu2020modern,upmem,Lee2021Hardware,Hadi2016Chameleon} enables memory-centric computing systems: processing units (general-purpose cores or specialized accelerators) are placed near or inside memory arrays. PIM is a practical solution to alleviate the data movement bottleneck of processor-centric systems, and in this work, we study the potential of real PIM systems in \gnnsN.
According to the location of PIM cores, PIM architectures could be classified into two categories: (i) \emph{near-rank} in which PIM cores are placed at the buffer chip of the DIMM and have access to all DRAM banks of the DIMM, and (ii) \emph{near-bank} in which each PIM core is tightly coupled with one (or a few) DRAM banks, and can access data placed in its local bank(s). Placing PIM cores at a lower level provides larger aggregated memory bandwidth, enabling higher levels of parallelism.
For example, UPMEM near-bank PIM can provide up to $\sim$80GB/s internal bandwidth per DIMM~\cite{Gomez2023Evaluating}, while TensorDIMM~\cite{kwon2019tensordimm} near-rank PIM only 22GB/s per DIMM. 
Therefore, several manufacturers~\cite{devaux2019,lee20221ynm,Lee2021Hardware} target the commercialization of near-bank PIM designs to enable high levels of (bank-level) parallelism and support \emph{thousands} of PIM cores.
UPMEM PIM~\cite{devaux2019} has already commercialized a PIM product that has a general-purpose core near each memory bank of a DDR4 DRAM chip. HBM-PIM~\cite{Lee2021Hardware} and AiM~\cite{lee20221ynm,He2020Newton} have been prototyped and evaluated in real systems. 
HBM-PIM proposes a SIMD unit with 16-bit floating-point support between every two banks in memory layers of HBM stack. AiM is a GDDR6-based PIM system with near-bank cores that support multiply-and-accumulate and activation operations.

\begin{figure}[h]
    \vspace{2pt}
    \centering
    \includegraphics[width=0.76\linewidth]{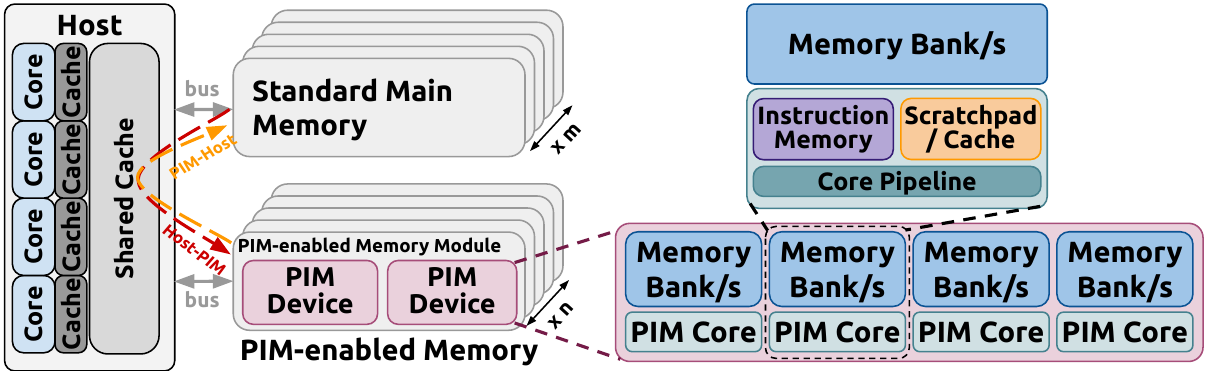}
    \vspace{5pt}
    \caption{Overview of a real near-bank PIM system. Host has access to $m$ standard and $n$ PIM-enabled  modules.}
    \label{fig:pim-baseline}
    \vspace{-2pt}
\end{figure}

These real-world PIM systems have some important common characteristics, shown in Fig.~\ref{fig:pim-baseline}. First, there is a Host processor (CPU or GPU) typically having a deep cache hierarchy, which is connected to standard main memory and PIM-enabled memory. Second, the PIM-enabled memory module has one (or a few) memory devices (rank of 2D DRAM or stacked layer of 3D-stacked DRAM). Each PIM device contains multiple processing elements (PIM cores), that have access to memory banks with higher bandwidth and lower latency than the Host cores. Third, the PIM cores (general-purpose cores, SIMD units, or specialized processors) run at only a few hundred megahertz, and have relatively small (or no) scratchpad or cache memory. Fourth, PIM cores may not be able to directly communicate with each other (UPMEM, HBM-PIM or AiM in different chips), and communication between them typically happens via the Host.

In our evaluation, we use UPMEM PIM~\cite{devaux2019}, the first PIM system that has been commercialized on real hardware. UPMEM PIM uses 2D DRAM arrays and combines them with general-purpose cores, called \emph{DPUs}, on the same chip. Each PIM-enabled module has two ranks (devices), each rank has 8 chips, and each chip has 8 DPUs. Each DPU is tightly coupled to a DRAM bank, has a 14-stage pipeline, and supports multiple threads (up to 24), called \emph{tasklets}.

DPUs have a 32-bit RISC-style general-purpose instruction set, and natively support in hardware 32-bit integer addition/subtraction and 8-bit/16-bit multiplication. More complex operations, e.g., 32-bit integer multiplication/division, and floating-point operations are software emulated~\cite{Gomez2022Benchmarking}. Each DPU has access to its own (1) 64MB DRAM bank, called MRAM, (2) 24KB instruction memory, and (3) 64KB scratchpad memory, called WRAM. The Host CPU can access the MRAM banks to copy input data (from main memory to MRAM, i.e., CPU-DPU) and retrieve results (from MRAM to main memory, i.e., DPU-CPU). The CPU-DPU/DPU-CPU transfers can be performed in parallel across multiple MRAM banks, if the size of data transferred from/to all MRAM banks is the same. 
There is no direct communication channel between DPUs, and 
inter-DPU communication takes place via the Host~\cite{Gilbert2024Scalabity}.

In our paper, we use generic terminology since our optimization strategies 
can generally apply to near-bank PIM systems (e.g., HBM-PIM or AiM), like the generic one shown in Fig.~\ref{fig:pim-baseline}, and not exclusively to the UPMEM PIM. Thus, we use the terms PIM device, PIM core, PIM thread, DRAM bank, scratchpad, and Host-PIM/PIM-Host data transfer, corresponding to PIM rank, DPU, tasklet, MRAM bank, WRAM, and CPU-DPU/DPU-CPU data transfer in UPMEM’s terminology.

\subsection{Prior PIM-Based \gnn Accelerators}\label{sec:background_prior_work}

A few prior works~\cite{zhou2022gnnear,tian2022GNMP,yun2023grande}  propose hardware accelerators for \gnn aggregation; however, they do not consider real-world PIM systems. They target near-rank PIM systems that 
have not been developed as proof-of-concept silicon yet, and  
typically provide lower memory bandwidth than near-bank PIM systems~\cite{Lee2021Hardware}, which have already been commercialized.
Moreover, these prior works focus on the ASIC design of the processing units, and implement simple parallelization strategies at the software level, which have been evaluated in software simulators (instead of a real system). Specifically, these works do not comprehensively evaluate the performance overheads of transferring data to/from PIM memory modules (i.e., Host-PIM and PIM-Host data transfer costs).

We find that applying the simple parallelization strategies of prior PIM-based \gnn accelerators~\cite{zhou2022gnnear,tian2022GNMP,yun2023grande} is not suitable or efficient for real near-bank PIM systems.
In detail, GNNear~\cite{zhou2022gnnear} equally distributes the graph's vertices across PIM cores, and the large dense feature matrix is copied at each PIM device of the system. Applying this approach at near-bank PIM systems would necessitate to replicate the large dense matrix of size $N$ (vertices) $\times$ $K$ (hidden size) at \emph{each} (or a few) PIM memory bank, and would cause out-of-memory errors for medium-/large-size graphs: e.g., assuming an UPMEM PIM bank of 64MB and 256 hidden size, GNNear's approach would support only small graphs with maximum $\sim$64K vertices. G-NMP~\cite{tian2022GNMP} equally distributes the feature matrix across PIM units. On the one hand, distributing the hidden size dimension $K$ of the feature matrix across PIM cores in near-bank PIM systems would leave \emph{many} PIM cores \emph{idle} causing low PIM utilization: \gnn layers typically have a much smaller hidden size (e.g., 128 or 256~\cite{kipf2016semi,xu2018powerful,hamilton2017inductive,Ye2023SparseTIR,Gong2022Graphite,Zheng2020DistDGL,Huang2021Understanding}) than the available PIM cores (thousands of cores) of near-bank PIM systems (e.g., UPMEM PIM system has 2560 PIM cores). On the other hand, distributing the vertex dimension $N$ of the feature matrix across the PIM cores would create a partial result of size $N$ $\times$ $K$ for the output feature matrix of aggregation at \emph{each} (or a few) PIM bank. Similarly to GNNear's scheme, this approach would cause out-of-memory errors for medium-/large-size graphs (e.g., graphs with only up to $\sim$64K vertices could be supported in UPMEM PIM), while also a large number of partial results would need to be merged by Host cores. 
Finally, considering all prior PIM-based works in \gnnsN,  GraNDe is the state-of-the-art work, and the most optimized in terms of its parallelization strategy.  GraNDe~\cite{yun2023grande} demonstrates that a 2D distribution scheme provides the highest on average performance in near-rank PIM systems. In this scheme, the vertex dimension $N$ of the feature matrix is distributed across PIM modules (e.g., PIM devices (ranks) in a near-bank PIM system) and the hidden size dimension $K$ of the feature matrix is distributed across PIM units of the same device (e.g., PIM cores of the same PIM device (rank)).
We evaluate this scheme in a real near-bank PIM system, and show that it achieves very low performance, being on average 6.3$\times$ (Fig.~\ref{fig:eval-full-infer}) worse in \gnn inference compared to our proposed \myName approach. This is because GraNDe is tailored for near-rank PIM systems, rather than near-bank PIM systems.

\section{\myNameN: Detailed Design}
\label{sec:mechanism}

\myName is a easy-to-use ML library to efficiently execute \gnns on real near-bank PIM systems. See common characteristics in \cref{sec:background_pim}. \myName improves system efficiency by running compute-bound and memory-bound kernels on processor-centric and memory-centric hardware, respectively (\cref{sec:MoA}), and providing highly efficient parallelization strategies in \gnn aggregation tailored for real PIM systems (\cref{sec:MoP}). Moreover, \myName adapts to the characteristics of the real-world graph without any programmer intervention 
via a lightweight tuning mechanism (\cref{sec:autotuning}) that automates the aggregation configuration. \myName enables high programming ease via a high-level ML-friendly interface (\cref{sec:system_integration}) that is integrated with state-of-the-art ML frameworks.

\subsection{Cooperative Acceleration Model (CoA)}\label{sec:MoA} 

\gnn execution alternates between sparse and dense operators: the aggregation step degenerates to an \spmm kernel, which is bottlenecked by memory bandwidth in processor-centric systems (Fig.~\ref{fig:roofline}), while the combination step mainly comprises compute-heavy kernels, e.g., GEMMs. We employ a \emph{Cooperative Acceleration} (\emph{CoA}) model that efficiently maps and executes each step to the best-fit underlying hardware. For \emph{each} \gnn layer, \myName executes the \spmm kernel of aggregation on PIM cores to leverage immense memory bandwidth available on PIM system, and the compute-heavy kernels of combination on Host (CPU or GPU) to exploit large processing capabilities available on processor-centric systems. Since aggregation and combination are repeated one after the other in multiple consecutive \gnn layers, the \textbf{key challenge} is how to minimize the overheads of passing the output of the one step as input to the next step. We discuss how we address it in the next subsection. Note that while PIM cores are running aggregation, Host cores are idle, until the dependent computation is finished, and vice versa. We leave for future work the extension of \myName to offload part of aggregation and combination computations on Host and PIM cores, respectively, to minimize idleness.

\subsection{Parallelism Fusion (PaF)}\label{sec:MoP}

\begin{figure}[h]
\vspace{3pt}
    \centering
    \includegraphics[width=\linewidth]{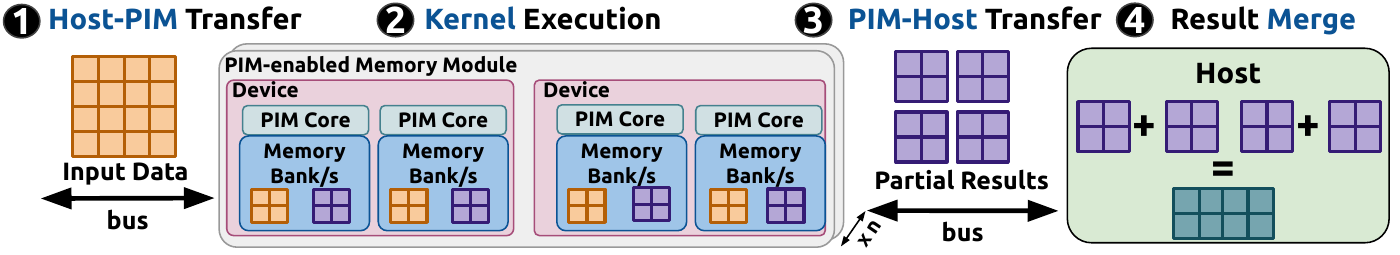}
    \vspace{-8pt}
    \caption{Execution of aggregation step on a real PIM system.}
    \label{fig:aggr-pim-execution}
    \vspace{2pt}
\end{figure}

Fig.~\ref{fig:aggr-pim-execution} shows the \gnn aggregation  execution on a real PIM system that can be broken down in four execution steps: (1) the time to transfer the input feature matrix of combination from Host into DRAM banks of PIM-enabled memory (\textbf{Host-PIM}), (2) the time to execute computational kernel on PIM cores (\textbf{Kernel}), (3) the time to retrieve from DRAM banks of PIM-enabled memory to the Host the results for the output (\textbf{PIM-Host}), and (4) the time to merge partial results and assemble the final output feature matrix on the Host (\textbf{Merge}). 
The graph (adjacency matrix) is pre-loaded into PIM-enabled memory \emph{once}, i.e., when reading the graph file from the disk and loading it to DRAM (pre-processing step). The same graph (adjacency matrix) is \emph{reused} across all layers.


\begin{figure}[h]
\vspace{2pt}
    \centering
    \includegraphics[width=0.7\linewidth]{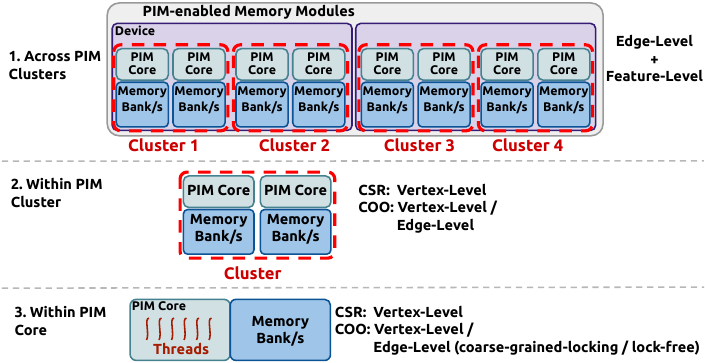}
    \vspace{4pt}
    \caption{ PaF overview.}
    \label{fig:paf-overview}
\end{figure}

We design \emph{Parallelism Fusion} (PaF) to mitigate data transfer and kernel time performance costs. We enable three levels of parallelism, each level implements a different strategy, shown in Fig.~\ref{fig:paf-overview}. 
The 
first-level strategy reduces data transfer overheads (Host-PIM and PIM-Host), thus addressing the aforementioned key challenge, while the second- and third-level strategies reduce computation overheads (kernel time) to achieve high PIM performance. This way 
PaF provides the sweet-spot and strives a balance between computation and data transfer costs. 
Moreover, in sparse computational kernels, including \spmmN, the execution behavior depends on the particular characteristics of the input given, i.e., the input graph in \gnnsN. Therefore, \myName enables various data partitioning approaches and load balance schemes, such that to tune the parallelization strategy based on the particular characteristics of the input graph, thus achieving performance that is as close as possible to the optimal performance for that given graph.

\noindent\textbf{Across PIM Clusters.}
We group the available PIM cores into clusters, named \textbf{PIM clusters}, and execute a part of the  \spmm aggregation at each cluster. A PIM device, i.e., a rank in 2D DRAM  (e.g., UPMEM PIM system) or a stacked layer in 3D DRAM (e.g., HBM-PIM system) can contain multiple PIM clusters, while all cores of the same cluster belong to the same PIM device (grouping cores located to different PIM devices is inefficient, since it would need multiple \emph{separate} Host-PIM \shep{and} PIM-Host transfers for the \emph{same} cluster, causing high transfer and launch overheads).

\begin{figure}[h]
\vspace{2pt}
    \centering
    \includegraphics[width=0.94\linewidth]{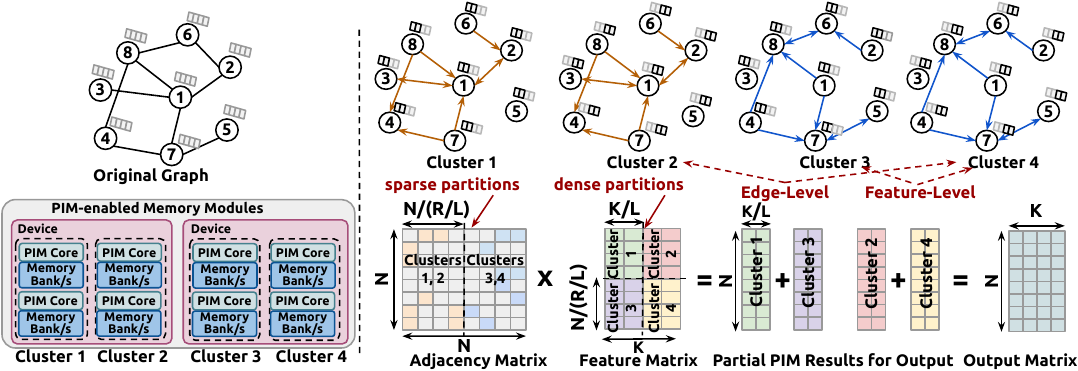}
    \vspace{4pt}
    \caption{Example of edge-level and feature-level parallelism across PIM clusters in \myNameN.}
    \label{fig:cluster-partition}
    \vspace{-8pt}
\end{figure}

We parallelize \spmm across PIM clusters via a hybrid \textbf{edge-level} and \textbf{feature-level} approach, as shown in Fig.~\ref{fig:cluster-partition}. Each cluster processes a subset of the graph's edges and a subset of the vertices' features to minimize Host-PIM \shep{and} PIM-Host transfer costs. We create vertical partitions (edge-level parallelism) on the adjacency sparse matrix, the number of which is henceforth referred to as \textbf{sparse partitions}. The adjacency matrix is distributed vertically across multiple PIM clusters, where each vertical partition is assigned to multiple PIM clusters. 
To minimize the amount of partial results produced for the final output matrix, and thereby minimizing the PIM-Host transfers, we combine edge-level parallelism with feature-level parallelism: we also create vertical partitions on the feature dense matrix, the number of which is henceforth referred to as \textbf{dense partitions}. Edge- and feature-level parallelism split the feature matrix both vertically and horizontally. This way the feature matrix is distributed into 2D tiles, the number of which is equal to the number of PIM clusters used, i.e., each 2D tile is assigned to a PIM cluster. Multiple PIM clusters process in parallel the 2D tiles of the feature matrix with their corresponding vertical partitions of the adjacency matrix, executing a part of the \spmmN. Multiple PIM clusters produce dense output matrices that correspond to partial results for the final output matrix. These partial results are merged in the Host cores (merge step) via matrix matrix addition.

Let's assume the number of sparse partitions is $s$ and the number of dense partitions is $d$. The sparse matrix is split into $s$ vertical partitions, and the feature matrix is split into $s$ horizontal partitions and $d$ vertical partitions, creating $s \times d$ 2D tiles.
Each PIM cluster will be assigned to process a 2D tile of the feature matrix with the corresponding vertical partition of the adjacency matrix.
The clusters that are assigned to 2D tiles of the same vertical partition in the feature matrix create partial results for the output matrix.
The values of sparse partitions $s$ and dense partitions $d$, and the number of PIM clusters per PIM device can be configured either manually by the user or automatically by the \myName tuner, as described in~\cref{sec:autotuning}. The PIM clusters per device multiplied with the number of PIM devices used in \spmm execution needs to be equal to the number $s \times d$.
By carefully selecting the values of sparse partitions $s$ and dense partitions $d$, as well as the number of PIM clusters per PIM device, PaF technique can significantly minimize the Host-PIM and PIM-Host data transfer costs.

Fig.~\ref{fig:cluster-partition} shows an example of two sparse partitions and two dense partitions and a PIM system of four PIM clusters. In this example, there are four 2D tiles in the feature matrix each assigned to a different PIM cluster. 
The clusters 1 and 2 are assigned to 2D tiles that belong in the first 2D tile row of the feature matrix, while the clusters 3 and 4 are assigned to 2D tiles that belong in the second 2D tile row of the feature matrix.
Therefore, in the adjacency matrix, clusters 1 and 2 are assigned to process the first vertical partition, i.e., the  vertical partition that is associated with the orange edges of the input graph, while clusters 3 and 4 are assigned to process the second vertical partition, i.e., the vertical partition that is associated with the blue edges of the graph.
The clusters  1 and 3 create partial results for the final output matrix corresponding to the first two columns of the final output matrix, while the cluster 2 and 4 create partial results for the final output matrix corresponding to the last two columns of the final output matrix.
Assuming a graph with $N$ vertices, hidden size $K$ and $R$ PIM clusters, when creating $L$ equal partitions on the feature matrix ($L<R$), each PIM cluster processes a feature matrix tile of size $(N/(R/L)) \times (K/L)$, which corresponds to the Host-PIM transfer cost for this cluster, and produces a partial output matrix of size $N \times (K/L)$, which corresponds to the PIM-Host transfer cost for this cluster.  
Although in the example of Fig.~\ref{fig:cluster-partition} 
the sparse and dense partitions have the same column width, \myName can support variable-sized vertical partitions on the adjacency and feature matrices. However, with variable-sized partitions, PIM clusters process variable-sized 2D feature tiles and produce variable-sized partial output results. Our exploratory evaluations showed that this approach incurs high load imbalance in Host-PIM/PIM-Host transfers, causing high overheads. Thus, in our evaluations we present equal-sized partitions. 

\begin{figure}[h]
    \vspace{2pt}
    \centering
    \includegraphics[width=0.9\linewidth]{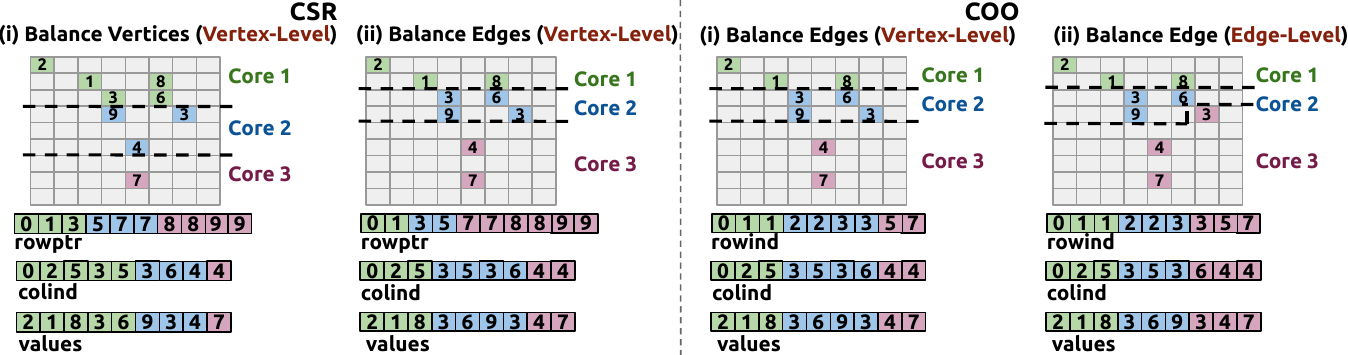}
    \vspace{4pt}
    \caption{Vertex- and edge-level parallelism across PIM cores within cluster. The gray cells represent zero values, while the green, blue and pink cells represent non-zero values (edges).}
    \label{fig:within-cluster-partition}
    \vspace{-8pt}
\end{figure}

\noindent\textbf{Within PIM Cluster.}
\myName encodes the adjacency matrix in CSR~\cite{Pooch1973Survey,bjorck1996numerical} and COO~\cite{Pooch1973Survey,Shubhabrata2007Scan} formats, the most widely-used compressed matrix storage formats for sparse matrices~\cite{Im1999Optimizing,Langr2016Evaluation,Pinar1999Improving,Mpakos2023Feature}.  We parallelize smaller \spmmN{s} across PIM cores of the same cluster, by enabling \emph{vertex-level} parallelism, if the adjacency matrix is stored in CSR, and either \emph{vertex-level} parallelism or \emph{edge-level} parallelism, if it is stored in COO. This way we provide compute balance across cores of the same cluster, minimizing the kernel time. The corresponding 2D feature matrix tile 
is replicated at each core of the same PIM cluster. Fig.~\ref{fig:within-cluster-partition} presents an example of parallelization across multiple cores of the \emph{same} cluster with CSR and COO formats.

The CSR format (Fig.~\ref{fig:within-cluster-partition} left) sequentially stores the edges (non-zero elements) in a vertex-wise (row) order. A column index array (\textsl{colind[]}) and a value array (\textsl{values[]}) store the column index and the value of each non-zero element, respectively. An array \textsl{rowptr[]}, stores the location of the first non-zero element of each row within the \textsl{values[]} array. An adjacent pair 
\textsl{rowptr[i, i+1]} stores the number of the non-zero elements of the i-th row. 
Since in CSR the adjacency matrix is stored in vertex-wise order, we perform \emph{vertex-level} parallelism across PIM cores of the same cluster: each core processes a subset of the vertices, i.e., consecutive rows in the adjacency matrix. We enable load balance across cores via two schemes: (a) equally balancing the number of vertices (rows) across PIM cores (Fig.~\ref{fig:within-cluster-partition} left i), or (b) equally balancing the number of edges (non-zero elements) across PIM cores at vertex (row) granularity (Fig.~\ref{fig:within-cluster-partition} left ii).

The COO format (Fig.~\ref{fig:within-cluster-partition} right) 
uses three arrays to store edges (non-zero elements): the row index (\textsl{rowind[]}), column index (\textsl{colind[]}) and value (\textsl{values[]}) arrays store the row index, column index and value of each non-zero element, respectively. Since in COO the adjacency matrix is stored in \emph{non-zero-element-wise order} (edge-wise order), we enable either \emph{vertex-level} parallelism, i.e., each PIM core of the cluster processes a subset of the vertices, or \emph{edge-level} parallelism, i.e., each PIM core processes a subset of the edges. We enable load balance across cores via two schemes: (a) equally balancing the number of edges (non-zero elements) across cores at a vertex (row) granularity (Fig.~\ref{fig:within-cluster-partition} right i), or (b) equally balancing the number of edges (non-zero elements) across cores by enabling splitting a vertex (row) across two (or more) neighboring cores to provide near-perfect edge-level balance (Fig.~\ref{fig:within-cluster-partition} right ii). In (b), 
when a vertex (row) is split between neighboring PIM cores, the cores produce partial results for the same row of the output matrix, which are 
merged by Host cores.

\noindent\textbf{Within PIM Core.} 
We enable a similar scheme across threads of a PIM core with that enabled across PIM cores of the same cluster (Fig.~\ref{fig:within-cluster-partition}). We enable high compute balance across threads of the same core to further minimize the kernel time.
In CSR, we perform \emph{vertex-level} parallelism by either (i) equally balancing the number of vertices (rows) across threads of a core, or (ii) equally balancing the number of edges (non-zero elements) at a vertex (row) granularity across threads. In COO, we either (i) equally balance the number of edges (non-zero elements) at a vertex (row) granularity across threads of a core (vertex-level parallelism), or (ii) equally balance the number of edges (non-zero elements) across threads, enabling splitting a vertex (row) across two (or more) PIM threads (edge-level parallelism). In the latter case, when a vertex (row) is split across two (or more) PIM threads, PIM threads perform write accesses to the \emph{same} elements of the output matrix, thus synchronization among threads is necessary. 
We  provide two synchronization schemes: 
\begin{itemize}[noitemsep,topsep=0pt,leftmargin=8pt]
\item \textbf{Coarse-grained locking}: one global mutex (lock) protects all the elements of the output matrix.
\item \textbf{Lock-free}: given that we assign consecutive rows to each thread (consider Fig.~\ref{fig:within-cluster-partition} COO (ii) with threads instead of cores), race conditions might arise only when a vertex (row) is split across two (or more) threads. These vertices (rows) are proportional to the number of threads, which are only a few per core. Thus, the number of partial results for the same final output element are a few, and threads can temporarily store partial results in the scratchpad memory (e.g., WRAM in the UPMEM PIM system). Then, only one single thread merges the partial results by reading them from scratchpad memory, and writes the final result to the 
DRAM bank/s with no synchronization.
\end{itemize}

\noindent\textbf{Kernel Implementation.} We describe how threads access the data involved in \spmmN. There are three types of data arrays: (i) the arrays that store the non-zero elements of the adjacency matrix, i.e., their values (\textsl{values[]}) and their positions (\textsl{rowptr[]}, \textsl{colind[]} for CSR, and \textsl{rowind[]}, \textsl{colind[]} for COO), (ii) the array that stores the elements of the feature matrix, and (iii) the array that stores the partial results for the output matrix. First, \spmm performs streaming memory accesses to the arrays that store the non-zero elements. To exploit PIM's immense internal bandwidth, each thread reads the non-zero elements (their values and positions) by fetching large chunks of bytes in a coarse-grained manner from DRAM bank to scratchpad (e.g., from MRAM to WRAM in the UPMEM PIM system).  
Then, it accesses data element by element via scratchpad. In the UPMEM PIM system, we fetch large chunks of 128-bytes/256-bytes, as suggested by prior work~\cite{Gomez2023Evaluating}. Second, \spmm processes the feature matrix elements at a row granularity, as a chunk of hidden size elements in the tile assigned to the PIM core, e.g., a chunk of $K/L$ elements in Fig.~\ref{fig:cluster-partition}. To exploit spatial locality, each thread reads the feature matrix elements by fetching chunks of tile hidden size $\times$  data type bytes from DRAM bank to scratchpad, and performs multiply-and-add. Third, threads temporarily store partial results for the elements of the same output matrix row in scratchpad, until all non-zero elements of the same row of the adjacency matrix are processed. This way we exploit temporal locality for multiple updates on the \emph{same} output matrix elements. Then, the produced results are written from scratchpad to DRAM bank/s as a chunk of tile hidden size $\times$  data type bytes.

\noindent\textbf{Merge Step.} \myName merges partial results created across PIM clusters and across cores within PIM cluster (merge step) on the Host cores.
In our CPU-PIM system, we use the OpenMP API~\cite{Dagum98OpenMP} to parallelize Merge, and perform (i)
2D block copy on the final output matrix for the partial results of the PIM clusters assigned to the first 2D block row of the feature matrix (Clusters 1,2 in Fig.~\ref{fig:cluster-partition}) and (ii) 2D block reduction (add) operation on the final output matrix (matrix-matrix addition) for the partial results of the PIM clusters assigned to the remaining 2D block rows of the feature matrix (Clusters 3,4 in Fig.~\ref{fig:cluster-partition}). 


\subsection{\myName  Tuner}\label{sec:autotuning}

\myNameN's PaF is designed to support various parallelization and load balancing strategies to efficiently cover various real-world graphs: as shown in prior works~\cite{kanellopoulos2019smash,giannoula2022sparsep,lenadora2024sensei,Ye2023SparseTIR}, the execution behavior of sparse kernels that process input data with diverse characteristics, such as the \spmm of \gnn aggregation that processes real-world graphs with varying neighboring degrees, diameters etc, depends on the particular characteristics of the input. Therefore, to enable high performance by adapting the PaF strategy to the particular graph's characteristics and avoid programmer's intervention, we integrate in \myName a lightweight tuner that selects the aggregation configuration to be used: the user selects the compression format (CSR/COO), and then the tuner predicts the best-performing aggregation configuration, i.e.,  the sparse and dense partitions, the groups per device, the selection of vertex- or edge-level parallelism across cores of a PIM cluster and across threads of a PIM core.

To enable the tuner, we first run a few microbenchmarks on the PIM system to collect information on hardware characteristics. These microbenchmarks run within less than $\sim$30 secs, and are executed only \emph{once} per PIM server to gather runtime characteristics of the underlying hardware. Our approach is similar to prior prediction and profiling tools for ML executions~\cite{geoffrey2021habitat,chen2018learning,zhai2023tlp,baghdadi2021deep,kaufman2021learned}. We devise four microbenchmarks. 1) The \textit{Host-PIM-BW-byte} and \textit{PIM-Host-BW-byte} measure the Host-PIM and PIM-Host bandwidth, respectively, when using multiple PIM devices and transferring $M$ bytes to/from each PIM core (\textit{PCore}). We vary $M$ from 64KB to 8MB, collecting 16 different byte sizes. 2) The \textit{Host-BW-byte} measures the Host bandwidth of standard DRAM modules, when copying $M$ bytes from one memory area (i.e., allocated matrix) to another area. We vary $M$ from 8B to 2KB and collect 9 different byte sizes. 3) \textit{FMA-PCore-chunk} is the fused multiply-add (FMA) throughput achieved by a PIM core (\textit{PCore}), i.e., number of FMA operations executed per second, when the PIM core performs FMA operations on data values that are transferred from the DRAM bank to scratchpad memory as chunks of $M$ elements. We vary $M$ from 2 to 512 elements collecting 9 different chunk sizes. 4) \textit{ADD-Host-block} is the addition (ADD) throughput achieved by Host cores, when accumulating (ADD) the $M$ elements of a block to a larger allocated matrix. We vary $M$ from 2 to 512 elements collecting 9 different block sizes.  After collecting data on hardware characteristics, the tuner estimates the execution time of an aggregation configuration using the following analytical models:

\vspace{6pt}
\noindent$T_{\text{total}} = T_{\text{Host-PIM}} + T_{\text{Kernel}} + T_{\text{PIM-Host}} + T_{\text{Merge}}$
\vspace{6pt}

\noindent$T_{\text{Host-PIM}} = \frac{\text{PCores x max-bytes-to-PCore}}{\text{Host-PIM-BW-byte (closest)}}$
\vspace{6pt}

\noindent$T_{\text{Kernel}} = \text{max-NNZs-PCore} \times \text{FMA-PCcore-chunk (closest)}$
\vspace{6pt}

\noindent$T_{\text{PIM-Host}} = \frac{\text{PCores $\times$ max-bytes-from-PCore}}{\text{PIM-Host-BW-byte (closest)}}$
\vspace{8pt}

\noindent$T_{\text{Merge}} = \frac{\text{dp x cluster-2Dtile-byte}}{\text{Host-BW-byte (closest)}} + \frac{\text{(sp-1) x dp x cluster-2Dtile}}{\text{ADD-Host-block (closest)}}$ 
\vspace{8pt}

The \textit{max-bytes-to/from-PCore} and \textit{max-NNZs-PCore} are the maximum bytes sent/received to/from a PIM core in Host-PIM/PIM-Host transfers and the maximum non-zero elements (\textit{NNZs}) processed by a PIM core, respectively. The \textit{sp} and \textit{dp} are the number of sparse and dense partitions created. The \textit{cluster-2Dtile} and \textit{cluster-2Dtile-byte} represent the number of elements and bytes of the 2D tile for partial results created by a PIM cluster (e.g., $N \times (K/L)$ in Fig.~\ref{fig:cluster-partition}), respectively. For the Host-PIM-BW, PIM-Host-BW and Host-BW, we use the collected bandwidth measurement associated with the data size that is closest to the data size of the aggregation configuration we are estimating. For FMA and ADD throughput, we use the collected throughput  measurement associated with the chunk of elements/block size that is closest to the chunk of elements/block size of the aggregation configuration we are estimating. 

Alg.~\ref{alg:autotuner} presents a brief description of the \myName tuner which iterates over all possible aggregation configurations and predicts the best-performing configuration. Specifically, the tuner iterates over all divisors of the available PIM devices in the system (line 5), creates 1, 2, or 4 clusters per PIM device, and it computes the number of dense partitions (line 7). If the number of dense partitions is larger than the hidden size, this configuration is not valid and is omitted (lines 9). Otherwise, the tuner iterates over the possible load balance strategies within the PIM cluster (line 10) and within the PIM core (line 11). For each current configuration, the tuner estimates the performance using the described analytical models (line 12), and keeps the lowest estimated execution time and its corresponding configuration in local variables (lines 13-15). When the tuner has examined all possible configurations, it returns the estimated best-performing configuration (line 16).

\begin{figure}[h]
\begin{minipage}{1.01\linewidth}

\begin{lstlisting}[language=Python]
def tune(graph, hidden_size, device_info):
    clst_cfg = ['ver', 'edg'] (*\textcolor{gray}{\# load balance within PIM cluster}*)
    core_cfg = ['ver', 'edg'] (*\textcolor{gray}{\# load balance within PIM core}*)
    best = inf, best-cfg = []
    for sp in divisors of num_pim_devices: (*\textcolor{gray}{\# sparse partitions}*)
        for grp in [1, 2, 4]: (*\textcolor{gray}{\# clusters per PIM device}*)
            dp = num_pim_devices / sp * grp (*\textcolor{gray}{\# dense partitions}*)
            if dp > hidden_size:
                continue
            for cl in clst_cfg:
                for cr in core_cfg:
                    Ttotal = predict(graph, hidden_size, grp, sp, dp, cl, cr, device_info)
                    if (Ttotal < best):
                        best = Ttotal
                        best-cfg = [sp, dp, grp, cl, cr]
    return best-cfg
\end{lstlisting}
\vspace{1pt}
\captionof{algorithm}{\myName tuner for the aggregation operator.}\label{alg:autotuner}
\end{minipage}
\vspace{-8pt}
\end{figure}

The \myName tuner is optional, providing flexibility to the programmer who may choose to utilize it or not. If the tuner is not used, the programmer is responsible for manually selecting and providing the desired aggregation configuration. In Alg.~\ref{alg:api} (an example of GCN inference), line 18 needs to be replaced with a manually tuned configuration. If the tuner is enabled, it takes around $\sim$33 secs and is executed \emph{once}, when reading the graph file from the disk and loading it into PIM devices in the pre-processing step. 
Then, users can submit multiple \gnn inference requests to query properties of vertices/edges or the existence of edges between graph's vertices~\cite{hamilton2017inductive,kipf2016semi,yang2016revisiting,velickovic2017graph,zeng2023serving}. 
Finally, the tuner's goal is to estimate the best-performing aggregation configuration by leveraging simple analytical models. In our evaluations (Fig.~\ref{fig:autotuner_csr}), we evaluate the efficiency of our tuner by comparing the performance achieved by the predicted configuration of tuner versus the best-performing manually tuned configuration (oracle prediction), and show that the simple analytical models used by the tuner are highly effective.

\subsection{\myName API and Integration}\label{sec:system_integration}

Combination comprises a small neural network, thus \myName leverages existing optimized ML kernels from PyTorch to execute the corresponding ML operators of \gnn combination on Host cores. We integrate \myName with PyTorch, as we explain in the next paragraph. Note that although we only have access to a CPU-PIM system for our evaluations, \myName can support GPU-PIM \gnn executions (GPU-PIM systems that are expected to be available in the market) by leveraging PyTorch's supported backends (CPU and GPU). 

To interact with PIM devices (e.g., Host-PIM/ PIM-Host transfers) in aggregation, Host code needs to be implemented. The Host code implements (developed in C language) the parallelization approaches and the corresponding data partitioning schemes proposed in ~\cref{sec:MoP}, when loading the graph into PIM-enabled memory modules (pre-processing step). 
The kernel code that PIM cores are running is implemented using the UPMEM PIM interface, since this is the only commercially available real PIM system. This interface is also written using the C language, and it can be easily ported to other PIM systems with similar interfaces to UPMEM. We create a PIM backend for PIM aggregation and expose this software runtime as a handy Python API so that programmers can easily use it via a high-level programming interface. We combine our Python-like API with PyTorch~\cite{paszke2019pytorch} to enable efficient CoA execution.  \myNameN's PIM aggregation can be also easily integrated to other ML frameworks, such as TensorFlow~\cite{abadi2016tensorflow}, Keras~\cite{gulli2017deep}, MXNet~\cite{chen2015mxnet} and Caffe~\cite{jia2014caffe}.

Alg.~\ref{alg:api} presents an example of GCN inference with \myNameN. Programmers need to (i) allocate PIM devices 
(line 15), (ii) load graph data into PIM-enabled memory (lines 17-19), (iii) create a \gnn model (lines 21-25), and (iv) run \gnn inference by configuring aggregation and combination to be executed on the PIM cores (line 9) and Host cores (line 11), respectively. The allocated PIM resources are released, when the program exits. 

\begin{figure}[h]
\begin{minipage}{0.78\linewidth}
\begin{lstlisting}[language=Python]
import torch, pygim as gyn
class GCNConv(torch.nn.Module): 
  def __init__(self, hidden_size):
    self.linear = torch.nn.Linear(hidden_size, hidden_size)

  def forward(self, graph_pim, in_dense): 
    (*\textcolor{gray}{\# Execute Aggregation in PIM}*)
    dense_parts = col_split(in_dense)
    out_dense = gyn.pim_run_aggr(graph_pim, dense_parts)
    (*\textcolor{gray}{\# Execute Combination in Host}*)
    out = self.linear(out_dense)
    return out

(*\textcolor{gray}{\# Allocate PIM Devices}*)
gyn.pim_init_devices(num_pim_devices, groups_per_device)
(*\textcolor{gray}{\# Load graph in PIM devices}*)
data = load_dataset()
graph_parts, config = gyn.tune(data.graph, hidden_size, device_info)
graph_pim = gyn.load_graph_pim(graph_parts, config)
(*\textcolor{gray}{\# Create GNN model}*)
model = torch.nn.Sequential([Linear(in_channels, hidden_size),
  GCNConv(hidden_size),
  GCNConv(hidden_size),
  GCNConv(hidden_size),
  Linear(hidden_size, out_channels) ])
model.forward(graph_pim, data.features)
\end{lstlisting}
\vspace{1pt}
\captionof{algorithm}{Example of GCN execution with \myName API.}\label{alg:api}
\end{minipage}
\vspace{-16pt}
\end{figure}

\section{Evaluation}\label{sec:evaluation}

\subsection{Methodology}\label{sec:methodology}

\noindent\textbf{System.} We use the UPMEM PIM architecture, a real-world PIM system. The system consists of a Host CPU (2-socket 
Intel Xeon with 8-cores each and a total of 32 threads at 2.10 GHz), standard DDR4 memory (128 GB), and 16 PIM DIMMs of 2 ranks (124.5 GB and 1992 PIM cores at 350 MHz), each rank has 64 cores. There are 56 faulty cores in the evaluated system that cannot be used, but they do not affect the correctness of our results (they are not used in our experiments).


\noindent\textbf{Models and Datasets.} We evaluate the GCN~\cite{kipf2016semi}, GIN~\cite{xu2018powerful} and SAGE~\cite{hamilton2017inductive} models. The multiplication of floating point data types is software emulated in the UPMEM
PIM system. Thus, we present detailed evaluations with 32-bit integer (\textbf{int32}) data type, since it has the same byte width with 32-bit float (\textbf{fp32}), its arithmetic operations are more effectively supported in UPMEM PIM hardware, and provides high accuracy (having int32 for both computation and memory representation results to less than 1\% accuracy drop in all models and datasets over fp32 using the quantization scheme of Ctranslate2~\cite{Ctranslate2}). 
Quantization~\cite{Ctranslate2,zhang2019qpytorch,loroch2017tensorquant} is orthogonal to our optimizations, and we expect that  future PIM systems (e.g., HBM-PIM) will provide native floating-point arithmetic support or optimized quantization schemes will provide high accuracy with fixed-precision data types. We evaluate real-world sparse matrices from the Sparse Matrix Suite Collection~\cite{davis2011florida}, when using one PIM core and one PIM cluster, and present large-scale experiments with three real-world graph datasets: ogbn-proteins~\cite{szklarczyk2019string}, Reddit~\cite{hamilton2017inductive} and AmazonProducts~\cite{zeng2019graphsaint}.  See also Appendix~\ref{sec:appendix-dataset} for detailed matrix and graph dataset characteristics.

\noindent\textbf{Comparison Points.}  We use the PyG library~\cite{PyG} for \gnn implementation. In \gnn inference, combination runs on the Host CPU with PyTorch's default backend implementation. In \gnn aggregation, we compare \myName with prior software schemes for PIM systems proposed in the literature. In \myName and other PIM-based software schemes,  when the kernel step of \gnn aggregation is running
on PIM cores, the host CPU cores are idle. We also compare \myName with a CPU-only scheme, that runs on \emph{same} system with PIM schemes, i.e., the Host CPU side of the UPMEM PIM server, similarly to the methodology of prior state-of-the-art PIM works~\cite{chen2023simplepim,Gilbert2024Scalabity,giannoula2022sparsep,Item2023TransPimLib}. 
Overall, we compare \myName with  four schemes:
\begin{itemize}[noitemsep,topsep=0pt,leftmargin=8pt]
\item \textbf{\CPUNameN}: the PyTorch's backend which is the state-of-the-art matmul operator from pytorch\_sparse library~\cite{Fey2019Fast}. We evaluate the latest default implementation of matmul that uses the optimized Intel MKL library. We run this scheme using all 32 threads of the 32-thread Intel Xeon CPU.
\item \textbf{GraNDe}~\cite{yun2023grande}: the best-performing PIM scheme of prior  PIM-based work for \gnnsN~\cite{yun2023grande} that equally distributes the vertex (row) dimension of the feature matrix across PIM devices, and then equally distributes the hidden size of the feature matrix across cores of the same PIM device.
\item \textbf{SP1} and \textbf{SP2}~\cite{giannoula2022sparsep}: two SpMV-based schemes of prior work~\cite{giannoula2022sparsep} for real PIM systems. SparseP~\cite{giannoula2022sparsep} proposes SpMV kernels for PIM systems, and shows that their optimized COO.nnz-lf kernel performs best, when using $\sim$2 PIM devices. We run aggregation as an SpMV execution: for each column of the feature matrix, we execute one SpMV kernel using either one PIM device (\textbf{SP1}) or two PIM devices (\textbf{SP2}), and parallelize multiple SpMVs for the multiple columns of  the feature matrix using multiple PIM devices.
\end{itemize}

\subsection{Within PIM Core Analysis}

\begin{figure}[h]
    \vspace{2pt}
    \centering
    \includegraphics[width=\linewidth]{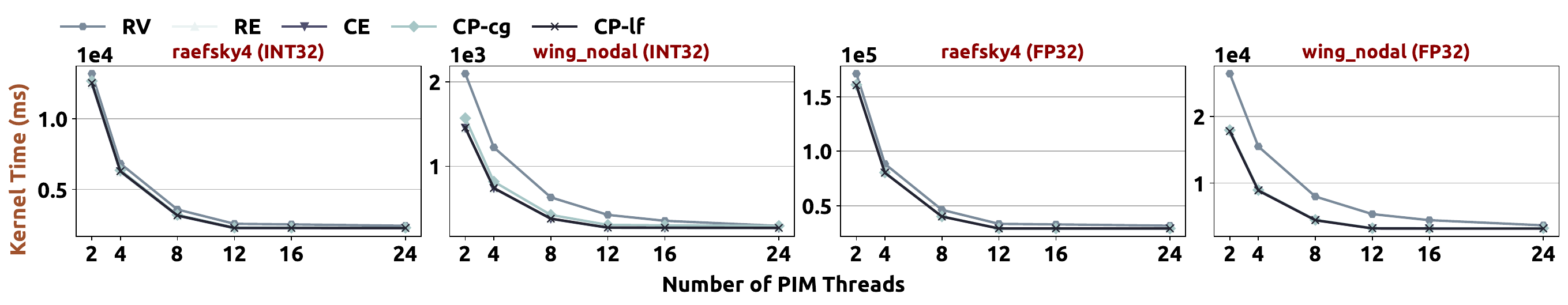}
    \vspace{-8pt}
    \caption{Scalability of all schemes with of a PIM core in  int32 (left) and fp32 (right) data types, as the number of threads of a PIM core increases.}
    \label{fig:1-dpu}
\end{figure}

We evaluate  \spmm  for int32 and fp32  data types with multiple threads of a PIM core. Fig.~\ref{fig:1-dpu} shows scalability of CSR when equally balancing the vertices (\textbf{RV}) or edges at vertex granularity (\textbf{RE}) across threads, and COO when equally balancing the edges at vertex granularity (\textbf{CE}) or via near-perfect edge balance using the coarse-grained locking (\textbf{CP-cg}) or lock-free (\textbf{CP-lf}) schemes.

We draw three findings. First, all schemes scale up to 16 threads, because the PIM core pipeline is fully utilized after 16 threads. 
In wing\_nodal with 16 threads, only one thread processes many more edges than the rest, thus RV slightly scales to 24 threads (by 0.09\%), because it 
exhibits better compute balance across threads.  Second, int32 data type provides at least one order of magnitude better performance than fp32 data type. The UPMEM PIM core does not support in hardware floating-point operations, while they are software emulated using integer arithmetic units. Thus, fp32 \spmm achieves much lower performance than int32 \spmm  due to the excessive amount of computations. Third, RV provides worse performance than other schemes, since balancing the vertices across threads incurs high edge (non-zero element) imbalance, thus causing high disparity in the amount of computations performed across threads (high compute imbalance).  

\begin{tcolorbox}
\noindent\textbf{Recommendation 1:} \\
PIM cores typically have low compute capabilities, thus we recommend programmers to design algorithms that minimize the amount of computations performed 
and support parallelization schemes that enable high compute balance across threads.
\end{tcolorbox}

\begin{tcolorbox}
\noindent\textbf{Recommendation 2:} \\
Programmers can leverage quantization in ML models, if PIM cores have limited precision and arithmetic operation support in hardware, and can design quantized data types that enable low compute requirements (e.g., replacing multiplications with logical/shift/add operations).
\end{tcolorbox}


\subsection{Within PIM Cluster Analysis}

\begin{figure}[h]
   \vspace{2pt}
    \centering
    \includegraphics[width=\linewidth]{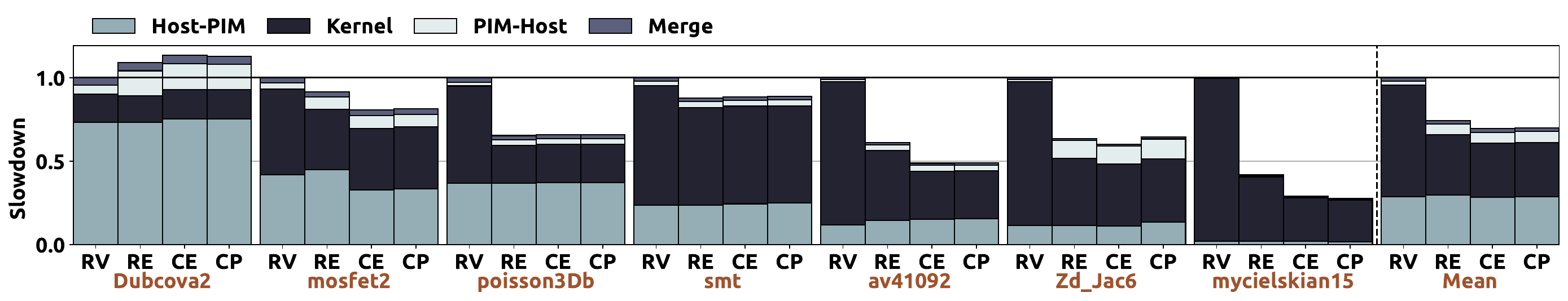}
    \vspace{-8pt}
    \caption{Comparison of various schemes using one PIM cluster of 64 PIM cores and various sparse matrices.}
    \label{fig:1-rank}
    \vspace{-2pt}
\end{figure}

Fig.~\ref{fig:1-rank} evaluates \spmm in one PIM cluster of 64 cores with int32 data type for CSR, when equally balancing the vertices (\textbf{RV}) or edges at vertex granularity (\textbf{RE}) across PIM cores, and for COO when equally balancing the edges at vertex granularity (\textbf{CE}) or via near-perfect edge balance (\textbf{CP}). Within each PIM core, we use 16 threads with edge-balance across threads in CSR and near-perfect edge balance across threads (lock-free synchronization) in COO. We present the execution time of  the
breakdown steps of Fig.~\ref{fig:aggr-pim-execution}, and sort matrices with increasing irregularity, i.e., standard deviation of non-zero elements among rows.

We draw three findings. First, the vertex-balance scheme (RV) incurs higher kernel time than edge-balance schemes (RE, CE, CP) by 1.96$\times$, because the latter provide high compute balance, i.e., similar number of edges (non-zeros) are processed across cores. Second, edge-balance schemes incur higher PIM-Host data transfer costs over vertex-balance by 2.63$\times$. In UPMEM PIM, PIM-Host data transfers can be performed in parallel across multiple cores, if the transfer sizes from all DRAM banks are the \emph{same}. To leverage parallel data transfers, we perform padding with empty bytes (zeros) at the granularity of a PIM device, when transferring data from/to Host. Edge-balance schemes have higher disparity in the number of vertices assigned to PIM cores, i.e., PIM cores produce different amount of partial results for the output matrix, thus they suffer from higher zero padding costs in PIM-Host data transfers. 
Based on first and second findings, we observe that if a single parallelization scheme, e.g., only vertex- or edge-parallelism, is used across all available PIM cores in the system (thousands of PIM cores), performance would be sub-optimal, since it would cause either high kernel (e.g., RV) or high data transfer time (e.g., RE, CE, CP). This is key experimental observation that inspired our PaF approach: PaF enables multiple parallelization strategies to trade off computation and data transfer costs in PIM executions. Third, most matrices have a power-law distribution~\cite{tang2015optimizing}, i.e., only a few vertices have a very large number of neighbors (edges), thus edge-balance schemes provide best end-to-end performance by significantly improving kernel time. Dubcova2 is a relatively regular matrix, thus the vertex-balance scheme provides enough compute balance across PIM cores, achieving 1.11$\times$ better total performance than edge-balance schemes.

\begin{tcolorbox}
\noindent\textbf{Recommendation 3:} \\
Commodity DRAM has multiple hierarchy levels (e.g., DIMM-, rank/layer-, bank group-, bank-level), each level has different characteristics in circuitry design. PIM architects 
can enable different hardware optimizations or accelerator cores at each level of hierarchy (e.g., adding processing capabilities both before and after the sense amplifiers of memory arrays). Then, system and software engineers can design different optimization techniques (e.g., different parallelization strategies) for \emph{each} different level of hardware hierarchy (similar to our proposed PaF approach) to enable high performance in PIM executions via hardware software co-design. 
\end{tcolorbox}

\begin{tcolorbox}
\noindent\textbf{Recommendation 4:} \\
Data transfers to/from PIM memory are typically expensive (since they are performed via the common memory bus), and are on the critical path in hybrid Host-PIM executions. Thus, hardware architects and system engineers can explore mechanisms for PIM systems that (i) overlap data transfers to/from PIM memory with computation on PIM cores, and (ii) minimize the zero padding amount needed in parallel Host-PIM/PIM-Host data transfers.
\end{tcolorbox}



\subsection{Across PIM Cluster Analysis}

\begin{figure}[h]
    \vspace{2pt}
    \centering
    \includegraphics[width=\linewidth]{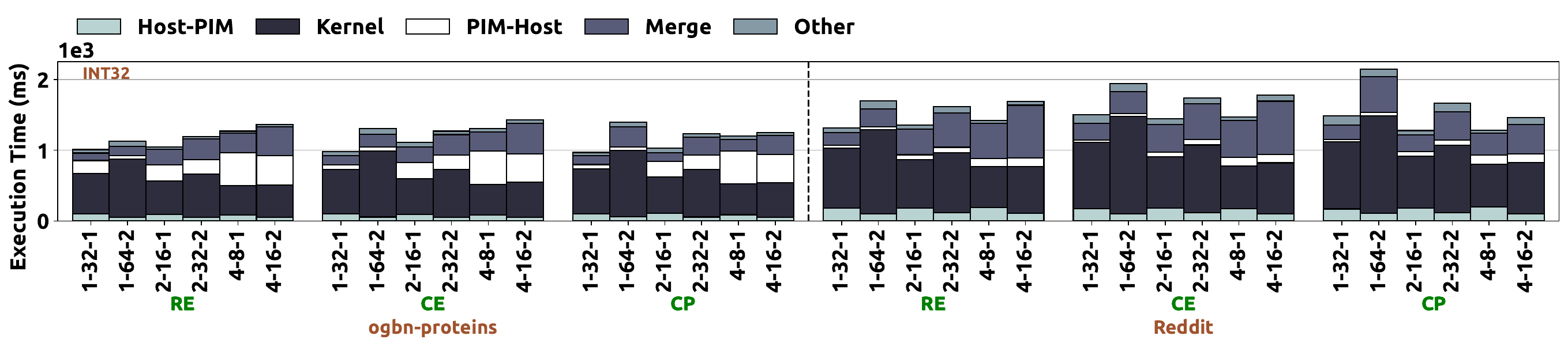}
    \caption{Performance of edge-balance schemes varying the number of sparse, the number of dense partitions, and the number ofPIM clusters per device.}
    \label{fig:multi-rank}
\end{figure}

We evaluate \spmm  using multiple PIM clusters and within cluster we select edge-balance schemes to minimize kernel time. Fig.~\ref{fig:multi-rank} presents the performance using real-world graphs, 128 hidden size, 
32 PIM devices, 
and a \emph{fixed} number of 1992 PIM cores, 
while varying the parallelization scheme used: each triple of values shows the number of sparse partitions, the number of dense partitions and the number of PIM clusters per PIM device, respectively. We show breakdown steps of Fig.~\ref{fig:aggr-pim-execution}, and the stacked bar ``Other'' corresponds to the time needed to partition the dense matrix.

We note four key points. First, having 2 PIM clusters per device with $\sim$28 cores per cluster increases the kernel time by 1.30$\times$ on average over having 1 cluster per device of $\sim$56 cores. Using a smaller number of cores per cluster results in higher compute costs, since \emph{each} PIM core processes a larger number of edges (non-zeros), executing many more computations. 
Second, creating a larger number of sparse partitions (e.g., 4 or larger) typically increases the PIM-Host data transfer and merge overheads, since PIM clusters create more partial results for the output matrix. 
Third, our proposed PaF strategy effectively provides low data transfer costs to/from PIM memory modules. We observe that in the best-performing configurations (e.g., 1-32-1 for ogbn-proteins dataset), the Host-PIM and PIM-Host data transfers account for $\sim$14\% of the total time, while most of the time ($\sim$67\%) is the actual sparse matrix matrix multiplication. 
Fourth, we identify two patterns: in ogbn-proteins, best performance is achieved using CP with 1 sparse partition, while in Reddit, best performance is achieved using CP with 2 sparse partitions (1.11$\times$ better over having 1 sparse partition). In ogbn-proteins, there is a high disparity in the number of vertices assigned to PIM cores, causing a large amount of zero padding in PIM-Host transfers. 
When increasing the sparse partitions from 1 to 2, the vertex disparity and amount of zero padding increase, thus incurring worse performance. 
Thus, we find  the graph's characteristics affect the best-performing parallelization strategy.
Our analysis shows that tuning mechanisms and ML compilers 
need to be developed to optimize performance of sparse workloads in  PIM systems based on the characteristics of each particular input.

\begin{tcolorbox}
\noindent\textbf{Recommendation 5:} \\
System performance of sparse workloads in PIM systems highly depends on the particular patterns of each input given. Therefore, software and system engineers can deploy intelligent heuristics, prediction and automation tools (similar to our proposed tuner) that tune the optimization strategies of sparse workloads at each particular given input at low cost, such that to provide high system performance on real PIM systems. 
\end{tcolorbox}

\begin{figure}[h]
    \vspace{3pt}
    \centering
    \includegraphics[width=\linewidth]{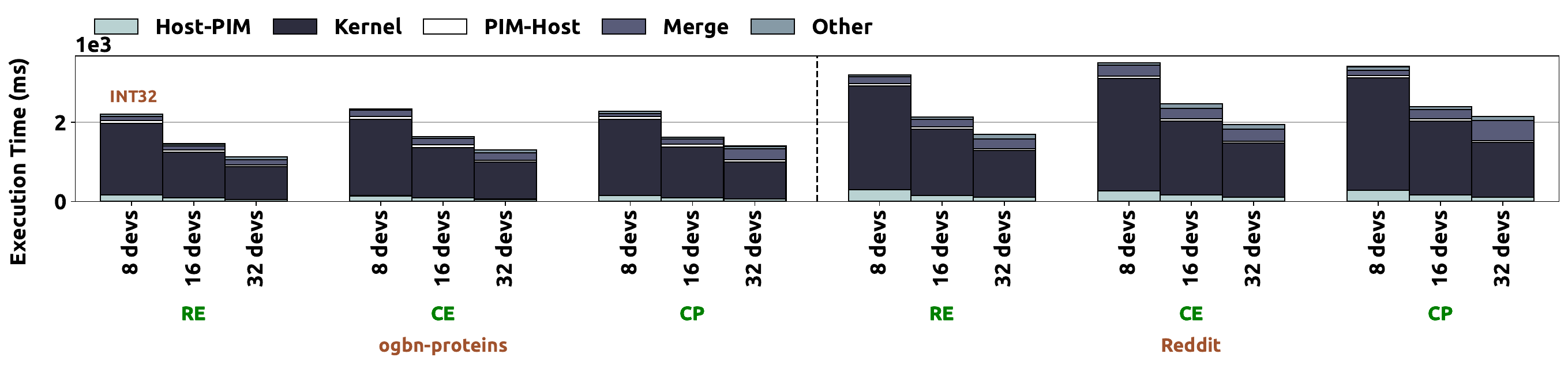}
    \caption{Scalability of edge-balance schemes, as the number of PIM devices (PIM cores) increases.}
    \label{fig:scalability}
    \vspace{-4pt}
\end{figure}

\noindent\textbf{Scalability of \myName  Implementations.} Fig.~\ref{fig:scalability} presents the scalability of \myNameN's edge-balance schemes, i.e., RE, CE and CP, using int32 data type  in \spmmN. In these experiments, we have 1 sparse partition, 128 hidden size  and 2 PIM clusters per PIM device, and increase the number of PIM devices: we evaluate 8, 16, and 32 PIM devices, i.e., the number of PIM cores increases from 456 up to 1992 (each PIM device has $\sim$56 PIM cores). We find that all \myNameN's edge-balance schemes scale well: when we double the number of PIM devices used (double the PIM cores used), the kernel time and total performance improve by on average 1.47$\times$ and 1.38$\times$, respectively. Thus, we conclude that \myName is a scalable \gnn library for real PIM systems with a very large number of PIM cores and PIM devices.

\subsection{\myName Tuner Efficiency}

Fig.~\ref{fig:autotuner_csr} evaluates the \myName tuner efficiency for CSR (See Fig.~\ref{fig:autotuner_coo} in Appendix for COO) by comparing the performance slowdown achieved by its predicted aggregation configuration (predicted) versus an oracle prediction using various datasets and hidden sizes. For the oracle prediction performance, we exhaustively collect the execution times of all possible configurations, and we present in Fig.~\ref{fig:autotuner_csr} the best-performing execution time among them (oracle).
The predicted aggregation configuration by the tuner achieves similar performance with the oracle configuration, being only 0.72\% and 1\% worse on average across all datasets and hidden sizes for CSR and COO, respectively.
Thus, \myName tuner effectively tunes the aggregation configuration in \gnn executions, eliminating the programmer's intervention and providing high performance.

\begin{figure}[h]
    \centering
    \includegraphics[width=\linewidth]{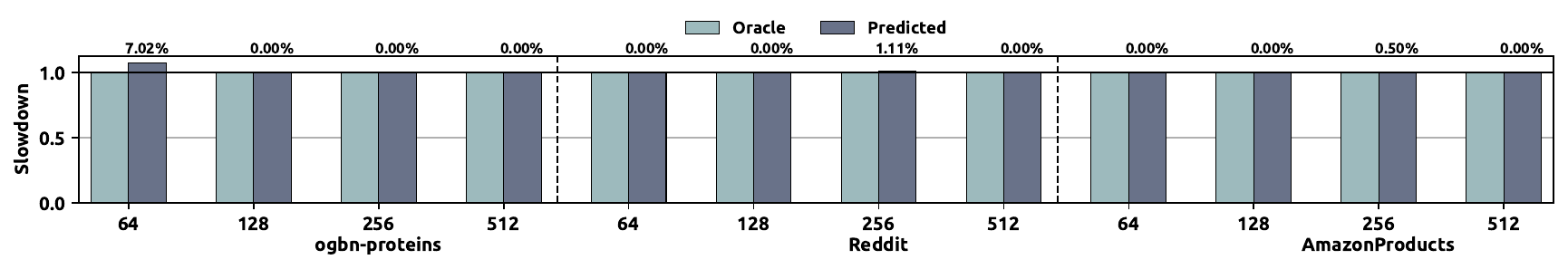}
    \vspace{-12pt}
    \caption{Slowdown of the predicted CSR aggregation configuration by tuner over oracle prediction.}
    \label{fig:autotuner_csr}
    \vspace{-8pt}
\end{figure}

\subsection{\gnn Aggregation Performance}\label{eval:aggregation-performance}

\begin{figure}[h]
    \vspace{2pt}
    \centering
    \includegraphics[width=\linewidth]{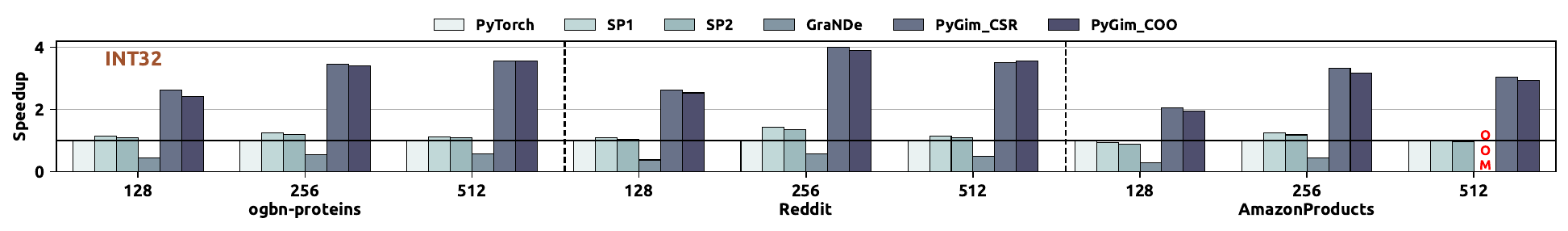}
    \caption{Performance of all comparison points in one aggregation, using various graphs and hidden sizes.}
    \label{fig:eval-aggregation}
\end{figure}

Fig.~\ref{fig:eval-aggregation} shows the performance of all comparison points described in ~\cref{sec:methodology}, in one aggregation operator using real-world graph datasets and common hidden sizes in feature matrix (x-axis). In PIM executions, we use 32 PIM devices ($\sim$56 cores per cluster). In \myNameN, we evaluate both CSR and COO schemes  and we enable the tuner to set the aggregation configuration. Please see Appendix~\ref{sec:append-energy-aggr} (Fig.~\ref{fig:eval-aggr-energy}) for energy consumption evaluation in \gnn aggregation.

We draw fourth findings. First, GraNDe's~\cite{yun2023grande} optimized scheme for simulated near-rank PIM systems achieves very low performance in real near-bank PIM systems, being up to 0.57$\times$ of \CPUName implementation. \myName provides significant performance benefits over GraNDe, because the GraNDe's parallelization strategy is tailored for near-rank PIM systems, rather than real near-bank PIM systems. Second, SP1 and SP2 achieve small speedups over \CPUName (on average 1.13$\times$), since they are optimized for SpMV kernel. Instead, aggregation by its nature performs \spmmN, thus \myName schemes provide significant performance speedups, on average \aggravgcpuspeedup$\times$ and up to \aggrmaxcpuspeedup$\times$ over \CPUName implementation. \myName outperforms prior PIM-based schemes (SP1, SP2, GraNDe) by on average \aggravgpimspeedup$\times$ and up to \aggrmaxpimspeedup$\times$. Third, we find that the best-performing \myName scheme selected by the tuner varies across datasets due to different connection characteristics between vertices of the graph.
Fourth, in Fig.~\ref{fig:eval-aggr-energy} (Appendix~\ref{sec:append-energy-aggr}), we show that \myName provides higher energy efficiency by on average 4.08$\times$ and 1.39$\times$ over prior PIM-based schemes (SP1, SP2, GraNDe)  and \CPUNameN, respectively.
Overall, \myName provides high performance and energy efficiency benefits in \gnn aggregation, significantly outperforming prior existing schemes across various graph datasets and hidden sizes.

\subsection{End-to-End \gnn Inference}\label{eval:inference-performance}

\begin{figure}[h]
    \centering
    \includegraphics[width=\linewidth]{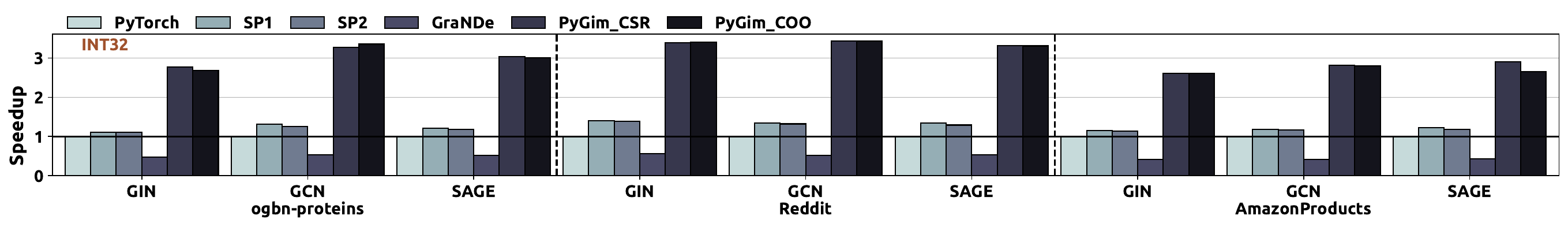}
    \caption{Performance of all comparison points in \gnn inference, using various graph datasets and models.}
    \label{fig:eval-full-infer}
\end{figure}

\noindent \textbf{Performance.} Fig.~\ref{fig:eval-full-infer} evaluates all comparison points in  \gnn inference using int32 data type and various graph datasets. We evaluate 3 \gnn models, each model has 3 layers of 256 hidden size. In PIM executions, we use 32 PIM devices, having in total 1992 cores. In \myNameN, we evaluate both CSR and COO schemes and enable the tuner to set the aggregation configuration.  All comparison points of  Fig.~\ref{fig:eval-full-infer} produce correct output data values and provide the same accuracy, as presented in Appendix~\ref{sec:append-infer-acc}. Note that \myName can be also used to execute \gnn training. Please also see Appendix~\ref{sec:append-training} for \gnn training results.

From Fig.~\ref{fig:eval-full-infer}, we find that \myName schemes provide significant performance speedups over  \CPUName  running on Host by \totalavgcpuspeedup$\times$ (up to \totalmaxcpuspeedup$\times$). \myName outperforms prior state-of-the-art PIM schemes, being 2.46$\times$ (up to 2.68$\times$) better compared to SP1 and SP2, and  6.3$\times$ (up to 7.2$\times$) better over GraNDe. 
Note that the UPMEM PIM core does not include a complete 32 $\times$ 32-bit multiplier to  efficiently support int32 data type in hardware: multiplications of 32-bit operands are implemented using bit shifting and addition and take $\sim$32 cycles. 
Thus, in Appendix~\ref{sec:append-endToEnd}, we also evaluate end-to-end \gnn inference using int8 and int16 data types, in which the arithmetic operations are natively supported by PIM hardware, as well as using the fp32 data type, in which the arithmetic operations are software emulated. 
PIM \gnn execution achieves low performance with fp32 values, since their arithmetic operations are software emulated in the UPMEM PIM hardware. However, ML-oriented PIM systems~\cite{Lee2021Hardware,Hadi2016Chameleon} are expected to be in the market, and 
natively support higher precision data types. Instead, Figs.~\ref{fig:eval-full-int8}, ~\ref{fig:eval-full-int16} in Appendix evaluate int8 and int16 values (their multiplications are natively supported by PIM hardware), and show that \myName provides superior speedups: \myName outperforms prior state-of-the-art PIM approaches by 3.59$\times$ (up to 9.89$\times$) and 3.60$\times$ (up to 8.90$\times$) for int8 and int16, respectively, and outperforms \CPUName running on Host  by on average 4.49$\times$ (up to 5.54$\times$) and 4.03$\times$ (up to 4.63$\times$) for int8 and int16, respectively. We conclude that \myName provides significant performance benefits in \gnn inference over prior approaches.

\begin{figure}[h]
    \centering
    \includegraphics[width=\linewidth]{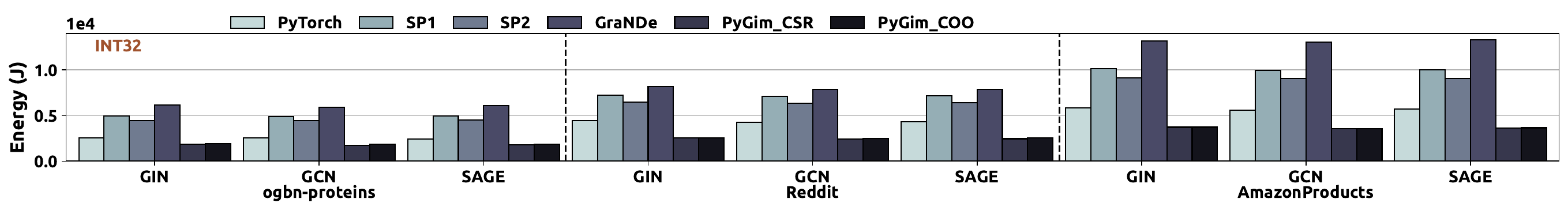}
    \caption{Energy consumption of all comparison points in the end-to-end inference, using various models.}
    \label{fig:eval-full-infer-energy}
\end{figure}

\noindent \textbf{Energy Consumption.} Fig.~\ref{fig:eval-full-infer-energy} presents the energy consumption (in Joules) of all comparison points in end-to-end \gnn inference using int32 data type, and various \gnn models and datasets. In PIM executions, we use 32 PIM devices, having in total 1992 cores, and \myNameN's tuner is enabled. 
We use Intel RAPL~\cite{rapl} to measure energy in CPU execution parts, which are (i) the whole \CPUName  scheme, and (ii) in PIM schemes, the combination operator as well as the load, retrieve, and merge steps of the aggregation operator. For the kernel step of aggregation, we measure the energy consumed in PIM-enabled chips using the methodology described in a recent paper~\cite{falevoz2023energy} written by the UPMEM PIM manufacturer: the power of each UPMEM PIM DIMM is 23.22W, thus the total energy of kernel time is conservatively calculated as the $kernel\_time$ $\times$ $\#PIM\_DIMMs$ $\times$ $power$.

We draw three findings. First, \myName provides significant energy benefits on average \totalavgpimenergyspeedup$\times$ (up to \totalmaxpimenergyspeedup$\times$) over prior PIM schemes.
Second, \myName improves energy efficiency over \CPUName scheme by on average \totalavgcpuenergyspeedup$\times$ (up to \totalmaxcpuenergyspeedup$\times$). 
Although \myName provides \totalavgcpuspeedup$\times$ better performance over \CPUNameN, it provides \totalavgcpuenergyspeedup$\times$ better energy efficiency, because the manufacturing process of PIM chips is still in an early stage. For example, UPMEM PIM chips have been manufactured with a larger technology node, i.e., at least 20nm, than the 14nm technology node used for CPU hardware (See Table~\ref{tab:system-characteristics}).
Future real PIM products could advance the technology node to become more energy-efficient.
Third, in aggregation, SP1 and SP2 schemes have a time-consuming kernel step executed 
on PIM cores. GraNDe has a more time-consuming merge step executed 
on Host. Thus, although SP1 and SP2 improve performance over GraNDe by on average 
2.56$\times$, they are on average 
1.25$\times$ better than than GraNDe in energy efficiency.

\vspace{-10pt}
\subsection{Evaluation of \gnn Aggregation in GPU Systems}\label{eval:gpu-comparison}

We propose a \gnn library for real near-bank PIM systems, and evaluate it over prior software parallelization schemes/libraries running on the same computing system, the UPMEM PIM server. 
To judge and compare different libraries tailored for different types of computing systems, we present the resource utilization as a representative metric: 
resource utilization measures how well the software maps and uses the available capabilities of the underlying hardware.
With this metric, we compare 
how well \myName library maps to the evaluated PIM system versus how well the PyTorch's  backend libraries map to GPU systems.
Resource utilization is defined as the number of operations performed divided by the execution time, and normalized as a percentage of the theoretical peak performance of the system. 
We focus our evaluations in \gnn aggregation operator, because \gnn combination uses implementations from existing ML frameworks (e.g., PyTorch) executed on Host side, and thus characterizing the efficiency of existing implementations 
in various Host systems is out of the scope of our work.
In \gnn aggregation, for fairness among all implementations, we calculate the number of operations performed as $edges \times hidden\_size$, which is the theoretical arithmetic operations of \spmmN.
Table~\ref{tab:system-characteristics} shows the characteristics of various computing systems. 
For peak performance and memory bandwidth, we use peakperf~\cite{peak-perf} and stream~\cite{stream}  in CPUs/GPUs, and the microbenchmarks from open-source works~\cite{Gomez2022Benchmarking,giannoula2022sparsep} for UPMEM PIM system.

\begin{table}[H]
\begin{center}
\centering
\resizebox{\linewidth}{!}{
\begin{tabular}{|l||c|c|c|c|c|c|c|c|}
    \hline
    \cellcolor{gray!15} & \cellcolor{gray!15} & \cellcolor{gray!15} & \cellcolor{gray!15}\raisebox{-0.20\height}{\textbf{INT32 Peak}} & 
    \cellcolor{gray!15}\raisebox{-0.20\height}{\textbf{FP32 Peak}} & 
   \cellcolor{gray!15}\raisebox{-0.20\height}{\textbf{Memory}} & \cellcolor{gray!15}\raisebox{-0.20\height}{\textbf{Total}} &
   \cellcolor{gray!15}\raisebox{-0.20\height}{\textbf{Technology}}\\
     \multirow{-2}{*}{\cellcolor{gray!15}\textbf{System}} & \multirow{-2}{*}{\cellcolor{gray!15}\textbf{Total Cores}} &  \multirow{-2}{*}{\cellcolor{gray!15}\textbf{Freq.}} & \cellcolor{gray!15}\textbf{Performance}
     & \cellcolor{gray!15}\textbf{Performance}
     & \cellcolor{gray!15}\textbf{Capacity} & \cellcolor{gray!15}\textbf{Bandwidth}  
     & \cellcolor{gray!15}\textbf{Node} 
      \\
    \hline \hline
    CPU Intel Xeon 4215 & 2x8 x86 cores& 2.5 GHz & 0.64 TOPS  & 1.28 TFLOPS & 128 GB & 23.1 GB/s & 14nm  \\ \hline 
    UPMEM PIM  & 1992 PIM cores & 350 MHz & 115.93 GOPS &  24.85 GFLOPS & 124.5 GB & 1.39 TB/s & at least 20nm\\ \hline 
    GPU GTX 1080 Ti & 3584 CUDA cores & 1.48 GHz & 13.25 TOPS & 13.25 TFLOPS   & 11 GB & 359.9 GB/s  & 16nm \\ \hline 
    GPU RTX 2080 Ti & 4352 CUDA cores & 1.35 GHz & 16.94 TOPS & 16.94 TFLOPS   & 11 GB & 558.1 GB/s  & 12nm  \\ \hline 
    GPU RTX 3090 & 10496 CUDA cores & 1.40 GHz & 17.79 TOPS & 35.58 TFLOPS   & 24 GB & 936.2 GB/s  & 8nm  \\ \hline 

\end{tabular}
}
\end{center}
\caption{Characteristics of CPU, PIM and GPU systems. 
}
\label{tab:system-characteristics}
\vspace{-2pt}
\end{table}

Table~\ref{tab:utilization-aggregation} shows the hardware utilization achieved by different libraries executed on the corresponding computing system in one aggregation operator using int32 and fp32 data types, various datasets, and with 256 hidden size. 
For CPU and GPU systems, we evaluate PyTorch's pytorch\_sparse library~\cite{Fey2019Fast}, which
employs \spmm implementations from Intel MKL library for CPUs, and optimized CUDA implementations for GPUs.
For the UPMEM PIM system, we evaluate \myName and
account for all breakdown steps of Fig.~\ref{fig:aggr-pim-execution}.

\begin{table}[h]
\vspace{2pt}
\begin{center}
\centering
\resizebox{\linewidth}{!}{
\begin{tabular}{|l||c|c|c|c|c|c|}
    \hline
    \cellcolor{gray!15}\raisebox{-0.20\height}{\textbf{Dataset and data type / }}  &  \cellcolor{gray!15}\raisebox{-0.20\height}{\textbf{OGBN}}  
    & \cellcolor{gray!15}\raisebox{-0.20\height}{\textbf{RDT}} 
    &  \cellcolor{gray!15}\raisebox{-0.20\height}{\textbf{AMZ}} 
    & \cellcolor{gray!15}\raisebox{-0.20\height}{\textbf{OGBN}} 
    &\cellcolor{gray!15}\raisebox{-0.20\height}{\textbf{RDT}} & \cellcolor{gray!15}\raisebox{-0.20\height}{\textbf{AMZ}}  \\
     \cellcolor{gray!15}\textbf{Software library}  & \cellcolor{gray!15}\textbf{INT32}
     & \cellcolor{gray!15}\textbf{INT32}
     & \cellcolor{gray!15}\textbf{INT32}
     & \cellcolor{gray!15}\textbf{FP32}
     & \cellcolor{gray!15}\textbf{FP32} & \cellcolor{gray!15}\textbf{FP32} \\
    \hline \hline
    pytorch\_sparse - Intel MKL (CPU Intel Xeon 4215)  & 0.74\% &  0.63\%   & 0.67\%  & 0.26\%  & 0.22\% & 0.20\% \\ \hline 
    pytorch\_sparse - CUDA (GPU GTX 1080 Ti)  & 2.15\% & 0.62\% & 0.71\%  & 2.02\% & 0.62\%  & 0.71\%  \\ \hline   
    pytorch\_sparse - CUDA (GPU RTX 2080 Ti)  & 1.45\% & 0.68\% & 0.71\%  & 1.45\% & 0.67\%  & 0.71\%  \\ \hline 
    pytorch\_sparse - CUDA (GPU RTX 3090)  & 3.03\% & 1.56\% & 1.32\%    & 1.58\% & 0.78\%  & 0.67\%   \\ \hline 
    \textbf{\myName (UPMEM PIM)} & \textbf{14.09\%} & \textbf{13.86\%} & \textbf{12.32\%} & \textbf{8.21\%}    & \textbf{9.13\%} & \textbf{8.84\%} \\ \hline 
\end{tabular}
}
\end{center}
\vspace{2pt}
\caption{Resource utilization in various systems for \gnn aggregation with 256 hidden size, the ogbn-proteins (\textbf{OGBN}), Reddit (\textbf{RDT}), and AmazonProducts (\textbf{AMZ}) datasets, and INT32 and FP32 data types.}
\label{tab:utilization-aggregation}
\end{table}

We make two observations. First, \myName achieves 12.9$\times$, 13.2$\times$, and 8.8$\times$ larger utilization on the UPMEM PIM system than that of the pytorch\_sparse CUDA library on GTX 1080 Ti, RTX 2080 Ti and RTX 3090 GPUs, respectively, and provides 29.4$\times$ larger utilization on the UPMEM PIM than that of pytorch\_sparse Intel MKL library on Intel Xeon CPU. Thus, \myName uses the PIM system more effectively than PyTorch's optimized backend libraries use the CPU and GPU systems. 
Second, across three GPU generations (Table~\ref{tab:system-characteristics}), GPU architects advance the technology node, and increase the number of cores and the available memory bandwidth.
However, the resource utilization in the memory-intensive \gnn aggregation still remains low in all GPUs. Comparing  RTX 2080 Ti over GTX 1080 Ti, both the compute throughput and the memory bandwidth increased by 1.30$\times$ and  1.55$\times$, respectively, however resource utilization in aggregation is similar: on average 0.95\% for RTX 2080  Ti  and 1.13\% for GTX 1080 Ti. 
Comparing RTX 3090 over RTX 2080 Ti, the fp32 compute throughput and memory bandwidth increased by 2.1$\times$ and 1.68$\times$, respectively, however  resource utilization in fp32 aggregation remains similarly low: on average 1.01\% for RTX 3090 and 0.94\% for RTX 2080 Ti. 
Resource utilization in int32 aggregation  is $\sim$2$\times$ compared that in fp32 aggregation in RTX 3090, since the int32 compute throughput from RTX 2080 Ti to RTX 3090 is similar (1.05$\times$), while memory  bandwidth increases by 1.68$\times$.
These observations show that GPU architectures are not necessarily evolving to better support memory-intensive workloads, such as \gnn aggregation.
Overall, we conclude that \myName running \gnn aggregation on real PIM systems provides a more cost-effective solution than \CPUName running \gnn on Host systems.




Finally, we also report the performance (seconds) and energy consumption (Joules) metrics to show the readers how much absolute performance and energy efficiency the evaluated UPMEM PIM system achieves on \gnn aggregation over commodity GPU systems.
UPMEM PIM with \myName library  is worse than GPUs with pytorch\_sparse library by 9.9$\times$, 10.5$\times$ and 25.8$\times$ in performance and by 5.0$\times$, 5.3$\times$ and 11.6$\times$  energy consumption over 1080Ti, 2080Ti, and 3090 GPUs, respectively, for int32 data type. 
In int8 data type (natively supported in UPMEM  hardware), UPMEM PIM with \myName provides better performance and energy efficiency compared to that provided in int32 data type: e.g., in Reddit data set with 256 hidden size, it is 2.5$\times$, 4.0$\times$, 9.5$\times$ worse in performance and 2.4$\times$, 2.9$\times$, 5.6$\times$ worse in energy consumption than 1080Ti, 2080Ti, and 3090 GPU with pytorch\_sparse library, respectively\footnote{In same aggregation configuration, \myName on UPMEM PIM improves performance and energy efficiency by 2.3$\times$ and 1.3$\times$ respectively, over an Intel Xeon 4314 CPU (10nm technology) with 2x16 cores (2.4GHz) and 227 GB/s memory bandwidth.}.

In Appendix~\ref{sec:appendix-gpu-comparison}, we present the detailed evaluation results over GPUs in \gnn aggregation and end-to-end \gnn inference.
Please note that these results are provided for completeness and \emph{not} for competition purposes.
Directly comparing this UPMEM PIM system  
over GPU systems is not a fair comparison. The evaluated UPMEM PIM system is available on the market only $\sim$1.5 years, it is in its first generation manufactured with a large technology node of at least 20nm, and is not yet well optimized for multiplication operations~\cite{Gomez2023Evaluating,hyun2024pathfinding}. Instead, GPU systems have been optimized for 15$+$ years, especially in multiplication operations, and there has been invested large financial budgets from major technology industry leaders to improve GPU architecture across its generations. 
Moreover, although our \emph{goal} in this work is to quantify the potential of a \emph{real} PIM system in \gnn executions, our proposed PaF optimizations cover near-bank PIM systems (See the described characteristics in §\ref{sec:background_pim}), and thus could be evaluated on other 
near-bank PIM systems with potentially better computation capabilities and energy efficiency than the evaluated UPMEM PIM system. 
UPMEM~\cite{upmem} has already announced a second generation product (expected to be released in the market), where PIM-enabled DIMMs are integrated with a more powerful CPU server (Ice Lake platform instead of Intel Xeon), the system will support 28 PIM DIMMs with $\sim$1.8$\times$ more cores, i.e., 3584 cores (instead of 16 PIM DIMMs with ~2K cores) and each PIM core will have $\sim$1.7$\times$ higher frequency, i.e., 600MHz core frequency (instead of 350MHz). 
The current evaluated UPMEM PIM also uses a relatively large technology node of \textbf{at least 20nm} for manufacturing (GPU RTX2080 Ti uses \textbf{12nm}), and advancing the technology node could improve energy efficiency. 
Additionally, according to~\cite{Li2024PIMDL},
the near-bank HBM-PIM and AiM (both have been prototyped) systems can achieve 1.2 TFLOPS and  1 TFLOPS compute throughput, which is 48.3$\times$ and 40.2$\times$ higher than that of the evaluated UPMEM PIM system.
Hence, running \gnn aggregation with our proposed PaF optimizations in upcoming real PIM systems could potentially lead to better performance and energy efficiency compared to the results reported in the above GPU comparisons.
Finally,  HBM-based PIM memory modules~\cite{lee20221ynm,He2020Newton,Lee2021Hardware} can be integrated into modern GPUs and provide much larger memory bandwidth for the PIM cores than the memory bandwidth available for GPU cores. In such HBM-based PIM systems, when executing \gnn aggregation in PIM cores, we could potentially expect important performance and energy benefits compared to executing \gnn aggregation on the GPU cores.



\section{Related Work}

\label{sec:related_work}

To our knowledge, our work is the first to design an ML library and tuner for \gnn executions on near-bank PIM systems, propose efficient \gnn aggregation schemes tailored for such systems, and extensively characterize \gnns on the first real-world PIM system. We briefly discuss prior work.

\noindent\textbf{PIM-Based Accelerators.} 
A few works~\cite{zhou2022gnnear,tian2022GNMP,yun2023grande,Chen2023MetaNMP} design PIM-based \gnn accelerators. 
Their custom microarchitecture designs for host and PIM cores are orthogonal to \myName software library. These works target \emph{near-rank} PIM architectures and use simulators for their evaluations. \myName targets \emph{near-bank} PIM systems, which typically 
provide much larger memory bandwidth than near-rank PIMs~\cite{lee20221ynm}, and evaluates \gnns on a real system. Finally, implementing  the software data mappings of these works to near-bank PIM systems would cause out-of-memory errors (See §~\ref{sec:background_prior_work}) or would be
inefficient: as shown in our evaluations,
\myName significantly outperforms the best-performing strategy of GraNDe~\cite{yun2023grande}.
Li et al.~\cite{li2023nmexplorer} design a tool to explore PIM design configurations (e.g., near-DIMM, near-rank, near-bank) for different application scenarios, which is orthogonal to our work. A few works~\cite{Lee2022SmartSAGE,Li2021Glist,Park2022Ginex} propose in-storage PIM designs that sample the large graphs inside the disk (SSDs) to reduce the amount of graph data sent from disk to host CPU/GPU cores. These works can work synergistically with ours: \myName can be used to efficiently process the produced smaller graph in DRAM.
Finally, a few  works~\cite{Li2024PIMDL,Park2024AttAcc,Li2024SpecPIM,Heo2024NeuPIM} design custom PIM-based architectures for large language models, however they do not support \gnns and the \spmm kernel.

\noindent\textbf{System Support and Software for PIM Systems.} Prior works~\cite{Gomez2022Benchmarking,giannoula2022sparsep,Diab2023Framework,Gomez2023Evaluating,Lim2023Design,Item2023TransPimLib,Das2022Implementation,Jibril2024Aggregation,chen2023simplepim,Shin2023PIMFlow,rhyner2024analysis,giannoula2022towards} design optimized  
linear algebra, graph processing, database, array iterators, ML training, bioinformatics, and image processing kernels  for PIM systems. The closest work to ours is SparseP~\cite{giannoula2022sparsep}, that is an efficient 
SpMV library for PIM systems. We use the best-performing SparseP kernel in \gnnsN,  show that  it has worse performance and energy efficiency than \myNameN.  
A few works~\cite{Noh2024PIDComm,Tian2024NDPBridge} propose efficient communication collectives for future PIM systems.
Moreover, a few works~\cite{hyun2024pathfinding,Gilbert2024Scalabity} 
discuss scalability issues and hardware limitations of real near-bank PIM systems, 
and propose architectural features for future PIM systems.
These prior works~\cite{hyun2024pathfinding,Gilbert2024Scalabity,Noh2024PIDComm,Tian2024NDPBridge} are orthogonal to \myNameN: \myName can be used in their proposed future PIM designs and/or leverage optimized communication collectives to enable high efficiency in \gnn executions.

\noindent\textbf{\gnns and \spmm in Commodity Systems.} Prior works optimize \gnns and \spmm on CPUs~\cite{Kjolstad2017TACO,Hong2019Adaptive,Akbudak2017Exploiting,yesil2022dense,Gong2022Graphite,Wang2014MKL,van2016blis},  GPUs~\cite{Dalton2015Optimizing,Liu2014Efficient,Niu2022TileSpGEMM,gale2020sparse,yang2018design,Huang2021Understanding,naumov2010cusparse,Kjolstad2017TACO,Ye2023SparseTIR} by leveraging the shared memory model of CPUs/GPUs and deep cache hierarchies (on-chip caches). Their optimizations cannot be applied in PIM systems, that have a distributed memory model and shallow cache hierarchy. Prior works~\cite{Koanantakool2016Communication,Bharadwaj2022Distributed,Tripathy2020Reducing,Md2021DistGNN,Zheng2020DistDGL,Ma2019Neugraph,Jia2020ImprovingTA,Selvitopi2021Distributed,Lin2023HyScale,qu2023tt,liu2023bgl,Charles2024TwoFace} optimize \gnns and \spmm on CPU-GPU, multi-CPU/multi-GPU systems by minimizing communication costs among cores, and/or overlapping computation with communication. However, real PIM systems may not support direct communication among PIM cores~\cite{Gilbert2024Scalabity}, and there are no real PIM systems that can overlap computation with communication across on PIM cores.
Thus, well-tuned \gnn and \spmm kernels for distributed processor-centric systems either cannot be directly applied in PIM systems, or their fine-grained inter-PIM-core communication (e.g., implemented over Host) would cause high performance overheads.

\noindent\textbf{Custom Accelerators for \gnns and \spmmN.} Prior works~\cite{auten2020hardware,liang2020engn,kiningham2022grip,li2021gcnax,song2021cambricon,stevens2021gnnerator,srivastava2020matraptor,pal2018outerspace,song2022sextans,kanellopoulos2019smash,gerogiannis2023spade,baek2021innersp} propose custom hardware accelerators for \gnns and \spmmN, but they target processor-centric systems with low available memory bandwidth. 
Instead, \myName provides software-level optimizations for \gnnsN, and targets 
memory-centric PIM systems.

\section{Conclusion}

We propose \myNameN, an efficient ML library for \gnns executions in PIM systems, and conduct a comprehensive characterization study of \gnns on a real-world PIM system. We design a hybrid \gnn execution on processor- and memory-centric computing 
systems,  intelligent parallelization techniques  for \gnn aggregation in near-bank PIM systems and a lightweight tuner to enable programming ease in \gnn deployment for PIM systems. In \gnn inference, \myName achieves \totalavgcpuspeedup$\times$ (up to \totalmaxcpuspeedup$\times$)  
and \totalavgpimspeedup$\times$ (up to \totalmaxpimspeedup$\times$) 
speedup over the state-of-the-art \CPUName and PIM-based schemes, respectively, and \totalavgcpuenergyspeedup $\times$  (up to \totalmaxcpuenergyspeedup$\times$) 
and \totalavgpimenergyspeedup $\times$ (up to \totalmaxpimenergyspeedup$\times$) 
higher energy efficiency than \CPUName and PIM-based schemes, respectively. In \gnn aggregation, \myName provides  on average 
\aggravggpuspeeduputilization $\times$ higher resource utilization in PIM system than that of the PyTorch CUDA library in GPUs.
We hope that our  parallelization strategies for \gnnsN, 
in-depth PIM analysis, and open-source library will enable further research on optimizing \gnns and other sparse ML models in memory-centric computing systems.

\begin{acks}
We thank UPMEM for generously providing hardware resources to perform this research. We thank the anonymous reviewers of SIGMETRICS 2025, and our shepherd, Bo Jiang, for their comments and suggestions. We thank Andreas Moshovos for valuable feedback and Bojian Zheng for technical support. We thank the SAFARI Research Group and EcoSystem members for providing a stimulating intellectual environment. Ivan Fernandez is partially supported by the Spanish Ministry of Science and Innovation (projects PID2019-107255GB-C21 and PID2019-107255GB-C22).
We acknowledge the generous gifts from our industrial partners, including Google, Huawei, Intel, Microsoft and AWS. This work is supported in part by
the Semiconductor Research Corporation (SRC), the ETH Future Computing Laboratory (EFCL), the European Union’s Horizon program for research and innovation [101047160 - BioPIM], and the AI Chip Center for Emerging Smart Systems, sponsored by InnoHK funding, Hong Kong SAR (ACCESS).
This paper is also supported in part by Vector Institute Research grants, the Canada Foundation for Innovation JELF grant, NSERC Discovery grant, AWS Machine Learning Research Award, Facebook Faculty Research Award, Google Scholar Research Award, and VMware Early Career Faculty Grant.
The \myName library is publicly available at \url{https://github.com/CMU-SAFARI/PyGim}.

\end{acks}

\newpage

\bibliographystyle{ACM-Reference-Format}
\bibliography{references}


\begin{thebibliography}{149}


\ifx \showCODEN    \undefined \def \showCODEN     #1{\unskip}     \fi
\ifx \showDOI      \undefined \def \showDOI       #1{#1}\fi
\ifx \showISBNx    \undefined \def \showISBNx     #1{\unskip}     \fi
\ifx \showISBNxiii \undefined \def \showISBNxiii  #1{\unskip}     \fi
\ifx \showISSN     \undefined \def \showISSN      #1{\unskip}     \fi
\ifx \showLCCN     \undefined \def \showLCCN      #1{\unskip}     \fi
\ifx \shownote     \undefined \def \shownote      #1{#1}          \fi
\ifx \showarticletitle \undefined \def \showarticletitle #1{#1}   \fi
\ifx \showURL      \undefined \def \showURL       {\relax}        \fi
\providecommand\bibfield[2]{#2}
\providecommand\bibinfo[2]{#2}
\providecommand\natexlab[1]{#1}
\providecommand\showeprint[2][]{arXiv:#2}

\bibitem[\protect\citeauthoryear{Abadi, Agarwal, Barham, Brevdo, Chen, Citro, Corrado, Davis, Dean, Devin, et~al\mbox{.}}{Abadi et~al\mbox{.}}{2016}]%
        {abadi2016tensorflow}
\bibfield{author}{\bibinfo{person}{Mart{\'\i}n Abadi}, \bibinfo{person}{Ashish Agarwal}, \bibinfo{person}{Paul Barham}, \bibinfo{person}{Eugene Brevdo}, \bibinfo{person}{Zhifeng Chen}, \bibinfo{person}{Craig Citro}, \bibinfo{person}{Greg~S Corrado}, \bibinfo{person}{Andy Davis}, \bibinfo{person}{Jeffrey Dean}, \bibinfo{person}{Matthieu Devin}, {et~al\mbox{.}}} \bibinfo{year}{2016}\natexlab{}.
\newblock \showarticletitle{{Tensorflow: Large-Scale Machine Learning on Heterogeneous Distributed Systems}}.
\newblock \bibinfo{journal}{\emph{arXiv}}, \bibinfo{year}{2016}.
\newblock


\bibitem[\protect\citeauthoryear{Ahn, Hong, Yoo, Mutlu, and Choi}{Ahn et~al\mbox{.}}{2015}]%
        {ahn2015scalable}
\bibfield{author}{\bibinfo{person}{Junwhan Ahn}, \bibinfo{person}{Sungpack Hong}, \bibinfo{person}{Sungjoo Yoo}, \bibinfo{person}{Onur Mutlu}, {and} \bibinfo{person}{Kiyoung Choi}} \bibinfo{year}{2015}\natexlab{}.
\newblock \showarticletitle{{A Scalable Processing-In-Memory Accelerator for Parallel Graph Processing}}, In \bibinfo{booktitle}{\emph{ISCA}}.
\newblock


\bibitem[\protect\citeauthoryear{Akbudak and Aykanat}{Akbudak and Aykanat}{2017}]%
        {Akbudak2017Exploiting}
\bibfield{author}{\bibinfo{person}{Kadir Akbudak} {and} \bibinfo{person}{Cevdet Aykanat}} \bibinfo{year}{2017}\natexlab{}.
\newblock \showarticletitle{{Exploiting Locality in Sparse Matrix-Matrix Multiplication on Many-Core Architectures}}.
\newblock \bibinfo{journal}{\emph{TPDS}}, \bibinfo{year}{2017}.
\newblock


\bibitem[\protect\citeauthoryear{Asghari-Moghaddam, Son, Ahn, and Kim}{Asghari-Moghaddam et~al\mbox{.}}{2016}]%
        {Hadi2016Chameleon}
\bibfield{author}{\bibinfo{person}{Hadi Asghari-Moghaddam}, \bibinfo{person}{Young~Hoon Son}, \bibinfo{person}{Jung~Ho Ahn}, {and} \bibinfo{person}{Nam~Sung Kim}} \bibinfo{year}{2016}\natexlab{}.
\newblock \showarticletitle{{Chameleon: Versatile and Practical Near-DRAM Acceleration Architecture for Large Memory Systems}}, In \bibinfo{booktitle}{\emph{MICRO}}.
\newblock


\bibitem[\protect\citeauthoryear{Auten, Tomei, and Kumar}{Auten et~al\mbox{.}}{2020}]%
        {auten2020hardware}
\bibfield{author}{\bibinfo{person}{Adam Auten}, \bibinfo{person}{Matthew Tomei}, {and} \bibinfo{person}{Rakesh Kumar}} \bibinfo{year}{2020}\natexlab{}.
\newblock \showarticletitle{{Hardware Acceleration of Graph Neural Networks}}, In \bibinfo{booktitle}{\emph{DAC}}.
\newblock


\bibitem[\protect\citeauthoryear{Baek, Hwang, Heo, Kim, and Huh}{Baek et~al\mbox{.}}{2021}]%
        {baek2021innersp}
\bibfield{author}{\bibinfo{person}{Daehyeon Baek}, \bibinfo{person}{Soojin Hwang}, \bibinfo{person}{Taekyung Heo}, \bibinfo{person}{Daehoon Kim}, {and} \bibinfo{person}{Jaehyuk Huh}} \bibinfo{year}{2021}\natexlab{}.
\newblock \showarticletitle{{InnerSP: A Memory Efficient Sparse Matrix Multiplication Accelerator With Locality-Aware Inner Product Processing}}, In \bibinfo{booktitle}{\emph{PACT}}.
\newblock


\bibitem[\protect\citeauthoryear{Baghdadi, Merouani, Leghettas, Abdous, Arbaoui, Benatchba, et~al\mbox{.}}{Baghdadi et~al\mbox{.}}{2021}]%
        {baghdadi2021deep}
\bibfield{author}{\bibinfo{person}{Riyadh Baghdadi}, \bibinfo{person}{Massinissa Merouani}, \bibinfo{person}{Mohamed-Hicham Leghettas}, \bibinfo{person}{Kamel Abdous}, \bibinfo{person}{Taha Arbaoui}, \bibinfo{person}{Karima Benatchba}, {et~al\mbox{.}}} \bibinfo{year}{2021}\natexlab{}.
\newblock \showarticletitle{{A Deep Learning Based Cost Model for Automatic Code Optimization}}.
\newblock \bibinfo{journal}{\emph{MLSys}}, \bibinfo{year}{2021}.
\newblock


\bibitem[\protect\citeauthoryear{Bharadwaj, Buluc, and Demmel}{Bharadwaj et~al\mbox{.}}{2022}]%
        {Bharadwaj2022Distributed}
\bibfield{author}{\bibinfo{person}{V. Bharadwaj}, \bibinfo{person}{A. Buluc}, {and} \bibinfo{person}{J. Demmel}} \bibinfo{year}{2022}\natexlab{}.
\newblock \showarticletitle{{Distributed-Memory Sparse Kernels for Machine Learning}}, In \bibinfo{booktitle}{\emph{IPDPS}}.
\newblock


\bibitem[\protect\citeauthoryear{Bishop}{Bishop}{1995}]%
        {bishop1995neural}
\bibfield{author}{\bibinfo{person}{Christopher~M Bishop}} \bibinfo{year}{1995}\natexlab{}.
\newblock \bibinfo{booktitle}{\emph{{Neural Networks for Pattern Recognition}}}.
\newblock \bibinfo{publisher}{Oxford University Press}.
\newblock


\bibitem[\protect\citeauthoryear{Bj{\"o}rck}{Bj{\"o}rck}{1996}]%
        {bjorck1996numerical}
\bibfield{author}{\bibinfo{person}{{\AA}ke Bj{\"o}rck}} \bibinfo{year}{1996}\natexlab{}.
\newblock \showarticletitle{{Numerical Methods for Least Squares Problems}}, In \bibinfo{booktitle}{\emph{SIAM}}.
\newblock


\bibitem[\protect\citeauthoryear{Block, Gerogiannis, Mendis, Azad, and Torrellas}{Block et~al\mbox{.}}{2024}]%
        {Charles2024TwoFace}
\bibfield{author}{\bibinfo{person}{Charles Block}, \bibinfo{person}{Gerasimos Gerogiannis}, \bibinfo{person}{Charith Mendis}, \bibinfo{person}{Ariful Azad}, {and} \bibinfo{person}{Josep Torrellas}} \bibinfo{year}{2024}\natexlab{}.
\newblock \showarticletitle{{Two-Face: Combining Collective and One-Sided Communication for Efficient Distributed SpMM}}, In \bibinfo{booktitle}{\emph{ASPLOS}}.
\newblock


\bibitem[\protect\citeauthoryear{Boroumand, Ghose, Kim, Ausavarungnirun, Shiu, Thakur, Kim, Kuusela, Knies, Ranganathan, and Mutlu}{Boroumand et~al\mbox{.}}{2018}]%
        {Boroumand2018Google}
\bibfield{author}{\bibinfo{person}{Amirali Boroumand}, \bibinfo{person}{Saugata Ghose}, \bibinfo{person}{Youngsok Kim}, \bibinfo{person}{Rachata Ausavarungnirun}, \bibinfo{person}{Eric Shiu}, \bibinfo{person}{Rahul Thakur}, \bibinfo{person}{Daehyun Kim}, \bibinfo{person}{Aki Kuusela}, \bibinfo{person}{Allan Knies}, \bibinfo{person}{Parthasarathy Ranganathan}, {and} \bibinfo{person}{Onur Mutlu}} \bibinfo{year}{2018}\natexlab{}.
\newblock \showarticletitle{{Google Workloads for Consumer Devices: Mitigating Data Movement Bottlenecks}}, In \bibinfo{booktitle}{\emph{ASPLOS}}.
\newblock


\bibitem[\protect\citeauthoryear{Chen, He, Jin, Zheng, Huang, Shen, and Liao}{Chen et~al\mbox{.}}{2023b}]%
        {Chen2023MetaNMP}
\bibfield{author}{\bibinfo{person}{Dan Chen}, \bibinfo{person}{Haiheng He}, \bibinfo{person}{Hai Jin}, \bibinfo{person}{Long Zheng}, \bibinfo{person}{Yu Huang}, \bibinfo{person}{Xinyang Shen}, {and} \bibinfo{person}{Xiaofei Liao}} \bibinfo{year}{2023}\natexlab{b}.
\newblock \showarticletitle{{MetaNMP: Leveraging Cartesian-Like Product to Accelerate HGNNs with Near-Memory Processing}}, In \bibinfo{booktitle}{\emph{ISCA}}.
\newblock


\bibitem[\protect\citeauthoryear{Chen, G{\'o}mez-Luna, El~Hajj, Guo, and Mutlu}{Chen et~al\mbox{.}}{2023a}]%
        {chen2023simplepim}
\bibfield{author}{\bibinfo{person}{Jinfan Chen}, \bibinfo{person}{Juan G{\'o}mez-Luna}, \bibinfo{person}{Izzat El~Hajj}, \bibinfo{person}{Yuxin Guo}, {and} \bibinfo{person}{Onur Mutlu}} \bibinfo{year}{2023}\natexlab{a}.
\newblock \showarticletitle{{SimplePIM: A Software Framework for Productive and Efficient Processing-in-Memory}}, In \bibinfo{booktitle}{\emph{PACT}}.
\newblock


\bibitem[\protect\citeauthoryear{Chen, Li, Li, Lin, Wang, Wang, Xiao, Xu, Zhang, and Zhang}{Chen et~al\mbox{.}}{2015}]%
        {chen2015mxnet}
\bibfield{author}{\bibinfo{person}{Tianqi Chen}, \bibinfo{person}{Mu Li}, \bibinfo{person}{Yutian Li}, \bibinfo{person}{Min Lin}, \bibinfo{person}{Naiyan Wang}, \bibinfo{person}{Minjie Wang}, \bibinfo{person}{Tianjun Xiao}, \bibinfo{person}{Bing Xu}, \bibinfo{person}{Chiyuan Zhang}, {and} \bibinfo{person}{Zheng Zhang}} \bibinfo{year}{2015}\natexlab{}.
\newblock \showarticletitle{{Mxnet: A Flexible and Efficient Machine Learning Library for Heterogeneous Distributed Systems}}.
\newblock \bibinfo{journal}{\emph{arXiv}}, \bibinfo{year}{2015}.
\newblock


\bibitem[\protect\citeauthoryear{Chen, Zheng, Yan, Jiang, Moreau, Ceze, Guestrin, and Krishnamurthy}{Chen et~al\mbox{.}}{2018}]%
        {chen2018learning}
\bibfield{author}{\bibinfo{person}{Tianqi Chen}, \bibinfo{person}{Lianmin Zheng}, \bibinfo{person}{Eddie Yan}, \bibinfo{person}{Ziheng Jiang}, \bibinfo{person}{Thierry Moreau}, \bibinfo{person}{Luis Ceze}, \bibinfo{person}{Carlos Guestrin}, {and} \bibinfo{person}{Arvind Krishnamurthy}} \bibinfo{year}{2018}\natexlab{}.
\newblock \showarticletitle{{Learning to Optimize Tensor Programs}}.
\newblock \bibinfo{journal}{\emph{NIPS}}, \bibinfo{year}{2018}.
\newblock


\bibitem[\protect\citeauthoryear{Chiang, Liu, Si, Li, Bengio, and Hsieh}{Chiang et~al\mbox{.}}{2019}]%
        {chiang2019cluster}
\bibfield{author}{\bibinfo{person}{Wei-Lin Chiang}, \bibinfo{person}{Xuanqing Liu}, \bibinfo{person}{Si Si}, \bibinfo{person}{Yang Li}, \bibinfo{person}{Samy Bengio}, {and} \bibinfo{person}{Cho-Jui Hsieh}} \bibinfo{year}{2019}\natexlab{}.
\newblock \showarticletitle{{Cluster-GCN: An Efficient Algorithm for Training Deep and Large Graph Convolutional Networks}}, In \bibinfo{booktitle}{\emph{SIGKDD}}.
\newblock


\bibitem[\protect\citeauthoryear{Cho, Kwon, Lym, and Erez}{Cho et~al\mbox{.}}{2020}]%
        {Cho2020Near}
\bibfield{author}{\bibinfo{person}{Benjamin~Y. Cho}, \bibinfo{person}{Yongkee Kwon}, \bibinfo{person}{Sangkug Lym}, {and} \bibinfo{person}{Mattan Erez}} \bibinfo{year}{2020}\natexlab{}.
\newblock \showarticletitle{{Near Data Acceleration with Concurrent Host Access}}, In \bibinfo{booktitle}{\emph{ISCA}}.
\newblock


\bibitem[\protect\citeauthoryear{Choe, Huang, Moreshet, Herlihy, and Bahar}{Choe et~al\mbox{.}}{2019}]%
        {choe2019concurrent}
\bibfield{author}{\bibinfo{person}{Jiwon Choe}, \bibinfo{person}{Amy Huang}, \bibinfo{person}{Tali Moreshet}, \bibinfo{person}{Maurice Herlihy}, {and} \bibinfo{person}{R.~Iris Bahar}} \bibinfo{year}{2019}\natexlab{}.
\newblock \showarticletitle{{Concurrent Data Structures with Near-Data-Processing: An Architecture-Aware Implementation}}, In \bibinfo{booktitle}{\emph{SPAA}}.
\newblock


\bibitem[\protect\citeauthoryear{Ctranslate2}{Ctranslate2}{2023}]%
        {Ctranslate2}
\bibfield{author}{\bibinfo{person}{Ctranslate2}} \bibinfo{year}{2023}\natexlab{}.
\newblock \showarticletitle{{Ctranslate2}}.
\newblock
\urldef\tempurl%
\url{https://github.com/OpenNMT/CTranslate2}
\showURL{%
\tempurl}


\bibitem[\protect\citeauthoryear{Dagum and Menon}{Dagum and Menon}{1998}]%
        {Dagum98OpenMP}
\bibfield{author}{\bibinfo{person}{Leonardo Dagum} {and} \bibinfo{person}{Ramesh Menon}} \bibinfo{year}{1998}\natexlab{}.
\newblock \showarticletitle{{OpenMP: An Industry-Standard API for Shared-Memory Programming}}, In \bibinfo{booktitle}{\emph{IEEE Comput. Sci. Eng.}}
\newblock


\bibitem[\protect\citeauthoryear{Dalton, Olson, and Bell}{Dalton et~al\mbox{.}}{2015}]%
        {Dalton2015Optimizing}
\bibfield{author}{\bibinfo{person}{Steven Dalton}, \bibinfo{person}{Luke Olson}, {and} \bibinfo{person}{Nathan Bell}} \bibinfo{year}{2015}\natexlab{}.
\newblock \showarticletitle{{Optimizing Sparse Matrix—Matrix Multiplication for the GPU}}.
\newblock \bibinfo{journal}{\emph{ACM Trans. Math. Softw.}}, \bibinfo{year}{2015}.
\newblock


\bibitem[\protect\citeauthoryear{Das, Sutradhar, Indovina, Dinakarrao, and Ganguly}{Das et~al\mbox{.}}{2022}]%
        {Das2022Implementation}
\bibfield{author}{\bibinfo{person}{Prangon Das}, \bibinfo{person}{Purab~Ranjan Sutradhar}, \bibinfo{person}{Mark Indovina}, \bibinfo{person}{Sai Manoj~Pudukotai Dinakarrao}, {and} \bibinfo{person}{Amlan Ganguly}} \bibinfo{year}{2022}\natexlab{}.
\newblock \showarticletitle{{Implementation and Evaluation of Deep Neural Networks in Commercially Available Processing in Memory Hardware}}, In \bibinfo{booktitle}{\emph{SOCC}}.
\newblock


\bibitem[\protect\citeauthoryear{Davis and Hu}{Davis and Hu}{2011}]%
        {davis2011florida}
\bibfield{author}{\bibinfo{person}{Timothy~A. Davis} {and} \bibinfo{person}{Yifan Hu}} \bibinfo{year}{2011}\natexlab{}.
\newblock \showarticletitle{{The University of Florida Sparse Matrix Collection}}, In \bibinfo{booktitle}{\emph{TOMS}}.
\newblock


\bibitem[\protect\citeauthoryear{{Devaux}}{{Devaux}}{2019}]%
        {devaux2019}
\bibfield{author}{\bibinfo{person}{F. {Devaux}}} \bibinfo{year}{2019}\natexlab{}.
\newblock \showarticletitle{{The True Processing In Memory Accelerator}}, In \bibinfo{booktitle}{\emph{Hot Chips}}.
\newblock


\bibitem[\protect\citeauthoryear{Diab, Nassereldine, Alser, Gómez~Luna, Mutlu, and El~Hajj}{Diab et~al\mbox{.}}{2023}]%
        {Diab2023Framework}
\bibfield{author}{\bibinfo{person}{Safaa Diab}, \bibinfo{person}{Amir Nassereldine}, \bibinfo{person}{Mohammed Alser}, \bibinfo{person}{Juan Gómez~Luna}, \bibinfo{person}{Onur Mutlu}, {and} \bibinfo{person}{Izzat El~Hajj}} \bibinfo{year}{2023}\natexlab{}.
\newblock \showarticletitle{{A Framework for High-Throughput Sequence Alignment Using Real Processing-in-Memory Systems}}.
\newblock \bibinfo{journal}{\emph{Bioinformatics}}, \bibinfo{year}{2023}.
\newblock


\bibitem[\protect\citeauthoryear{Drumond, Daglis, Mirzadeh, Ustiugov, Picorel, Falsafi, Grot, and Pnevmatikatos}{Drumond et~al\mbox{.}}{2017}]%
        {Drumond2017mondrian}
\bibfield{author}{\bibinfo{person}{Mario Drumond}, \bibinfo{person}{Alexandros Daglis}, \bibinfo{person}{Nooshin Mirzadeh}, \bibinfo{person}{Dmitrii Ustiugov}, \bibinfo{person}{Javier Picorel}, \bibinfo{person}{Babak Falsafi}, \bibinfo{person}{Boris Grot}, {and} \bibinfo{person}{Dionisios Pnevmatikatos}} \bibinfo{year}{2017}\natexlab{}.
\newblock \showarticletitle{{The Mondrian Data Engine}}, In \bibinfo{booktitle}{\emph{ISCA}}.
\newblock


\bibitem[\protect\citeauthoryear{Falevoz and Legriel}{Falevoz and Legriel}{2023}]%
        {falevoz2023energy}
\bibfield{author}{\bibinfo{person}{Yann Falevoz} {and} \bibinfo{person}{Julien Legriel}} \bibinfo{year}{2023}\natexlab{}.
\newblock \showarticletitle{{Energy Efficiency Impact of Processing in Memory: A Comprehensive Review of Workloads on the UPMEM Architecture}}, In \bibinfo{booktitle}{\emph{Euro-PAR}}. Springer.
\newblock


\bibitem[\protect\citeauthoryear{Fan, Ma, Li, He, Zhao, Tang, and Yin}{Fan et~al\mbox{.}}{2019}]%
        {fan2019graph}
\bibfield{author}{\bibinfo{person}{Wenqi Fan}, \bibinfo{person}{Yao Ma}, \bibinfo{person}{Qing Li}, \bibinfo{person}{Yuan He}, \bibinfo{person}{Eric Zhao}, \bibinfo{person}{Jiliang Tang}, {and} \bibinfo{person}{Dawei Yin}} \bibinfo{year}{2019}\natexlab{}.
\newblock \showarticletitle{{Graph Neural Networks for Social Recommendation}}, In \bibinfo{booktitle}{\emph{The World Wide Web Conference}}.
\newblock


\bibitem[\protect\citeauthoryear{Fernandez, Giannoula, Manglik, Quislant, Ghiasi, Gómez-Luna, Gutierrez, Plata, and Mutlu}{Fernandez et~al\mbox{.}}{2024}]%
        {fernandez2024Matsa}
\bibfield{author}{\bibinfo{person}{Ivan Fernandez}, \bibinfo{person}{Christina Giannoula}, \bibinfo{person}{Aditya Manglik}, \bibinfo{person}{Ricardo Quislant}, \bibinfo{person}{Nika~Mansouri Ghiasi}, \bibinfo{person}{Juan Gómez-Luna}, \bibinfo{person}{Eladio Gutierrez}, \bibinfo{person}{Oscar Plata}, {and} \bibinfo{person}{Onur Mutlu}} \bibinfo{year}{2024}\natexlab{}.
\newblock \showarticletitle{{MATSA: An MRAM-Based Energy-Efficient Accelerator for Time Series Analysis}}.
\newblock \bibinfo{journal}{\emph{IEEE Access}}, \bibinfo{year}{2024}.
\newblock


\bibitem[\protect\citeauthoryear{Fernandez, Quislant, Giannoula, Alser, G{\'o}mez-Luna, Guti{\'e}rrez, Plata, and Mutlu}{Fernandez et~al\mbox{.}}{2020}]%
        {fernandez2020natsa}
\bibfield{author}{\bibinfo{person}{Ivan Fernandez}, \bibinfo{person}{Ricardo Quislant}, \bibinfo{person}{Christina Giannoula}, \bibinfo{person}{Mohammed Alser}, \bibinfo{person}{Juan G{\'o}mez-Luna}, \bibinfo{person}{Eladio Guti{\'e}rrez}, \bibinfo{person}{Oscar Plata}, {and} \bibinfo{person}{Onur Mutlu}} \bibinfo{year}{2020}\natexlab{}.
\newblock \showarticletitle{{NATSA: A Near-Data Processing Accelerator for Time Series Analysis}}, In \bibinfo{booktitle}{\emph{ICCD}}.
\newblock


\bibitem[\protect\citeauthoryear{Fey and Lenssen}{Fey and Lenssen}{2019}]%
        {Fey2019Fast}
\bibfield{author}{\bibinfo{person}{Matthias Fey} {and} \bibinfo{person}{Jan~E. Lenssen}} \bibinfo{year}{2019}\natexlab{}.
\newblock \showarticletitle{{Fast Graph Representation Learning with PyTorch Geometric}}, In \bibinfo{booktitle}{\emph{ICLR}}.
\newblock


\bibitem[\protect\citeauthoryear{Gale, Zaharia, Young, and Elsen}{Gale et~al\mbox{.}}{[n.\,d.]}]%
        {gale2020sparse}
\bibfield{author}{\bibinfo{person}{Trevor Gale}, \bibinfo{person}{Matei Zaharia}, \bibinfo{person}{Cliff Young}, {and} \bibinfo{person}{Erich Elsen}} \bibinfo{year}{[n.\,d.]}\natexlab{}.
\newblock \showarticletitle{{Sparse GPU Kernels for Deep Learning}}, In \bibinfo{booktitle}{\emph{SC}}.
\newblock


\bibitem[\protect\citeauthoryear{Gao, Ayers, and Kozyrakis}{Gao et~al\mbox{.}}{2015}]%
        {Gao2015Practical}
\bibfield{author}{\bibinfo{person}{Mingyu Gao}, \bibinfo{person}{Grant Ayers}, {and} \bibinfo{person}{Christos Kozyrakis}} \bibinfo{year}{2015}\natexlab{}.
\newblock \showarticletitle{{Practical Near-Data Processing for In-Memory Analytics Frameworks}}, In \bibinfo{booktitle}{\emph{PACT}}.
\newblock


\bibitem[\protect\citeauthoryear{Geoffrey, Gao, Golikov, and Pekhimenko}{Geoffrey et~al\mbox{.}}{2021}]%
        {geoffrey2021habitat}
\bibfield{author}{\bibinfo{person}{X~Yu Geoffrey}, \bibinfo{person}{Yubo Gao}, \bibinfo{person}{Pavel Golikov}, {and} \bibinfo{person}{Gennady Pekhimenko}} \bibinfo{year}{2021}\natexlab{}.
\newblock \showarticletitle{{Habitat: A Runtime-Based Computational Performance Predictor for Deep Neural Network Training}}, In \bibinfo{booktitle}{\emph{ATC}}.
\newblock


\bibitem[\protect\citeauthoryear{Gerogiannis, Yesil, Lenadora, Cao, Mendis, and Torrellas}{Gerogiannis et~al\mbox{.}}{2023}]%
        {gerogiannis2023spade}
\bibfield{author}{\bibinfo{person}{Gerasimos Gerogiannis}, \bibinfo{person}{Serif Yesil}, \bibinfo{person}{Damitha Lenadora}, \bibinfo{person}{Dingyuan Cao}, \bibinfo{person}{Charith Mendis}, {and} \bibinfo{person}{Josep Torrellas}} \bibinfo{year}{2023}\natexlab{}.
\newblock \showarticletitle{{SPADE: A Flexible and Scalable Accelerator for SpMM and SDDMM}}, In \bibinfo{booktitle}{\emph{ISCA}}.
\newblock


\bibitem[\protect\citeauthoryear{Ghose, Boroumand, Kim, Gómez-Luna, and Mutlu}{Ghose et~al\mbox{.}}{2019}]%
        {Ghose2019Workload}
\bibfield{author}{\bibinfo{person}{Saugata Ghose}, \bibinfo{person}{Amirali Boroumand}, \bibinfo{person}{Jeremie Kim}, \bibinfo{person}{Juan Gómez-Luna}, {and} \bibinfo{person}{Onur Mutlu}} \bibinfo{year}{2019}\natexlab{}.
\newblock \showarticletitle{{Processing-in-Memory: A Workload-Driven Perspective}}, In \bibinfo{booktitle}{\emph{IBM JRD}}.
\newblock


\bibitem[\protect\citeauthoryear{Giannoula, Fernandez, G\'{o}mez-Luna, Koziris, Goumas, and Mutlu}{Giannoula et~al\mbox{.}}{2022a}]%
        {giannoula2022towards}
\bibfield{author}{\bibinfo{person}{Christina Giannoula}, \bibinfo{person}{Ivan Fernandez}, \bibinfo{person}{Juan G\'{o}mez-Luna}, \bibinfo{person}{Nectarios Koziris}, \bibinfo{person}{Georgios Goumas}, {and} \bibinfo{person}{Onur Mutlu}} \bibinfo{year}{2022}\natexlab{a}.
\newblock \showarticletitle{{Towards Efficient Sparse Matrix Vector Multiplication on Real Processing-In-Memory Architectures}}, In \bibinfo{booktitle}{\emph{SIGMETRICS}}.
\newblock


\bibitem[\protect\citeauthoryear{Giannoula, Fernandez, Luna, Koziris, Goumas, and Mutlu}{Giannoula et~al\mbox{.}}{2022b}]%
        {giannoula2022sparsep}
\bibfield{author}{\bibinfo{person}{Christina Giannoula}, \bibinfo{person}{Ivan Fernandez}, \bibinfo{person}{Juan~G{\'o}mez Luna}, \bibinfo{person}{Nectarios Koziris}, \bibinfo{person}{Georgios Goumas}, {and} \bibinfo{person}{Onur Mutlu}} \bibinfo{year}{2022}\natexlab{b}.
\newblock \showarticletitle{{SparseP: Towards Efficient Sparse Matrix Vector Multiplication on Real Processing-in-Memory Architectures}}.
\newblock \bibinfo{journal}{\emph{POMACS}}, \bibinfo{year}{2022}.
\newblock


\bibitem[\protect\citeauthoryear{Giannoula, Vijaykumar, Papadopoulou, Karakostas, Fernandez, G{\'o}mez-Luna, Orosa, Koziris, Goumas, and Mutlu}{Giannoula et~al\mbox{.}}{2021}]%
        {giannoula2021syncron}
\bibfield{author}{\bibinfo{person}{Christina Giannoula}, \bibinfo{person}{Nandita Vijaykumar}, \bibinfo{person}{Nikela Papadopoulou}, \bibinfo{person}{Vasileios Karakostas}, \bibinfo{person}{Ivan Fernandez}, \bibinfo{person}{Juan G{\'o}mez-Luna}, \bibinfo{person}{Lois Orosa}, \bibinfo{person}{Nectarios Koziris}, \bibinfo{person}{Georgios Goumas}, {and} \bibinfo{person}{Onur Mutlu}} \bibinfo{year}{2021}\natexlab{}.
\newblock \showarticletitle{{SynCron: Efficient Synchronization Support for Near-Data-Processing Architectures}}, In \bibinfo{booktitle}{\emph{HPCA}}.
\newblock


\bibitem[\protect\citeauthoryear{Gokhale, Lloyd, and Hajas}{Gokhale et~al\mbox{.}}{2015}]%
        {Gokhale2015Near}
\bibfield{author}{\bibinfo{person}{Maya Gokhale}, \bibinfo{person}{Scott Lloyd}, {and} \bibinfo{person}{Chris Hajas}} \bibinfo{year}{2015}\natexlab{}.
\newblock \showarticletitle{{Near Memory Data Structure Rearrangement}}, In \bibinfo{booktitle}{\emph{MEMSYS}}.
\newblock


\bibitem[\protect\citeauthoryear{G{\'{o}}mez{-}Luna, Hajj, Fernandez, Giannoula, Oliveira, and Mutlu}{G{\'{o}}mez{-}Luna et~al\mbox{.}}{2021}]%
        {Gomez2021Benchmarking}
\bibfield{author}{\bibinfo{person}{Juan G{\'{o}}mez{-}Luna}, \bibinfo{person}{Izzat~El Hajj}, \bibinfo{person}{Ivan Fernandez}, \bibinfo{person}{Christina Giannoula}, \bibinfo{person}{Geraldo~F. Oliveira}, {and} \bibinfo{person}{Onur Mutlu}} \bibinfo{year}{2021}\natexlab{}.
\newblock \showarticletitle{{Benchmarking a New Paradigm: An Experimental Analysis of a Real Processing-in-Memory Architecture}}, In \bibinfo{booktitle}{\emph{CoRR}}.
\newblock
\urldef\tempurl%
\url{https://arxiv.org/abs/2105.03814}
\showURL{%
\tempurl}


\bibitem[\protect\citeauthoryear{Gong, Ji, Yao, Fletcher, Hughes, and Torrellas}{Gong et~al\mbox{.}}{2022}]%
        {Gong2022Graphite}
\bibfield{author}{\bibinfo{person}{Zhangxiaowen Gong}, \bibinfo{person}{Houxiang Ji}, \bibinfo{person}{Yao Yao}, \bibinfo{person}{Christopher~W. Fletcher}, \bibinfo{person}{Christopher~J. Hughes}, {and} \bibinfo{person}{Josep Torrellas}} \bibinfo{year}{2022}\natexlab{}.
\newblock \showarticletitle{{Graphite: Optimizing Graph Neural Networks on CPUs through Cooperative Software-Hardware Techniques}}, In \bibinfo{booktitle}{\emph{ISCA}}.
\newblock


\bibitem[\protect\citeauthoryear{Group}{Group}{2022}]%
        {PyGimLibrary}
\bibfield{author}{\bibinfo{person}{SAFARI~Research Group}} \bibinfo{year}{2022}\natexlab{}.
\newblock \showarticletitle{{PyGim Software Package}}.
\newblock
\urldef\tempurl%
\url{https://github.com/CMU-SAFARI/PyGim}
\showURL{%
\tempurl}


\bibitem[\protect\citeauthoryear{Gu, Moreira, Edelsohn, and Azad}{Gu et~al\mbox{.}}{2020}]%
        {Gu2020Bandwidth}
\bibfield{author}{\bibinfo{person}{Zhixiang Gu}, \bibinfo{person}{Jose Moreira}, \bibinfo{person}{David Edelsohn}, {and} \bibinfo{person}{Ariful Azad}} \bibinfo{year}{2020}\natexlab{}.
\newblock \showarticletitle{{Bandwidth Optimized Parallel Algorithms for Sparse Matrix-Matrix Multiplication Using Propagation Blocking}}, In \bibinfo{booktitle}{\emph{SPAA}}.
\newblock


\bibitem[\protect\citeauthoryear{Gulli and Pal}{Gulli and Pal}{2017}]%
        {gulli2017deep}
\bibfield{author}{\bibinfo{person}{Antonio Gulli} {and} \bibinfo{person}{Sujit Pal}} \bibinfo{year}{2017}\natexlab{}.
\newblock \bibinfo{booktitle}{\emph{{Deep Learning with Keras}}}.
\newblock \bibinfo{publisher}{Packt Publishing Ltd}.
\newblock


\bibitem[\protect\citeauthoryear{Gómez-Luna, Guo, Brocard, Legriel, Cimadomo, Oliveira, Singh, and Mutlu}{Gómez-Luna et~al\mbox{.}}{2023}]%
        {Gomez2023Evaluating}
\bibfield{author}{\bibinfo{person}{Juan Gómez-Luna}, \bibinfo{person}{Yuxin Guo}, \bibinfo{person}{Sylvan Brocard}, \bibinfo{person}{Julien Legriel}, \bibinfo{person}{Remy Cimadomo}, \bibinfo{person}{Geraldo~F. Oliveira}, \bibinfo{person}{Gagandeep Singh}, {and} \bibinfo{person}{Onur Mutlu}} \bibinfo{year}{2023}\natexlab{}.
\newblock \showarticletitle{{Evaluating Machine Learning Workloads on Memory-Centric Computing Systems}}, In \bibinfo{booktitle}{\emph{ISPASS}}.
\newblock


\bibitem[\protect\citeauthoryear{Gómez-Luna, Hajj, Fernandez, Giannoula, Oliveira, and Mutlu}{Gómez-Luna et~al\mbox{.}}{2022}]%
        {Gomez2022Benchmarking}
\bibfield{author}{\bibinfo{person}{Juan Gómez-Luna}, \bibinfo{person}{Izzat~El Hajj}, \bibinfo{person}{Ivan Fernandez}, \bibinfo{person}{Christina Giannoula}, \bibinfo{person}{Geraldo~F. Oliveira}, {and} \bibinfo{person}{Onur Mutlu}} \bibinfo{year}{2022}\natexlab{}.
\newblock \showarticletitle{{Benchmarking a New Paradigm: Experimental Analysis and Characterization of a Real Processing-in-Memory System}}.
\newblock \bibinfo{journal}{\emph{IEEE Access}}, \bibinfo{year}{2022}.
\newblock


\bibitem[\protect\citeauthoryear{Hamilton, Ying, and Leskovec}{Hamilton et~al\mbox{.}}{2017}]%
        {hamilton2017inductive}
\bibfield{author}{\bibinfo{person}{Will Hamilton}, \bibinfo{person}{Zhitao Ying}, {and} \bibinfo{person}{Jure Leskovec}} \bibinfo{year}{2017}\natexlab{}.
\newblock \showarticletitle{{Inductive Representation Learning on Large Graphs}}.
\newblock \bibinfo{journal}{\emph{NIPS}}  \bibinfo{volume}{30}, \bibinfo{year}{2017}.
\newblock


\bibitem[\protect\citeauthoryear{He, Song, Kim, Jeong, Kim, Park, Thottethodi, and Vijaykumar}{He et~al\mbox{.}}{2020}]%
        {He2020Newton}
\bibfield{author}{\bibinfo{person}{Mingxuan He}, \bibinfo{person}{Choungki Song}, \bibinfo{person}{Ilkon Kim}, \bibinfo{person}{Chunseok Jeong}, \bibinfo{person}{Seho Kim}, \bibinfo{person}{Il Park}, \bibinfo{person}{Mithuna Thottethodi}, {and} \bibinfo{person}{T.~N. Vijaykumar}} \bibinfo{year}{2020}\natexlab{}.
\newblock \showarticletitle{{Newton: A DRAM-Maker’s Accelerator-in-Memory (AiM) Architecture for Machine Learning}}, In \bibinfo{booktitle}{\emph{MICRO}}.
\newblock


\bibitem[\protect\citeauthoryear{Heo, Lee, Cho, Choi, Lee, Ham, Kim, Mahajan, and Park}{Heo et~al\mbox{.}}{2024}]%
        {Heo2024NeuPIM}
\bibfield{author}{\bibinfo{person}{Guseul Heo}, \bibinfo{person}{Sangyeop Lee}, \bibinfo{person}{Jaehong Cho}, \bibinfo{person}{Hyunmin Choi}, \bibinfo{person}{Sanghyeon Lee}, \bibinfo{person}{Hyungkyu Ham}, \bibinfo{person}{Gwangsun Kim}, \bibinfo{person}{Divya Mahajan}, {and} \bibinfo{person}{Jongse Park}} \bibinfo{year}{2024}\natexlab{}.
\newblock \showarticletitle{{NeuPIMs: NPU-PIM Heterogeneous Acceleration for Batched LLM Inferencing}}, In \bibinfo{booktitle}{\emph{ASPLOS}}.
\newblock


\bibitem[\protect\citeauthoryear{Hong, Sukumaran-Rajam, Nisa, Singh, and Sadayappan}{Hong et~al\mbox{.}}{2019}]%
        {Hong2019Adaptive}
\bibfield{author}{\bibinfo{person}{Changwan Hong}, \bibinfo{person}{Aravind Sukumaran-Rajam}, \bibinfo{person}{Israt Nisa}, \bibinfo{person}{Kunal Singh}, {and} \bibinfo{person}{P. Sadayappan}} \bibinfo{year}{2019}\natexlab{}.
\newblock \showarticletitle{{Adaptive Sparse Tiling for Sparse Matrix Multiplication}}, In \bibinfo{booktitle}{\emph{PpopP}}.
\newblock


\bibitem[\protect\citeauthoryear{Hsieh, Khan, Vijaykumar, Chang, Boroumand, Ghose, and Mutlu}{Hsieh et~al\mbox{.}}{2016}]%
        {Hsieh2016accelerating}
\bibfield{author}{\bibinfo{person}{Kevin Hsieh}, \bibinfo{person}{Samira Khan}, \bibinfo{person}{Nandita Vijaykumar}, \bibinfo{person}{Kevin Chang}, \bibinfo{person}{Amirali Boroumand}, \bibinfo{person}{Saugata Ghose}, {and} \bibinfo{person}{Onur Mutlu}} \bibinfo{year}{2016}\natexlab{}.
\newblock \showarticletitle{{Accelerating Pointer Chasing in 3D-stacked Memory: Challenges, Mechanisms, Evaluation}}, In \bibinfo{booktitle}{\emph{ICCD}}.
\newblock


\bibitem[\protect\citeauthoryear{Huang, Zhai, Zheng, Yi, and Shen}{Huang et~al\mbox{.}}{2021}]%
        {Huang2021Understanding}
\bibfield{author}{\bibinfo{person}{Kezhao Huang}, \bibinfo{person}{Jidong Zhai}, \bibinfo{person}{Zhen Zheng}, \bibinfo{person}{Youngmin Yi}, {and} \bibinfo{person}{Xipeng Shen}} \bibinfo{year}{2021}\natexlab{}.
\newblock \showarticletitle{{Understanding and Bridging the Gaps in Current GNN Performance Optimizations}}, In \bibinfo{booktitle}{\emph{PpopP}}.
\newblock


\bibitem[\protect\citeauthoryear{Hyun, Kim, Lee, and Rhu}{Hyun et~al\mbox{.}}{2024}]%
        {hyun2024pathfinding}
\bibfield{author}{\bibinfo{person}{Bongjoon Hyun}, \bibinfo{person}{Taehun Kim}, \bibinfo{person}{Dongjae Lee}, {and} \bibinfo{person}{Minsoo Rhu}} \bibinfo{year}{2024}\natexlab{}.
\newblock \showarticletitle{{Pathfinding Future PIM Architectures by Demystifying a Commercial PIM Technology}}, In \bibinfo{booktitle}{\emph{HPCA}}.
\newblock


\bibitem[\protect\citeauthoryear{Im and Yelick}{Im and Yelick}{1999}]%
        {Im1999Optimizing}
\bibfield{author}{\bibinfo{person}{Eun{-}Jin Im} {and} \bibinfo{person}{Katherine~A. Yelick}} \bibinfo{year}{1999}\natexlab{}.
\newblock \showarticletitle{{Optimizing Sparse Matrix Vector Multiplication on SMP}}, In \bibinfo{booktitle}{\emph{{PPSC}}}.
\newblock


\bibitem[\protect\citeauthoryear{Item, Oliveira, Gómez-Luna, Sadrosadati, Guo, and Mutlu}{Item et~al\mbox{.}}{2023}]%
        {Item2023TransPimLib}
\bibfield{author}{\bibinfo{person}{Maurus Item}, \bibinfo{person}{Geraldo~F. Oliveira}, \bibinfo{person}{Juan Gómez-Luna}, \bibinfo{person}{Mohammad Sadrosadati}, \bibinfo{person}{Yuxin Guo}, {and} \bibinfo{person}{Onur Mutlu}} \bibinfo{year}{2023}\natexlab{}.
\newblock \showarticletitle{{TransPimLib: Efficient Transcendental Functions for Processing-in-Memory Systems}}, In \bibinfo{booktitle}{\emph{ISPASS}}.
\newblock


\bibitem[\protect\citeauthoryear{Jia, Shelhamer, Donahue, Karayev, Long, Girshick, Guadarrama, and Darrell}{Jia et~al\mbox{.}}{2014}]%
        {jia2014caffe}
\bibfield{author}{\bibinfo{person}{Yangqing Jia}, \bibinfo{person}{Evan Shelhamer}, \bibinfo{person}{Jeff Donahue}, \bibinfo{person}{Sergey Karayev}, \bibinfo{person}{Jonathan Long}, \bibinfo{person}{Ross Girshick}, \bibinfo{person}{Sergio Guadarrama}, {and} \bibinfo{person}{Trevor Darrell}} \bibinfo{year}{2014}\natexlab{}.
\newblock \showarticletitle{{Caffe: Convolutional Architecture for Fast Feature Embedding}}, In \bibinfo{booktitle}{\emph{Proceedings of the 22nd ACM International Conference on Multimedia}}.
\newblock


\bibitem[\protect\citeauthoryear{Jia, Lin, Gao, Zaharia, and Aiken}{Jia et~al\mbox{.}}{2020}]%
        {Jia2020ImprovingTA}
\bibfield{author}{\bibinfo{person}{Zhihao Jia}, \bibinfo{person}{Sina Lin}, \bibinfo{person}{Mingyu Gao}, \bibinfo{person}{Matei~A. Zaharia}, {and} \bibinfo{person}{Alexander Aiken}} \bibinfo{year}{2020}\natexlab{}.
\newblock \showarticletitle{{Improving the Accuracy, Scalability, and Performance of Graph Neural Networks with Roc}}, In \bibinfo{booktitle}{\emph{MLSys}}.
\newblock


\bibitem[\protect\citeauthoryear{Jibril, Al-Sayeh, and Sattler}{Jibril et~al\mbox{.}}{2024}]%
        {Jibril2024Aggregation}
\bibfield{author}{\bibinfo{person}{Muhammad~Attahir Jibril}, \bibinfo{person}{Hani Al-Sayeh}, {and} \bibinfo{person}{Kai-Uwe Sattler}} \bibinfo{year}{2024}\natexlab{}.
\newblock \showarticletitle{{Accelerating Aggregation Using a Real Processing-in-Memory System}}, In \bibinfo{booktitle}{\emph{ICDE}}.
\newblock


\bibitem[\protect\citeauthoryear{Jonatan, Cho, Son, Wu, Livesay, Mora, Shivdikar, Abell\'{a}n, Joshi, Kaeli, and Kim}{Jonatan et~al\mbox{.}}{2024}]%
        {Gilbert2024Scalabity}
\bibfield{author}{\bibinfo{person}{Gilbert Jonatan}, \bibinfo{person}{Haeyoon Cho}, \bibinfo{person}{Hyojun Son}, \bibinfo{person}{Xiangyu Wu}, \bibinfo{person}{Neal Livesay}, \bibinfo{person}{Evelio Mora}, \bibinfo{person}{Kaustubh Shivdikar}, \bibinfo{person}{Jos\'{e}~L. Abell\'{a}n}, \bibinfo{person}{Ajay Joshi}, \bibinfo{person}{David Kaeli}, {and} \bibinfo{person}{John Kim}} \bibinfo{year}{2024}\natexlab{}.
\newblock \showarticletitle{{Scalability Limitations of Processing-in-Memory Using Real System Evaluations}}.
\newblock \bibinfo{journal}{\emph{Proc. ACM Meas. Anal. Comput. Syst.}}, \bibinfo{year}{2024}.
\newblock


\bibitem[\protect\citeauthoryear{Kanellopoulos, Vijaykumar, Giannoula, Azizi, Koppula, Ghiasi, Shahroodi, Luna, and Mutlu}{Kanellopoulos et~al\mbox{.}}{2019}]%
        {kanellopoulos2019smash}
\bibfield{author}{\bibinfo{person}{Konstantinos Kanellopoulos}, \bibinfo{person}{Nandita Vijaykumar}, \bibinfo{person}{Christina Giannoula}, \bibinfo{person}{Roknoddin Azizi}, \bibinfo{person}{Skanda Koppula}, \bibinfo{person}{Nika~Mansouri Ghiasi}, \bibinfo{person}{Taha Shahroodi}, \bibinfo{person}{Juan~Gomez Luna}, {and} \bibinfo{person}{Onur Mutlu}} \bibinfo{year}{2019}\natexlab{}.
\newblock \showarticletitle{{Smash: Co-Designing Software Compression and Hardware-Accelerated Indexing for Efficient Sparse Matrix Operations}}, In \bibinfo{booktitle}{\emph{MICRO}}.
\newblock


\bibitem[\protect\citeauthoryear{Kaufman, Phothilimthana, Zhou, Mendis, Roy, Sabne, and Burrows}{Kaufman et~al\mbox{.}}{2021}]%
        {kaufman2021learned}
\bibfield{author}{\bibinfo{person}{Sam Kaufman}, \bibinfo{person}{Phitchaya Phothilimthana}, \bibinfo{person}{Yanqi Zhou}, \bibinfo{person}{Charith Mendis}, \bibinfo{person}{Sudip Roy}, \bibinfo{person}{Amit Sabne}, {and} \bibinfo{person}{Mike Burrows}} \bibinfo{year}{2021}\natexlab{}.
\newblock \showarticletitle{{A Learned Performance Model for Tensor Processing Units}}.
\newblock \bibinfo{journal}{\emph{MLSys}}, \bibinfo{year}{2021}.
\newblock


\bibitem[\protect\citeauthoryear{Ke, Gupta, Wu, Cho, Hempstead, Reagen, Zhang, Brooks, Chandra, Diril, et~al\mbox{.}}{Ke et~al\mbox{.}}{2020}]%
        {ke2019recnmp}
\bibfield{author}{\bibinfo{person}{Liu Ke}, \bibinfo{person}{Udit Gupta}, \bibinfo{person}{Carole-Jean Wu}, \bibinfo{person}{Benjamin~Youngjae Cho}, \bibinfo{person}{Mark Hempstead}, \bibinfo{person}{Brandon Reagen}, \bibinfo{person}{Xuan Zhang}, \bibinfo{person}{David Brooks}, \bibinfo{person}{Vikas Chandra}, \bibinfo{person}{Utku Diril}, {et~al\mbox{.}}} \bibinfo{year}{2020}\natexlab{}.
\newblock \showarticletitle{{RecNMP: Accelerating Personalized Recommendation with Near-Memory Processing}}, In \bibinfo{booktitle}{\emph{ISCA}}.
\newblock


\bibitem[\protect\citeauthoryear{Khan, Hirki, Niemi, Nurminen, and Ou}{Khan et~al\mbox{.}}{2018}]%
        {rapl}
\bibfield{author}{\bibinfo{person}{Kashif~Nizam Khan}, \bibinfo{person}{Mikael Hirki}, \bibinfo{person}{Tapio Niemi}, \bibinfo{person}{Jukka~K Nurminen}, {and} \bibinfo{person}{Zhonghong Ou}} \bibinfo{year}{2018}\natexlab{}.
\newblock \showarticletitle{{Rapl in Action: Experiences in Using RAPL for Power Measurements}}, In \bibinfo{booktitle}{\emph{TOMPECS}}.
\newblock


\bibitem[\protect\citeauthoryear{Kiningham, Levis, and R{\'e}}{Kiningham et~al\mbox{.}}{2022}]%
        {kiningham2022grip}
\bibfield{author}{\bibinfo{person}{Kevin Kiningham}, \bibinfo{person}{Philip Levis}, {and} \bibinfo{person}{Christopher R{\'e}}} \bibinfo{year}{2022}\natexlab{}.
\newblock \showarticletitle{{GRIP: A Graph Neural Network Accelerator Architecture}}.
\newblock \bibinfo{journal}{\emph{IEEE Trans. Comput.}}, \bibinfo{year}{2022}.
\newblock


\bibitem[\protect\citeauthoryear{Kipf and Welling}{Kipf and Welling}{2016}]%
        {kipf2016semi}
\bibfield{author}{\bibinfo{person}{Thomas~N Kipf} {and} \bibinfo{person}{Max Welling}} \bibinfo{year}{2016}\natexlab{}.
\newblock \showarticletitle{{Semi-Supervised Classification with Graph Convolutional Networks}}.
\newblock \bibinfo{journal}{\emph{arXiv}}, \bibinfo{year}{2016}.
\newblock


\bibitem[\protect\citeauthoryear{Kjolstad, Kamil, Chou, Lugato, and Amarasinghe}{Kjolstad et~al\mbox{.}}{2017}]%
        {Kjolstad2017TACO}
\bibfield{author}{\bibinfo{person}{Fredrik Kjolstad}, \bibinfo{person}{Shoaib Kamil}, \bibinfo{person}{Stephen Chou}, \bibinfo{person}{David Lugato}, {and} \bibinfo{person}{Saman Amarasinghe}} \bibinfo{year}{2017}\natexlab{}.
\newblock \showarticletitle{{The Tensor Algebra Compiler}}.
\newblock \bibinfo{journal}{\emph{Proc. ACM Program. Lang.}}, \bibinfo{year}{2017}.
\newblock


\bibitem[\protect\citeauthoryear{Koanantakool, Azad, Buluç, Morozov, Oh, Oliker, and Yelick}{Koanantakool et~al\mbox{.}}{2016}]%
        {Koanantakool2016Communication}
\bibfield{author}{\bibinfo{person}{Penporn Koanantakool}, \bibinfo{person}{Ariful Azad}, \bibinfo{person}{Aydin Buluç}, \bibinfo{person}{Dmitriy Morozov}, \bibinfo{person}{Sang-Yun Oh}, \bibinfo{person}{Leonid Oliker}, {and} \bibinfo{person}{Katherine Yelick}} \bibinfo{year}{2016}\natexlab{}.
\newblock \showarticletitle{{Communication-Avoiding Parallel Sparse-Dense Matrix-Matrix Multiplication}}, In \bibinfo{booktitle}{\emph{IPDPS}}.
\newblock


\bibitem[\protect\citeauthoryear{Kwon, Lee, and Rhu}{Kwon et~al\mbox{.}}{2019a}]%
        {tensordimm}
\bibfield{author}{\bibinfo{person}{Youngeun Kwon}, \bibinfo{person}{Yunjae Lee}, {and} \bibinfo{person}{Minsoo Rhu}} \bibinfo{year}{2019}\natexlab{a}.
\newblock \showarticletitle{{TensorDIMM: A Practical Near-Memory Processing Architecture for Embeddings and Tensor Operations in Deep Learning}}, In \bibinfo{booktitle}{\emph{MICRO}}.
\newblock


\bibitem[\protect\citeauthoryear{Kwon, Lee, and Rhu}{Kwon et~al\mbox{.}}{2019b}]%
        {kwon2019tensordimm}
\bibfield{author}{\bibinfo{person}{Youngeun Kwon}, \bibinfo{person}{Yunjae Lee}, {and} \bibinfo{person}{Minsoo Rhu}} \bibinfo{year}{2019}\natexlab{b}.
\newblock \showarticletitle{{Tensordimm: A Practical Near-Memory Processing Architecture for Embeddings and Tensor Operations in Deep Learning}}, In \bibinfo{booktitle}{\emph{MICRO}}.
\newblock


\bibitem[\protect\citeauthoryear{Kwon, Lee, Lee, Kwon, Ryu, Son, Seongil, Yu, Lee, Kim, Cho, Kim, Choi, Shin, Kim, Phuah, Kim, Song, Choi, Kim, Kim, Kim, Wang, Kang, Ro, Seo, Song, Youn, Sohn, and Kim}{Kwon et~al\mbox{.}}{2021}]%
        {Kwon2021Function}
\bibfield{author}{\bibinfo{person}{Young-Cheon Kwon}, \bibinfo{person}{Suk~Han Lee}, \bibinfo{person}{Jaehoon Lee}, \bibinfo{person}{Sang-Hyuk Kwon}, \bibinfo{person}{Je~Min Ryu}, \bibinfo{person}{Jong-Pil Son}, \bibinfo{person}{O Seongil}, \bibinfo{person}{Hak-Soo Yu}, \bibinfo{person}{Haesuk Lee}, \bibinfo{person}{Soo~Young Kim}, \bibinfo{person}{Youngmin Cho}, \bibinfo{person}{Jin~Guk Kim}, \bibinfo{person}{Jongyoon Choi}, \bibinfo{person}{Hyun-Sung Shin}, \bibinfo{person}{Jin Kim}, \bibinfo{person}{BengSeng Phuah}, \bibinfo{person}{HyoungMin Kim}, \bibinfo{person}{Myeong~Jun Song}, \bibinfo{person}{Ahn Choi}, \bibinfo{person}{Daeho Kim}, \bibinfo{person}{SooYoung Kim}, \bibinfo{person}{Eun-Bong Kim}, \bibinfo{person}{David Wang}, \bibinfo{person}{Shinhaeng Kang}, \bibinfo{person}{Yuhwan Ro}, \bibinfo{person}{Seungwoo Seo}, \bibinfo{person}{JoonHo Song}, \bibinfo{person}{Jaeyoun Youn}, \bibinfo{person}{Kyomin Sohn}, {and} \bibinfo{person}{Nam~Sung Kim}} \bibinfo{year}{2021}\natexlab{}.
\newblock \showarticletitle{{25.4 A 20nm 6GB Function-In-Memory DRAM, Based on HBM2 with a 1.2TFLOPS Programmable Computing Unit Using Bank-Level Parallelism, for Machine Learning Applications}}, In \bibinfo{booktitle}{\emph{ISSCC}}.
\newblock


\bibitem[\protect\citeauthoryear{Langr and Tvrdík}{Langr and Tvrdík}{2016}]%
        {Langr2016Evaluation}
\bibfield{author}{\bibinfo{person}{Daniel Langr} {and} \bibinfo{person}{Pavel Tvrdík}} \bibinfo{year}{2016}\natexlab{}.
\newblock \showarticletitle{{Evaluation Criteria for Sparse Matrix Storage Formats}}, In \bibinfo{booktitle}{\emph{TPDS}}.
\newblock


\bibitem[\protect\citeauthoryear{Lee, Kang, Lee, Kim, Lee, Seo, Yoon, Lee, Lim, Shin, Kim, Seongil, Iyer, Wang, Sohn, and Kim}{Lee et~al\mbox{.}}{2021}]%
        {Lee2021Hardware}
\bibfield{author}{\bibinfo{person}{Sukhan Lee}, \bibinfo{person}{Shin-haeng Kang}, \bibinfo{person}{Jaehoon Lee}, \bibinfo{person}{Hyeonsu Kim}, \bibinfo{person}{Eojin Lee}, \bibinfo{person}{Seungwoo Seo}, \bibinfo{person}{Hosang Yoon}, \bibinfo{person}{Seungwon Lee}, \bibinfo{person}{Kyounghwan Lim}, \bibinfo{person}{Hyunsung Shin}, \bibinfo{person}{Jinhyun Kim}, \bibinfo{person}{O Seongil}, \bibinfo{person}{Anand Iyer}, \bibinfo{person}{David Wang}, \bibinfo{person}{Kyomin Sohn}, {and} \bibinfo{person}{Nam~Sung Kim}} \bibinfo{year}{2021}\natexlab{}.
\newblock \showarticletitle{{Hardware Architecture and Software Stack for PIM Based on Commercial DRAM Technology: Industrial Product}}, In \bibinfo{booktitle}{\emph{ISCA}}.
\newblock


\bibitem[\protect\citeauthoryear{Lee, Kim, Oh, Park, Hong, Ka, Hwang, Park, Kang, Kim, et~al\mbox{.}}{Lee et~al\mbox{.}}{2022b}]%
        {lee20221ynm}
\bibfield{author}{\bibinfo{person}{Seongju Lee}, \bibinfo{person}{Kyuyoung Kim}, \bibinfo{person}{Sanghoon Oh}, \bibinfo{person}{Joonhong Park}, \bibinfo{person}{Gimoon Hong}, \bibinfo{person}{Dongyoon Ka}, \bibinfo{person}{Kyudong Hwang}, \bibinfo{person}{Jeongje Park}, \bibinfo{person}{Kyeongpil Kang}, \bibinfo{person}{Jungyeon Kim}, {et~al\mbox{.}}} \bibinfo{year}{2022}\natexlab{b}.
\newblock \showarticletitle{{A 1ynm 1.25 V 8GB, 16GB/s/Pin GDDR6-Based Accelerator-in-Memory Supporting 1Tflops MAC Operation and Various Activation Functions for Deep-Learning Applications}}, In \bibinfo{booktitle}{\emph{ISSCC}}.
\newblock


\bibitem[\protect\citeauthoryear{Lee, Chung, and Rhu}{Lee et~al\mbox{.}}{2022a}]%
        {Lee2022SmartSAGE}
\bibfield{author}{\bibinfo{person}{Yunjae Lee}, \bibinfo{person}{Jinha Chung}, {and} \bibinfo{person}{Minsoo Rhu}} \bibinfo{year}{2022}\natexlab{a}.
\newblock \showarticletitle{{SmartSAGE: Training Large-Scale Graph Neural Networks Using In-Storage Processing Architectures}}, In \bibinfo{booktitle}{\emph{ISCA}}.
\newblock


\bibitem[\protect\citeauthoryear{Lenadora, Sathia, Gerogiannis, Yesil, Torrellas, and Mendis}{Lenadora et~al\mbox{.}}{2024}]%
        {lenadora2024sensei}
\bibfield{author}{\bibinfo{person}{Damitha Lenadora}, \bibinfo{person}{Vimarsh Sathia}, \bibinfo{person}{Gerasimos Gerogiannis}, \bibinfo{person}{Serif Yesil}, \bibinfo{person}{Josep Torrellas}, {and} \bibinfo{person}{Charith Mendis}} \bibinfo{year}{2024}\natexlab{}.
\newblock \showarticletitle{{SENSEi: Input-Sensitive Compilation for Accelerating GNNs}}, In \bibinfo{booktitle}{\emph{arXiv}}.
\newblock


\bibitem[\protect\citeauthoryear{Li, Wang, Liu, Liang, Li, and Li}{Li et~al\mbox{.}}{2021b}]%
        {Li2021Glist}
\bibfield{author}{\bibinfo{person}{Cangyuan Li}, \bibinfo{person}{Ying Wang}, \bibinfo{person}{Cheng Liu}, \bibinfo{person}{Shengwen Liang}, \bibinfo{person}{Huawei Li}, {and} \bibinfo{person}{Xiaowei Li}} \bibinfo{year}{2021}\natexlab{b}.
\newblock \showarticletitle{{GLIST: Towards In-Storage Graph Learning}}, In \bibinfo{booktitle}{\emph{ATC}}.
\newblock


\bibitem[\protect\citeauthoryear{Li, Zhou, Li, Sun, and Niu}{Li et~al\mbox{.}}{2023}]%
        {li2023nmexplorer}
\bibfield{author}{\bibinfo{person}{Cong Li}, \bibinfo{person}{Zhe Zhou}, \bibinfo{person}{Xingchen Li}, \bibinfo{person}{Guangyu Sun}, {and} \bibinfo{person}{Dimin Niu}} \bibinfo{year}{2023}\natexlab{}.
\newblock \showarticletitle{{NMExplorer: An Efficient Exploration Framework for DIMM-Based Near-Memory Tensor Reduction}}, In \bibinfo{booktitle}{\emph{DAC}}.
\newblock


\bibitem[\protect\citeauthoryear{Li, Zhou, Wang, Yang, Cao, Yang, Liang, and Sun}{Li et~al\mbox{.}}{2024a}]%
        {Li2024PIMDL}
\bibfield{author}{\bibinfo{person}{Cong Li}, \bibinfo{person}{Zhe Zhou}, \bibinfo{person}{Yang Wang}, \bibinfo{person}{Fan Yang}, \bibinfo{person}{Ting Cao}, \bibinfo{person}{Mao Yang}, \bibinfo{person}{Yun Liang}, {and} \bibinfo{person}{Guangyu Sun}} \bibinfo{year}{2024}\natexlab{a}.
\newblock \showarticletitle{{PIM-DL: Expanding the Applicability of Commodity DRAM-PIMs for Deep Learning via Algorithm-System Co-Optimization}}, In \bibinfo{booktitle}{\emph{ASPLOS}}.
\newblock


\bibitem[\protect\citeauthoryear{Li, Zhou, Zheng, Zhang, Liang, and Sun}{Li et~al\mbox{.}}{2024b}]%
        {Li2024SpecPIM}
\bibfield{author}{\bibinfo{person}{Cong Li}, \bibinfo{person}{Zhe Zhou}, \bibinfo{person}{Size Zheng}, \bibinfo{person}{Jiaxi Zhang}, \bibinfo{person}{Yun Liang}, {and} \bibinfo{person}{Guangyu Sun}} \bibinfo{year}{2024}\natexlab{b}.
\newblock \showarticletitle{{SpecPIM: Accelerating Speculative Inference on PIM-Enabled System via Architecture-Dataflow Co-Exploration}}, In \bibinfo{booktitle}{\emph{ASPLOS}}.
\newblock


\bibitem[\protect\citeauthoryear{Li, Louri, Karanth, and Bunescu}{Li et~al\mbox{.}}{2021a}]%
        {li2021gcnax}
\bibfield{author}{\bibinfo{person}{Jiajun Li}, \bibinfo{person}{Ahmed Louri}, \bibinfo{person}{Avinash Karanth}, {and} \bibinfo{person}{Razvan Bunescu}} \bibinfo{year}{2021}\natexlab{a}.
\newblock \showarticletitle{{GCNAX: A Flexible and Energy-Efficient Accelerator for Graph Convolutional Neural Networks}}, In \bibinfo{booktitle}{\emph{HPCA}}.
\newblock


\bibitem[\protect\citeauthoryear{Liang, Wang, Liu, He, Huawei, Xu, and Li}{Liang et~al\mbox{.}}{2020}]%
        {liang2020engn}
\bibfield{author}{\bibinfo{person}{Shengwen Liang}, \bibinfo{person}{Ying Wang}, \bibinfo{person}{Cheng Liu}, \bibinfo{person}{Lei He}, \bibinfo{person}{LI Huawei}, \bibinfo{person}{Dawen Xu}, {and} \bibinfo{person}{Xiaowei Li}} \bibinfo{year}{2020}\natexlab{}.
\newblock \showarticletitle{{Engn: A High-Throughput and Energy-Efficient Accelerator for Large Graph Neural Networks}}.
\newblock \bibinfo{journal}{\emph{IEEE Trans. Comput.}}, \bibinfo{year}{2020}.
\newblock


\bibitem[\protect\citeauthoryear{Lim, Lee, Choi, Lee, Park, Kim, Lee, and Kim}{Lim et~al\mbox{.}}{2023}]%
        {Lim2023Design}
\bibfield{author}{\bibinfo{person}{Chaemin Lim}, \bibinfo{person}{Suhyun Lee}, \bibinfo{person}{Jinwoo Choi}, \bibinfo{person}{Jounghoo Lee}, \bibinfo{person}{Seongyeon Park}, \bibinfo{person}{Hanjun Kim}, \bibinfo{person}{Jinho Lee}, {and} \bibinfo{person}{Youngsok Kim}} \bibinfo{year}{2023}\natexlab{}.
\newblock \showarticletitle{{Design and Analysis of a Processing-in-DIMM Join Algorithm: A Case Study with UPMEM DIMMs}}.
\newblock \bibinfo{journal}{\emph{Proc. ACM Manag. Data}}, \bibinfo{year}{2023}.
\newblock


\bibitem[\protect\citeauthoryear{Lin and Prasanna}{Lin and Prasanna}{2023}]%
        {Lin2023HyScale}
\bibfield{author}{\bibinfo{person}{Y. Lin} {and} \bibinfo{person}{V. Prasanna}} \bibinfo{year}{2023}\natexlab{}.
\newblock \showarticletitle{{HyScale-GNN: A Scalable Hybrid GNN Training System on Single-Node Heterogeneous Architecture}}, In \bibinfo{booktitle}{\emph{IPDPS}}.
\newblock


\bibitem[\protect\citeauthoryear{Liu, Zhao, Ogleari, Li, and Zhao}{Liu et~al\mbox{.}}{2018}]%
        {Liu2018Processing}
\bibfield{author}{\bibinfo{person}{Jiawen Liu}, \bibinfo{person}{Hengyu Zhao}, \bibinfo{person}{Matheus~Almeida Ogleari}, \bibinfo{person}{Dong Li}, {and} \bibinfo{person}{Jishen Zhao}} \bibinfo{year}{2018}\natexlab{}.
\newblock \showarticletitle{{Processing-in-Memory for Energy-Efficient Neural Network Training: A Heterogeneous Approach}}, In \bibinfo{booktitle}{\emph{MICRO}}.
\newblock


\bibitem[\protect\citeauthoryear{Liu, Chen, Li, Wu, Zhu, He, Peng, Chen, Chen, and Guo}{Liu et~al\mbox{.}}{2023}]%
        {liu2023bgl}
\bibfield{author}{\bibinfo{person}{Tianfeng Liu}, \bibinfo{person}{Yangrui Chen}, \bibinfo{person}{Dan Li}, \bibinfo{person}{Chuan Wu}, \bibinfo{person}{Yibo Zhu}, \bibinfo{person}{Jun He}, \bibinfo{person}{Yanghua Peng}, \bibinfo{person}{Hongzheng Chen}, \bibinfo{person}{Hongzhi Chen}, {and} \bibinfo{person}{Chuanxiong Guo}} \bibinfo{year}{2023}\natexlab{}.
\newblock \showarticletitle{{BGL: GPU-Efficient GNN Training by Optimizing Graph Data I/O and Preprocessing}}, In \bibinfo{booktitle}{\emph{NSDI}}.
\newblock


\bibitem[\protect\citeauthoryear{Liu and Vinter}{Liu and Vinter}{2014}]%
        {Liu2014Efficient}
\bibfield{author}{\bibinfo{person}{Weifeng Liu} {and} \bibinfo{person}{Brian Vinter}} \bibinfo{year}{2014}\natexlab{}.
\newblock \showarticletitle{{An Efficient GPU General Sparse Matrix-Matrix Multiplication for Irregular Data}}, In \bibinfo{booktitle}{\emph{IPDPS}}.
\newblock


\bibitem[\protect\citeauthoryear{Liu, Calciu, Herlihy, and Mutlu}{Liu et~al\mbox{.}}{2017}]%
        {liu2017concurrent}
\bibfield{author}{\bibinfo{person}{Zhiyu Liu}, \bibinfo{person}{Irina Calciu}, \bibinfo{person}{Maurice Herlihy}, {and} \bibinfo{person}{Onur Mutlu}} \bibinfo{year}{2017}\natexlab{}.
\newblock \showarticletitle{{Concurrent Data Structures for Near-Memory Computing}}, In \bibinfo{booktitle}{\emph{SPAA}}.
\newblock


\bibitem[\protect\citeauthoryear{Loroch, Wehn, Pfreundt, and Keuper}{Loroch et~al\mbox{.}}{2017}]%
        {loroch2017tensorquant}
\bibfield{author}{\bibinfo{person}{Dominik~Marek Loroch}, \bibinfo{person}{Norbert Wehn}, \bibinfo{person}{Franz-Josef Pfreundt}, {and} \bibinfo{person}{Janis Keuper}} \bibinfo{year}{2017}\natexlab{}.
\newblock \showarticletitle{{TensorQuant - A Simulation Toolbox for Deep Neural Network Quantization}}, In \bibinfo{booktitle}{\emph{arXiv}}.
\newblock


\bibitem[\protect\citeauthoryear{Ma, Yang, Miao, Xue, Wu, Zhou, and Dai}{Ma et~al\mbox{.}}{2019}]%
        {Ma2019Neugraph}
\bibfield{author}{\bibinfo{person}{Lingxiao Ma}, \bibinfo{person}{Zhi Yang}, \bibinfo{person}{Youshan Miao}, \bibinfo{person}{Jilong Xue}, \bibinfo{person}{Ming Wu}, \bibinfo{person}{Lidong Zhou}, {and} \bibinfo{person}{Yafei Dai}} \bibinfo{year}{2019}\natexlab{}.
\newblock \showarticletitle{{Neugraph: Parallel Deep Neural Network Computation on Large Graphs}}, In \bibinfo{booktitle}{\emph{ATC}}.
\newblock


\bibitem[\protect\citeauthoryear{Md, Misra, Ma, Mohanty, Georganas, Heinecke, Kalamkar, Ahmed, and Avancha}{Md et~al\mbox{.}}{2021}]%
        {Md2021DistGNN}
\bibfield{author}{\bibinfo{person}{Vasimuddin Md}, \bibinfo{person}{Sanchit Misra}, \bibinfo{person}{Guixiang Ma}, \bibinfo{person}{Ramanarayan Mohanty}, \bibinfo{person}{Evangelos Georganas}, \bibinfo{person}{Alexander Heinecke}, \bibinfo{person}{Dhiraj Kalamkar}, \bibinfo{person}{Nesreen~K. Ahmed}, {and} \bibinfo{person}{Sasikanth Avancha}} \bibinfo{year}{2021}\natexlab{}.
\newblock \showarticletitle{{DistGNN: Scalable Distributed Training for Large-Scale Graph Neural Networks}}, In \bibinfo{booktitle}{\emph{SC}}.
\newblock


\bibitem[\protect\citeauthoryear{Mohamed, Nov{\'a}{\v{c}}ek, and Nounu}{Mohamed et~al\mbox{.}}{2020}]%
        {mohamed2020discovering}
\bibfield{author}{\bibinfo{person}{Sameh~K Mohamed}, \bibinfo{person}{V{\'\i}t Nov{\'a}{\v{c}}ek}, {and} \bibinfo{person}{Aayah Nounu}} \bibinfo{year}{2020}\natexlab{}.
\newblock \showarticletitle{{Discovering Protein Drug Targets Using Knowledge Graph Embeddings}}.
\newblock \bibinfo{journal}{\emph{Bioinformatics}}, \bibinfo{year}{2020}.
\newblock


\bibitem[\protect\citeauthoryear{Mpakos, Galanopoulos, Anastasiadis, Papadopoulou, Koziris, and Goumas}{Mpakos et~al\mbox{.}}{2023}]%
        {Mpakos2023Feature}
\bibfield{author}{\bibinfo{person}{P. Mpakos}, \bibinfo{person}{D. Galanopoulos}, \bibinfo{person}{P. Anastasiadis}, \bibinfo{person}{N. Papadopoulou}, \bibinfo{person}{N. Koziris}, {and} \bibinfo{person}{G. Goumas}} \bibinfo{year}{2023}\natexlab{}.
\newblock \showarticletitle{{Feature-Based SpMV Performance Analysis on Contemporary Devices}}, In \bibinfo{booktitle}{\emph{IPDPS}}.
\newblock


\bibitem[\protect\citeauthoryear{Mutlu, Ghose, G{\'{o}}mez{-}Luna, and Ausavarungnirun}{Mutlu et~al\mbox{.}}{2019}]%
        {Mutlu2019Processing}
\bibfield{author}{\bibinfo{person}{Onur Mutlu}, \bibinfo{person}{Saugata Ghose}, \bibinfo{person}{Juan G{\'{o}}mez{-}Luna}, {and} \bibinfo{person}{Rachata Ausavarungnirun}} \bibinfo{year}{2019}\natexlab{}.
\newblock \showarticletitle{{Processing Data Where It Makes Sense: Enabling In-Memory Computation}}, In \bibinfo{booktitle}{\emph{MICPRO}}.
\newblock


\bibitem[\protect\citeauthoryear{Mutlu, Ghose, G{\'o}mez-Luna, and Ausavarungnirun}{Mutlu et~al\mbox{.}}{2021a}]%
        {mutlu2020modern}
\bibfield{author}{\bibinfo{person}{Onur Mutlu}, \bibinfo{person}{Saugata Ghose}, \bibinfo{person}{Juan G{\'o}mez-Luna}, {and} \bibinfo{person}{Rachata Ausavarungnirun}} \bibinfo{year}{2021}\natexlab{a}.
\newblock \showarticletitle{{A Modern Primer on Processing in Memory}}, In \bibinfo{booktitle}{\emph{{Emerging Computing: From Devices to Systems - Looking Beyond Moore and Von Neumann}}}.
\newblock
\urldef\tempurl%
\url{https://arxiv.org/pdf/2012.03112.pdf}
\showURL{%
\tempurl}


\bibitem[\protect\citeauthoryear{Mutlu, Ghose, G{\'o}mez-Luna, and Ausavarungnirun}{Mutlu et~al\mbox{.}}{2021b}]%
        {Mutlu2020AMP}
\bibfield{author}{\bibinfo{person}{Onur Mutlu}, \bibinfo{person}{Saugata Ghose}, \bibinfo{person}{Juan G{\'o}mez-Luna}, {and} \bibinfo{person}{R. Ausavarungnirun}} \bibinfo{year}{2021}\natexlab{b}.
\newblock \showarticletitle{{A Modern Primer on Processing in Memory}}.
\newblock \bibinfo{journal}{\emph{Emerging Computing: From Devices to Systems - Looking Beyond Moore and Von Neumann}}, \bibinfo{year}{2021}.
\newblock


\bibitem[\protect\citeauthoryear{Nair, Antao, Bertolli, Bose, Brunheroto, Chen, Cher, Costa, Doi, Evangelinos, and et~al.}{Nair et~al\mbox{.}}{2015}]%
        {Nair2015Active}
\bibfield{author}{\bibinfo{person}{R. Nair}, \bibinfo{person}{S.~F. Antao}, \bibinfo{person}{C. Bertolli}, \bibinfo{person}{P. Bose}, \bibinfo{person}{J.~R. Brunheroto}, \bibinfo{person}{T. Chen}, \bibinfo{person}{C.-Y. Cher}, \bibinfo{person}{C.~H.~A. Costa}, \bibinfo{person}{J. Doi}, \bibinfo{person}{C. Evangelinos}, {and} \bibinfo{person}{et al.}} \bibinfo{year}{2015}\natexlab{}.
\newblock \showarticletitle{{Active Memory Cube: A Processing-in-Memory Architecture for Exascale Systems}}, In \bibinfo{booktitle}{\emph{IBM JRD}}.
\newblock


\bibitem[\protect\citeauthoryear{Naumov, Chien, Vandermersch, and Kapasi}{Naumov et~al\mbox{.}}{2010}]%
        {naumov2010cusparse}
\bibfield{author}{\bibinfo{person}{Maxim Naumov}, \bibinfo{person}{L Chien}, \bibinfo{person}{Philippe Vandermersch}, {and} \bibinfo{person}{Ujval Kapasi}} \bibinfo{year}{2010}\natexlab{}.
\newblock \showarticletitle{{Cusparse Library}}, In \bibinfo{booktitle}{\emph{GPU Technology Conference}}.
\newblock


\bibitem[\protect\citeauthoryear{Niu, Lu, Ji, Song, Jin, and Liu}{Niu et~al\mbox{.}}{2022}]%
        {Niu2022TileSpGEMM}
\bibfield{author}{\bibinfo{person}{Yuyao Niu}, \bibinfo{person}{Zhengyang Lu}, \bibinfo{person}{Haonan Ji}, \bibinfo{person}{Shuhui Song}, \bibinfo{person}{Zhou Jin}, {and} \bibinfo{person}{Weifeng Liu}} \bibinfo{year}{2022}\natexlab{}.
\newblock \showarticletitle{{TileSpGEMM: A Tiled Algorithm for Parallel Sparse General Matrix-Matrix Multiplication on GPUs}}, In \bibinfo{booktitle}{\emph{PpopP}}.
\newblock


\bibitem[\protect\citeauthoryear{Noh, Hong, Lim, Park, Kim, Kim, Kim, and Lee}{Noh et~al\mbox{.}}{2024}]%
        {Noh2024PIDComm}
\bibfield{author}{\bibinfo{person}{S. Noh}, \bibinfo{person}{J. Hong}, \bibinfo{person}{C. Lim}, \bibinfo{person}{S. Park}, \bibinfo{person}{J. Kim}, \bibinfo{person}{H. Kim}, \bibinfo{person}{Y. Kim}, {and} \bibinfo{person}{J. Lee}} \bibinfo{year}{2024}\natexlab{}.
\newblock \showarticletitle{{PID-Comm: A Fast and Flexible Collective Communication Framework for Commodity Processing-in-DIMM Devices}}, In \bibinfo{booktitle}{\emph{ISCA}}.
\newblock


\bibitem[\protect\citeauthoryear{Pal, Beaumont, Park, Amarnath, Feng, Chakrabarti, Kim, Blaauw, Mudge, and Dreslinski}{Pal et~al\mbox{.}}{2018}]%
        {pal2018outerspace}
\bibfield{author}{\bibinfo{person}{Subhankar Pal}, \bibinfo{person}{Jonathan Beaumont}, \bibinfo{person}{Dong-Hyeon Park}, \bibinfo{person}{Aporva Amarnath}, \bibinfo{person}{Siying Feng}, \bibinfo{person}{Chaitali Chakrabarti}, \bibinfo{person}{Hun-Seok Kim}, \bibinfo{person}{David Blaauw}, \bibinfo{person}{Trevor Mudge}, {and} \bibinfo{person}{Ronald Dreslinski}} \bibinfo{year}{2018}\natexlab{}.
\newblock \showarticletitle{{Outerspace: An Outer Product Based Sparse Matrix Multiplication Accelerator}}, In \bibinfo{booktitle}{\emph{HPCA}}.
\newblock


\bibitem[\protect\citeauthoryear{Park, Choi, Kyung, Kim, Kwon, Kim, and Ahn}{Park et~al\mbox{.}}{2024}]%
        {Park2024AttAcc}
\bibfield{author}{\bibinfo{person}{Jaehyun Park}, \bibinfo{person}{Jaewan Choi}, \bibinfo{person}{Kwanhee Kyung}, \bibinfo{person}{Michael~Jaemin Kim}, \bibinfo{person}{Yongsuk Kwon}, \bibinfo{person}{Nam~Sung Kim}, {and} \bibinfo{person}{Jung~Ho Ahn}} \bibinfo{year}{2024}\natexlab{}.
\newblock \showarticletitle{{AttAcc! Unleashing the Power of PIM for Batched Transformer-Based Generative Model Inference}}, In \bibinfo{booktitle}{\emph{ASPLOS}}.
\newblock


\bibitem[\protect\citeauthoryear{Park, Min, and Lee}{Park et~al\mbox{.}}{2022}]%
        {Park2022Ginex}
\bibfield{author}{\bibinfo{person}{Yeonhong Park}, \bibinfo{person}{Sunhong Min}, {and} \bibinfo{person}{Jae~W. Lee}} \bibinfo{year}{2022}\natexlab{}.
\newblock \showarticletitle{{Ginex: SSD-Enabled Billion-Scale Graph Neural Network Training on a Single Machine via Provably Optimal In-Memory Caching}}.
\newblock \bibinfo{journal}{\emph{Proc. VLDB Endow.}}, \bibinfo{year}{2022}.
\newblock


\bibitem[\protect\citeauthoryear{Paszke, Gross, Massa, Lerer, Bradbury, Chanan, Killeen, Lin, Gimelshein, Antiga, et~al\mbox{.}}{Paszke et~al\mbox{.}}{2019}]%
        {paszke2019pytorch}
\bibfield{author}{\bibinfo{person}{Adam Paszke}, \bibinfo{person}{Sam Gross}, \bibinfo{person}{Francisco Massa}, \bibinfo{person}{Adam Lerer}, \bibinfo{person}{James Bradbury}, \bibinfo{person}{Gregory Chanan}, \bibinfo{person}{Trevor Killeen}, \bibinfo{person}{Zeming Lin}, \bibinfo{person}{Natalia Gimelshein}, \bibinfo{person}{Luca Antiga}, {et~al\mbox{.}}} \bibinfo{year}{2019}\natexlab{}.
\newblock \showarticletitle{{Pytorch: An Imperative Style, High-Performance Deep Learning Library}}.
\newblock \bibinfo{journal}{\emph{NIPS}}, \bibinfo{year}{2019}.
\newblock


\bibitem[\protect\citeauthoryear{PeakPerf}{PeakPerf}{2021}]%
        {peak-perf}
\bibfield{author}{\bibinfo{person}{PeakPerf}} \bibinfo{year}{2021}\natexlab{}.
\newblock \showarticletitle{{PeakPerf}}.
\newblock
\urldef\tempurl%
\url{https://github.com/Dr-Noob/peakperf.git}
\showURL{%
\tempurl}


\bibitem[\protect\citeauthoryear{Pinar and Heath}{Pinar and Heath}{1999}]%
        {Pinar1999Improving}
\bibfield{author}{\bibinfo{person}{Ali Pinar} {and} \bibinfo{person}{Michael~T. Heath}} \bibinfo{year}{1999}\natexlab{}.
\newblock \showarticletitle{{Improving Performance of Sparse Matrix-Vector Multiplication}}, In \bibinfo{booktitle}{\emph{SC}}.
\newblock


\bibitem[\protect\citeauthoryear{Pooch and Nieder}{Pooch and Nieder}{1973}]%
        {Pooch1973Survey}
\bibfield{author}{\bibinfo{person}{Udo~W. Pooch} {and} \bibinfo{person}{Al Nieder}} \bibinfo{year}{1973}\natexlab{}.
\newblock \showarticletitle{{A Survey of Indexing Techniques for Sparse Matrices}}, In \bibinfo{booktitle}{\emph{ACM Comput. Surv.}}
\newblock


\bibitem[\protect\citeauthoryear{PyG}{PyG}{2024}]%
        {PyG}
\bibfield{author}{\bibinfo{person}{PyG}} \bibinfo{year}{2024}\natexlab{}.
\newblock \showarticletitle{{PyG Website}}.
\newblock
\urldef\tempurl%
\url{https://pyg.org/}
\showURL{%
\tempurl}


\bibitem[\protect\citeauthoryear{Qian, Abualshour, Li, Thabet, and Ghanem}{Qian et~al\mbox{.}}{2021}]%
        {qian2021pu}
\bibfield{author}{\bibinfo{person}{Guocheng Qian}, \bibinfo{person}{Abdulellah Abualshour}, \bibinfo{person}{Guohao Li}, \bibinfo{person}{Ali Thabet}, {and} \bibinfo{person}{Bernard Ghanem}} \bibinfo{year}{2021}\natexlab{}.
\newblock \showarticletitle{{Pu-GCN: Point Cloud Upsampling Using Graph Convolutional Networks}}, In \bibinfo{booktitle}{\emph{CVPR}}.
\newblock


\bibitem[\protect\citeauthoryear{Qu, Niu, Li, Zheng, and Xie}{Qu et~al\mbox{.}}{2023}]%
        {qu2023tt}
\bibfield{author}{\bibinfo{person}{Zheng Qu}, \bibinfo{person}{Dimin Niu}, \bibinfo{person}{Shuangchen Li}, \bibinfo{person}{Hongzhong Zheng}, {and} \bibinfo{person}{Yuan Xie}} \bibinfo{year}{2023}\natexlab{}.
\newblock \showarticletitle{{TT-GNN: Efficient On-Chip Graph Neural Network Training via Embedding Reformation and Hardware Optimization}}, In \bibinfo{booktitle}{\emph{MICRO}}.
\newblock


\bibitem[\protect\citeauthoryear{Rhyner, Luo, Gómez-Luna, Sadrosadati, Jiang, Olgun, Gupta, Zhang, and Mutlu}{Rhyner et~al\mbox{.}}{2024}]%
        {rhyner2024analysis}
\bibfield{author}{\bibinfo{person}{Steve Rhyner}, \bibinfo{person}{Haocong Luo}, \bibinfo{person}{Juan Gómez-Luna}, \bibinfo{person}{Mohammad Sadrosadati}, \bibinfo{person}{Jiawei Jiang}, \bibinfo{person}{Ataberk Olgun}, \bibinfo{person}{Harshita Gupta}, \bibinfo{person}{Ce Zhang}, {and} \bibinfo{person}{Onur Mutlu}} \bibinfo{year}{2024}\natexlab{}.
\newblock \showarticletitle{{Analysis of Distributed Optimization Algorithms on a Real Processing-In-Memory System}}, In \bibinfo{booktitle}{\emph{PACT}}.
\newblock


\bibitem[\protect\citeauthoryear{Selvitopi, Brock, Nisa, Tripathy, Yelick, and Bulu\c{c}}{Selvitopi et~al\mbox{.}}{2021}]%
        {Selvitopi2021Distributed}
\bibfield{author}{\bibinfo{person}{Oguz Selvitopi}, \bibinfo{person}{Benjamin Brock}, \bibinfo{person}{Israt Nisa}, \bibinfo{person}{Alok Tripathy}, \bibinfo{person}{Katherine Yelick}, {and} \bibinfo{person}{Ayd\i{}n Bulu\c{c}}} \bibinfo{year}{2021}\natexlab{}.
\newblock \showarticletitle{{Distributed-Memory Parallel Algorithms for Sparse Times Tall-Skinny-Dense Matrix Multiplication}}, In \bibinfo{booktitle}{\emph{ICS}}.
\newblock


\bibitem[\protect\citeauthoryear{Sengupta, Harris, Zhang, and Owens}{Sengupta et~al\mbox{.}}{2007}]%
        {Shubhabrata2007Scan}
\bibfield{author}{\bibinfo{person}{Shubhabrata Sengupta}, \bibinfo{person}{Mark Harris}, \bibinfo{person}{Yao Zhang}, {and} \bibinfo{person}{John~D. Owens}} \bibinfo{year}{2007}\natexlab{}.
\newblock \showarticletitle{{Scan Primitives for GPU Computing}}, In \bibinfo{booktitle}{\emph{GH}}.
\newblock


\bibitem[\protect\citeauthoryear{Shang, Chen, and Bi}{Shang et~al\mbox{.}}{2021}]%
        {shang2021discrete}
\bibfield{author}{\bibinfo{person}{Chao Shang}, \bibinfo{person}{Jie Chen}, {and} \bibinfo{person}{Jinbo Bi}} \bibinfo{year}{2021}\natexlab{}.
\newblock \showarticletitle{{Discrete Graph Structure Learning for Forecasting Multiple Time Series}}, In \bibinfo{booktitle}{\emph{ICLR}}.
\newblock


\bibitem[\protect\citeauthoryear{Shin, Park, Cho, and Sung}{Shin et~al\mbox{.}}{2023}]%
        {Shin2023PIMFlow}
\bibfield{author}{\bibinfo{person}{Yongwon Shin}, \bibinfo{person}{Juseong Park}, \bibinfo{person}{Sungjun Cho}, {and} \bibinfo{person}{Hyojin Sung}} \bibinfo{year}{2023}\natexlab{}.
\newblock \showarticletitle{{PIMFlow: Compiler and Runtime Support for CNN Models on Processing-in-Memory DRAM}}, In \bibinfo{booktitle}{\emph{CGO}}.
\newblock


\bibitem[\protect\citeauthoryear{Song, Chi, Sohrabizadeh, Choi, Lau, and Cong}{Song et~al\mbox{.}}{2022}]%
        {song2022sextans}
\bibfield{author}{\bibinfo{person}{Linghao Song}, \bibinfo{person}{Yuze Chi}, \bibinfo{person}{Atefeh Sohrabizadeh}, \bibinfo{person}{Young-kyu Choi}, \bibinfo{person}{Jason Lau}, {and} \bibinfo{person}{Jason Cong}} \bibinfo{year}{2022}\natexlab{}.
\newblock \showarticletitle{{Sextans: A Streaming Accelerator for General-Purpose Sparse-Matrix Dense-Matrix Multiplication}}, In \bibinfo{booktitle}{\emph{SIGDA}}.
\newblock


\bibitem[\protect\citeauthoryear{Song, Zhi, Fan, Zhang, Zeng, Li, Hu, Du, Guo, and Chen}{Song et~al\mbox{.}}{2021}]%
        {song2021cambricon}
\bibfield{author}{\bibinfo{person}{Xinkai Song}, \bibinfo{person}{Tian Zhi}, \bibinfo{person}{Zhe Fan}, \bibinfo{person}{Zhenxing Zhang}, \bibinfo{person}{Xi Zeng}, \bibinfo{person}{Wei Li}, \bibinfo{person}{Xing Hu}, \bibinfo{person}{Zidong Du}, \bibinfo{person}{Qi Guo}, {and} \bibinfo{person}{Yunji Chen}} \bibinfo{year}{2021}\natexlab{}.
\newblock \showarticletitle{{Cambricon-G: A Polyvalent Energy-Efficient Accelerator for Dynamic Graph Neural Networks}}.
\newblock \bibinfo{journal}{\emph{TCAD}}, \bibinfo{year}{2021}.
\newblock


\bibitem[\protect\citeauthoryear{Srivastava, Jin, Liu, Albonesi, and Zhang}{Srivastava et~al\mbox{.}}{2020}]%
        {srivastava2020matraptor}
\bibfield{author}{\bibinfo{person}{Nitish Srivastava}, \bibinfo{person}{Hanchen Jin}, \bibinfo{person}{Jie Liu}, \bibinfo{person}{David Albonesi}, {and} \bibinfo{person}{Zhiru Zhang}} \bibinfo{year}{2020}\natexlab{}.
\newblock \showarticletitle{{Matraptor: A Sparse-Sparse Matrix Multiplication Accelerator Based on Row-Wise Product}}, In \bibinfo{booktitle}{\emph{MICRO}}.
\newblock


\bibitem[\protect\citeauthoryear{Stevens, Das, Avancha, Kaul, and Raghunathan}{Stevens et~al\mbox{.}}{2021}]%
        {stevens2021gnnerator}
\bibfield{author}{\bibinfo{person}{Jacob~R Stevens}, \bibinfo{person}{Dipankar Das}, \bibinfo{person}{Sasikanth Avancha}, \bibinfo{person}{Bharat Kaul}, {and} \bibinfo{person}{Anand Raghunathan}} \bibinfo{year}{2021}\natexlab{}.
\newblock \showarticletitle{{GNNerator: A Hardware/Software Framework for Accelerating Graph Neural Networks}}, In \bibinfo{booktitle}{\emph{DAC}}.
\newblock


\bibitem[\protect\citeauthoryear{Stokes, Yang, Swanson, Jin, Cubillos-Ruiz, Donghia, MacNair, French, Carfrae, Bloom-Ackermann, et~al\mbox{.}}{Stokes et~al\mbox{.}}{2020}]%
        {stokes2020deep}
\bibfield{author}{\bibinfo{person}{Jonathan~M Stokes}, \bibinfo{person}{Kevin Yang}, \bibinfo{person}{Kyle Swanson}, \bibinfo{person}{Wengong Jin}, \bibinfo{person}{Andres Cubillos-Ruiz}, \bibinfo{person}{Nina~M Donghia}, \bibinfo{person}{Craig~R MacNair}, \bibinfo{person}{Shawn French}, \bibinfo{person}{Lindsey~A Carfrae}, \bibinfo{person}{Zohar Bloom-Ackermann}, {et~al\mbox{.}}} \bibinfo{year}{2020}\natexlab{}.
\newblock \showarticletitle{{A Deep Learning Approach to Antibiotic Discovery}}.
\newblock \bibinfo{journal}{\emph{Cell}}, \bibinfo{year}{2020}.
\newblock


\bibitem[\protect\citeauthoryear{Strati, Giannoula, Siakavaras, Goumas, and Koziris}{Strati et~al\mbox{.}}{2019}]%
        {Strati2019Adaptive}
\bibfield{author}{\bibinfo{person}{Foteini Strati}, \bibinfo{person}{Christina Giannoula}, \bibinfo{person}{Dimitrios Siakavaras}, \bibinfo{person}{Georgios Goumas}, {and} \bibinfo{person}{Nectarios Koziris}} \bibinfo{year}{2019}\natexlab{}.
\newblock \showarticletitle{{An Adaptive Concurrent Priority Queue for NUMA Architectures}}, In \bibinfo{booktitle}{\emph{International Conference on Computing Frontiers}}.
\newblock


\bibitem[\protect\citeauthoryear{stream}{stream}{2021}]%
        {stream}
\bibfield{author}{\bibinfo{person}{stream}} \bibinfo{year}{2021}\natexlab{}.
\newblock \showarticletitle{{STREAM}}.
\newblock
\urldef\tempurl%
\url{https://github.com/jeffhammond/STREAM.git}
\showURL{%
\tempurl}


\bibitem[\protect\citeauthoryear{Szklarczyk, Gable, Lyon, Junge, Wyder, Huerta-Cepas, Simonovic, Doncheva, Morris, Bork, et~al\mbox{.}}{Szklarczyk et~al\mbox{.}}{2019}]%
        {szklarczyk2019string}
\bibfield{author}{\bibinfo{person}{Damian Szklarczyk}, \bibinfo{person}{Annika~L Gable}, \bibinfo{person}{David Lyon}, \bibinfo{person}{Alexander Junge}, \bibinfo{person}{Stefan Wyder}, \bibinfo{person}{Jaime Huerta-Cepas}, \bibinfo{person}{Milan Simonovic}, \bibinfo{person}{Nadezhda~T Doncheva}, \bibinfo{person}{John~H Morris}, \bibinfo{person}{Peer Bork}, {et~al\mbox{.}}} \bibinfo{year}{2019}\natexlab{}.
\newblock \showarticletitle{{STRING v11: Protein--Protein Association Networks with Increased Coverage, Supporting Functional Discovery in Genome-Wide Experimental Datasets}}.
\newblock \bibinfo{journal}{\emph{Nucleic Acids Research}}, \bibinfo{year}{2019}.
\newblock


\bibitem[\protect\citeauthoryear{Tang, Zhao, Lu, Liang, Huyng, Li, and Goh}{Tang et~al\mbox{.}}{2015}]%
        {tang2015optimizing}
\bibfield{author}{\bibinfo{person}{Wai~Teng Tang}, \bibinfo{person}{Ruizhe Zhao}, \bibinfo{person}{Mian Lu}, \bibinfo{person}{Yun Liang}, \bibinfo{person}{Huynh~Phung Huyng}, \bibinfo{person}{Xibai Li}, {and} \bibinfo{person}{Rick Siow~Mong Goh}} \bibinfo{year}{2015}\natexlab{}.
\newblock \showarticletitle{{Optimizing and Auto-Tuning Scale-Free Sparse Matrix-Vector Multiplication on Intel Xeon Phi}}, In \bibinfo{booktitle}{\emph{CGO}}.
\newblock


\bibitem[\protect\citeauthoryear{Thorpe, Qiao, Eyolfson, Teng, Hu, Jia, Wei, Vora, Netravali, Kim, and Xu}{Thorpe et~al\mbox{.}}{2021}]%
        {Thorpe2021Dorylus}
\bibfield{author}{\bibinfo{person}{John Thorpe}, \bibinfo{person}{Yifan Qiao}, \bibinfo{person}{Jonathan Eyolfson}, \bibinfo{person}{Shen Teng}, \bibinfo{person}{Guanzhou Hu}, \bibinfo{person}{Zhihao Jia}, \bibinfo{person}{Jinliang Wei}, \bibinfo{person}{Keval Vora}, \bibinfo{person}{Ravi Netravali}, \bibinfo{person}{Miryung Kim}, {and} \bibinfo{person}{Guoqing~Harry Xu}} \bibinfo{year}{2021}\natexlab{}.
\newblock \showarticletitle{{Dorylus: Affordable, Scalable, and Accurate GNN Training with Distributed CPU Servers and Serverless Threads}}, In \bibinfo{booktitle}{\emph{OSDI}}.
\newblock


\bibitem[\protect\citeauthoryear{Tian, Li, Jiang, Cai, and Gao}{Tian et~al\mbox{.}}{2024}]%
        {Tian2024NDPBridge}
\bibfield{author}{\bibinfo{person}{Boyu Tian}, \bibinfo{person}{Yiwei Li}, \bibinfo{person}{Li Jiang}, \bibinfo{person}{Shuangyu Cai}, {and} \bibinfo{person}{Mingyu Gao}} \bibinfo{year}{2024}\natexlab{}.
\newblock \showarticletitle{{NDPBridge: Enabling Cross-Bank Coordination in Near-DRAM-Bank Processing Architectures}}, In \bibinfo{booktitle}{\emph{ISCA}}.
\newblock


\bibitem[\protect\citeauthoryear{Tian, Wang, Zhao, Wu, Zhang, Lu, Wang, and Jin}{Tian et~al\mbox{.}}{2022}]%
        {tian2022GNMP}
\bibfield{author}{\bibinfo{person}{Teng Tian}, \bibinfo{person}{Xiaotian Wang}, \bibinfo{person}{Letian Zhao}, \bibinfo{person}{Wei Wu}, \bibinfo{person}{Xuecang Zhang}, \bibinfo{person}{Fangmin Lu}, \bibinfo{person}{Tianqi Wang}, {and} \bibinfo{person}{Xi Jin}} \bibinfo{year}{2022}\natexlab{}.
\newblock \showarticletitle{{G-NMP: Accelerating Graph Neural Networks with DIMM-Based Near-Memory Processing}}.
\newblock \bibinfo{journal}{\emph{Journal of Systems Architecture}}, \bibinfo{year}{2022}.
\newblock


\bibitem[\protect\citeauthoryear{Tripathy, Yelick, and Bulu\c{c}}{Tripathy et~al\mbox{.}}{2020}]%
        {Tripathy2020Reducing}
\bibfield{author}{\bibinfo{person}{Alok Tripathy}, \bibinfo{person}{Katherine Yelick}, {and} \bibinfo{person}{Ayd\i{}n Bulu\c{c}}} \bibinfo{year}{2020}\natexlab{}.
\newblock \showarticletitle{{Reducing Communication in Graph Neural Network Training}}, In \bibinfo{booktitle}{\emph{SC}}.
\newblock


\bibitem[\protect\citeauthoryear{UPMEM}{UPMEM}{2020}]%
        {upmem}
\bibfield{author}{\bibinfo{person}{UPMEM}} \bibinfo{year}{2020}\natexlab{}.
\newblock \showarticletitle{{UPMEM Website}}.
\newblock
\urldef\tempurl%
\url{https://www.upmem.com}
\showURL{%
\tempurl}


\bibitem[\protect\citeauthoryear{Van~Zee, Smith, Marker, Low, Geijn, Igual, Smelyanskiy, Zhang, Kistler, Austel, et~al\mbox{.}}{Van~Zee et~al\mbox{.}}{2016}]%
        {van2016blis}
\bibfield{author}{\bibinfo{person}{Field~G Van~Zee}, \bibinfo{person}{Tyler~M Smith}, \bibinfo{person}{Bryan Marker}, \bibinfo{person}{Tze~Meng Low}, \bibinfo{person}{Robert A Van~De Geijn}, \bibinfo{person}{Francisco~D Igual}, \bibinfo{person}{Mikhail Smelyanskiy}, \bibinfo{person}{Xianyi Zhang}, \bibinfo{person}{Michael Kistler}, \bibinfo{person}{Vernon Austel}, {et~al\mbox{.}}} \bibinfo{year}{2016}\natexlab{}.
\newblock \showarticletitle{{The BLIS Framework: Experiments in Portability}}.
\newblock \bibinfo{journal}{\emph{TOMS}}, \bibinfo{year}{2016}.
\newblock


\bibitem[\protect\citeauthoryear{Velickovic, Cucurull, Casanova, Romero, Lio, Bengio, et~al\mbox{.}}{Velickovic et~al\mbox{.}}{2017}]%
        {velickovic2017graph}
\bibfield{author}{\bibinfo{person}{Petar Velickovic}, \bibinfo{person}{Guillem Cucurull}, \bibinfo{person}{Arantxa Casanova}, \bibinfo{person}{Adriana Romero}, \bibinfo{person}{Pietro Lio}, \bibinfo{person}{Yoshua Bengio}, {et~al\mbox{.}}} \bibinfo{year}{2017}\natexlab{}.
\newblock \showarticletitle{{Graph Attention Networks}}.
\newblock \bibinfo{journal}{\emph{Stat}}, \bibinfo{year}{2017}.
\newblock


\bibitem[\protect\citeauthoryear{Wang, Zhang, Shen, Zhang, Lu, Wu, and Wang}{Wang et~al\mbox{.}}{2014}]%
        {Wang2014MKL}
\bibfield{author}{\bibinfo{person}{Endong Wang}, \bibinfo{person}{Qing Zhang}, \bibinfo{person}{Bo Shen}, \bibinfo{person}{Guangyong Zhang}, \bibinfo{person}{Xiaowei Lu}, \bibinfo{person}{Qing Wu}, {and} \bibinfo{person}{Yajuan Wang}} \bibinfo{year}{2014}\natexlab{}.
\newblock \bibinfo{booktitle}{\emph{{Intel Math Kernel Library}}}.
\newblock


\bibitem[\protect\citeauthoryear{Wu, Tang, Zhu, Wang, Xie, and Tan}{Wu et~al\mbox{.}}{2019}]%
        {wu2019session}
\bibfield{author}{\bibinfo{person}{Shu Wu}, \bibinfo{person}{Yuyuan Tang}, \bibinfo{person}{Yanqiao Zhu}, \bibinfo{person}{Liang Wang}, \bibinfo{person}{Xing Xie}, {and} \bibinfo{person}{Tieniu Tan}} \bibinfo{year}{2019}\natexlab{}.
\newblock \showarticletitle{{Session-Based Recommendation with Graph Neural Networks}}, In \bibinfo{booktitle}{\emph{AAAI}}.
\newblock


\bibitem[\protect\citeauthoryear{Xu, Hu, Leskovec, and Jegelka}{Xu et~al\mbox{.}}{2018}]%
        {xu2018powerful}
\bibfield{author}{\bibinfo{person}{Keyulu Xu}, \bibinfo{person}{Weihua Hu}, \bibinfo{person}{Jure Leskovec}, {and} \bibinfo{person}{Stefanie Jegelka}} \bibinfo{year}{2018}\natexlab{}.
\newblock \showarticletitle{{How Powerful Are Graph Neural Networks?}}
\newblock \bibinfo{journal}{\emph{arXiv}}, \bibinfo{year}{2018}.
\newblock


\bibitem[\protect\citeauthoryear{Yang, Bulu{\c{c}}, and Owens}{Yang et~al\mbox{.}}{2018}]%
        {yang2018design}
\bibfield{author}{\bibinfo{person}{Carl Yang}, \bibinfo{person}{Ayd{\i}n Bulu{\c{c}}}, {and} \bibinfo{person}{John~D. Owens}} \bibinfo{year}{2018}\natexlab{}.
\newblock \showarticletitle{{Design Principles for Sparse Matrix Multiplication on the GPU}}, In \bibinfo{booktitle}{\emph{Euro-PAR}}.
\newblock


\bibitem[\protect\citeauthoryear{Yang, Cohen, and Salakhudinov}{Yang et~al\mbox{.}}{2016}]%
        {yang2016revisiting}
\bibfield{author}{\bibinfo{person}{Zhilin Yang}, \bibinfo{person}{William Cohen}, {and} \bibinfo{person}{Ruslan Salakhudinov}} \bibinfo{year}{2016}\natexlab{}.
\newblock \showarticletitle{{Revisiting Semi-Supervised Learning with Graph Embeddings}}, In \bibinfo{booktitle}{\emph{ICML}}.
\newblock


\bibitem[\protect\citeauthoryear{Ye, Lai, Shao, Chen, and Ceze}{Ye et~al\mbox{.}}{2023}]%
        {Ye2023SparseTIR}
\bibfield{author}{\bibinfo{person}{Zihao Ye}, \bibinfo{person}{Ruihang Lai}, \bibinfo{person}{Junru Shao}, \bibinfo{person}{Tianqi Chen}, {and} \bibinfo{person}{Luis Ceze}} \bibinfo{year}{2023}\natexlab{}.
\newblock \showarticletitle{{SparseTIR: Composable Abstractions for Sparse Compilation in Deep Learning}}, In \bibinfo{booktitle}{\emph{ASPLOS}}.
\newblock


\bibitem[\protect\citeauthoryear{Yesil, Moreira, and Torrellas}{Yesil et~al\mbox{.}}{2022}]%
        {yesil2022dense}
\bibfield{author}{\bibinfo{person}{Serif Yesil}, \bibinfo{person}{Jos{\'e}~E Moreira}, {and} \bibinfo{person}{Josep Torrellas}} \bibinfo{year}{2022}\natexlab{}.
\newblock \showarticletitle{{Dense Dynamic Blocks: Optimizing SpMM for Processors with Vector and Matrix Units Using Machine Learning Techniques}}, In \bibinfo{booktitle}{\emph{ICS}}.
\newblock


\bibitem[\protect\citeauthoryear{Ying, He, Chen, Eksombatchai, Hamilton, and Leskovec}{Ying et~al\mbox{.}}{2018}]%
        {ying2018graph}
\bibfield{author}{\bibinfo{person}{Rex Ying}, \bibinfo{person}{Ruining He}, \bibinfo{person}{Kaifeng Chen}, \bibinfo{person}{Pong Eksombatchai}, \bibinfo{person}{William~L Hamilton}, {and} \bibinfo{person}{Jure Leskovec}} \bibinfo{year}{2018}\natexlab{}.
\newblock \showarticletitle{{Graph Convolutional Neural Networks for Web-Scale Recommender Systems}}, In \bibinfo{booktitle}{\emph{SIGKDD}}.
\newblock


\bibitem[\protect\citeauthoryear{Yun, Nam, Park, Kim, Ahn, and Lee}{Yun et~al\mbox{.}}{2023}]%
        {yun2023grande}
\bibfield{author}{\bibinfo{person}{Sungmin Yun}, \bibinfo{person}{Hwayong Nam}, \bibinfo{person}{Jaehyun Park}, \bibinfo{person}{Byeongho Kim}, \bibinfo{person}{Jung~Ho Ahn}, {and} \bibinfo{person}{Eojin Lee}} \bibinfo{year}{2023}\natexlab{}.
\newblock \showarticletitle{{GraNDe: Efficient Near-Data Processing Architecture for Graph Neural Networks}}.
\newblock \bibinfo{journal}{\emph{IEEE Trans. Comput.}}, \bibinfo{year}{2023}.
\newblock


\bibitem[\protect\citeauthoryear{Zeng, Zhou, Srivastava, Kannan, and Prasanna}{Zeng et~al\mbox{.}}{2019}]%
        {zeng2019graphsaint}
\bibfield{author}{\bibinfo{person}{Hanqing Zeng}, \bibinfo{person}{Hongkuan Zhou}, \bibinfo{person}{Ajitesh Srivastava}, \bibinfo{person}{Rajgopal Kannan}, {and} \bibinfo{person}{Viktor Prasanna}} \bibinfo{year}{2019}\natexlab{}.
\newblock \showarticletitle{{GraphSaint: Graph Sampling Based Inductive Learning Method}}.
\newblock \bibinfo{journal}{\emph{arXiv}}, \bibinfo{year}{2019}.
\newblock


\bibitem[\protect\citeauthoryear{Zeng, Chen, Huang, Luo, Zhang, and Zhou}{Zeng et~al\mbox{.}}{2023}]%
        {zeng2023serving}
\bibfield{author}{\bibinfo{person}{Liekang Zeng}, \bibinfo{person}{Xu Chen}, \bibinfo{person}{Peng Huang}, \bibinfo{person}{Ke Luo}, \bibinfo{person}{Xiaoxi Zhang}, {and} \bibinfo{person}{Zhi Zhou}} \bibinfo{year}{2023}\natexlab{}.
\newblock \showarticletitle{{Serving Graph Neural Networks With Distributed Fog Servers for Smart IoT Services}}.
\newblock \bibinfo{journal}{\emph{IEEE/ACM Transactions on Networking}}, \bibinfo{year}{2023}.
\newblock


\bibitem[\protect\citeauthoryear{Zhai, Zhang, Liu, Chu, Peng, Ji, and Zhang}{Zhai et~al\mbox{.}}{2023}]%
        {zhai2023tlp}
\bibfield{author}{\bibinfo{person}{Yi Zhai}, \bibinfo{person}{Yu Zhang}, \bibinfo{person}{Shuo Liu}, \bibinfo{person}{Xiaomeng Chu}, \bibinfo{person}{Jie Peng}, \bibinfo{person}{Jianmin Ji}, {and} \bibinfo{person}{Yanyong Zhang}} \bibinfo{year}{2023}\natexlab{}.
\newblock \showarticletitle{{TLP: A Deep Learning-Based Cost Model for Tensor Program Tuning}}, In \bibinfo{booktitle}{\emph{ASPLOS}}.
\newblock


\bibitem[\protect\citeauthoryear{Zhang, Zhuo, Wang, Gao, Wu, Chen, Kozyrakis, and Qian}{Zhang et~al\mbox{.}}{2018}]%
        {Zhang2018GraphP}
\bibfield{author}{\bibinfo{person}{Mingxing Zhang}, \bibinfo{person}{Youwei Zhuo}, \bibinfo{person}{Chao Wang}, \bibinfo{person}{Mingyu Gao}, \bibinfo{person}{Yongwei Wu}, \bibinfo{person}{Kang Chen}, \bibinfo{person}{Christos Kozyrakis}, {and} \bibinfo{person}{Xuehai Qian}} \bibinfo{year}{2018}\natexlab{}.
\newblock \showarticletitle{{GraphP: Reducing Communication for PIM-Based Graph Processing with Efficient Data Partition}}, In \bibinfo{booktitle}{\emph{HPCA}}.
\newblock


\bibitem[\protect\citeauthoryear{Zhang, Lin, Yang, and Sa}{Zhang et~al\mbox{.}}{2019}]%
        {zhang2019qpytorch}
\bibfield{author}{\bibinfo{person}{Tianyi Zhang}, \bibinfo{person}{Zhiqiu Lin}, \bibinfo{person}{Guandao Yang}, {and} \bibinfo{person}{Christopher~De Sa}} \bibinfo{year}{2019}\natexlab{}.
\newblock \showarticletitle{{QPyTorch: A Low-Precision Arithmetic Simulation Framework}}, In \bibinfo{booktitle}{\emph{arXiv}}.
\newblock


\bibitem[\protect\citeauthoryear{Zheng, Ma, Wang, Zhou, Su, Song, Gan, Zhang, and Karypis}{Zheng et~al\mbox{.}}{2020}]%
        {Zheng2020DistDGL}
\bibfield{author}{\bibinfo{person}{D. Zheng}, \bibinfo{person}{C. Ma}, \bibinfo{person}{M. Wang}, \bibinfo{person}{J. Zhou}, \bibinfo{person}{Q. Su}, \bibinfo{person}{X. Song}, \bibinfo{person}{Q. Gan}, \bibinfo{person}{Z. Zhang}, {and} \bibinfo{person}{G. Karypis}} \bibinfo{year}{2020}\natexlab{}.
\newblock \showarticletitle{{DistDGL: Distributed Graph Neural Network Training for Billion-Scale Graphs}}, In \bibinfo{booktitle}{\emph{IA3}}.
\newblock


\bibitem[\protect\citeauthoryear{Zhou, Li, Wei, Wang, and Sun}{Zhou et~al\mbox{.}}{2022}]%
        {zhou2022gnnear}
\bibfield{author}{\bibinfo{person}{Zhe Zhou}, \bibinfo{person}{Cong Li}, \bibinfo{person}{Xuechao Wei}, \bibinfo{person}{Xiaoyang Wang}, {and} \bibinfo{person}{Guangyu Sun}} \bibinfo{year}{2022}\natexlab{}.
\newblock \showarticletitle{{Gnnear: Accelerating Full-Batch Training of Graph Neural Networks with Near-Memory Processing}}, In \bibinfo{booktitle}{\emph{PACT}}.
\newblock


\bibitem[\protect\citeauthoryear{Zhuo, Wang, Zhang, Wang, Niu, Wang, and Qian}{Zhuo et~al\mbox{.}}{2019}]%
        {zhuo2019graphq}
\bibfield{author}{\bibinfo{person}{Youwei Zhuo}, \bibinfo{person}{Chao Wang}, \bibinfo{person}{Mingxing Zhang}, \bibinfo{person}{Rui Wang}, \bibinfo{person}{Dimin Niu}, \bibinfo{person}{Yanzhi Wang}, {and} \bibinfo{person}{Xuehai Qian}} \bibinfo{year}{2019}\natexlab{}.
\newblock \showarticletitle{{GraphQ: Scalable PIM-based Graph Processing}}, In \bibinfo{booktitle}{\emph{MICRO}}.
\newblock


\end{thebibliography}
\clearpage
\appendix
\newpage

\section{Appendix}

\subsection{\myName Tuner Efficiency}\label{sec:appendix-autotuner}

\begin{figure}[h]
\vspace{-1pt}
    \centering
    \includegraphics[width=\linewidth]{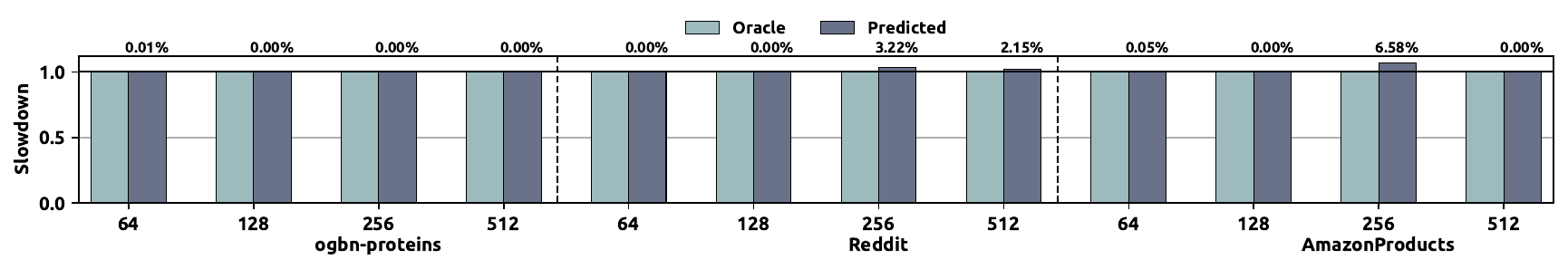}
    \caption{Performance slowdown of the predicted COO aggregation configuration by tuner over oracle prediction.}
    \label{fig:autotuner_coo}
    \vspace{-4pt}
\end{figure}

Fig.~\ref{fig:autotuner_coo} evaluates the \myName tuner efficiency for COO format by comparing the performance slowdown achieved by the predicted aggregation configuration (predicted) of tuner versus the oracle prediction using various datasets and hidden sizes. For the oracle prediction performance, we exhaustively iterate and collect the execution times of all possible aggregation configurations, then we select and present in Fig.~\ref{fig:autotuner_coo} the best-performing execution time among them (oracle).
The selected aggregation configuration by the tuner achieves only 1\% worse performance over the oracle-predicted configuration on average across all datasets and hidden sizes, when using COO format for \gnn aggregation.
We conclude that \myName tuner effectively tunes the aggregation configuration in \gnn executions for both CSR (Fig.~\ref{fig:autotuner_csr}) and COO formats.

\subsection{\gnn Aggregation Energy Consumption}\label{sec:append-energy-aggr}

\begin{figure}[h]
    \centering
    \includegraphics[width=\linewidth]{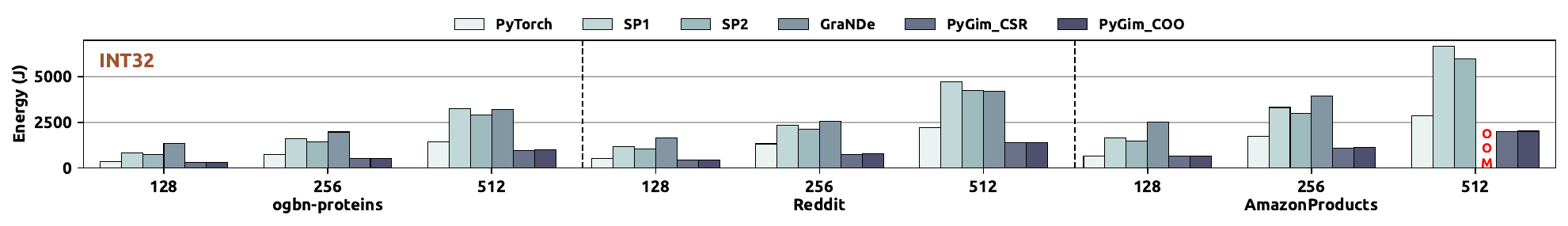}
    \caption{Energy consumption of all comparison in the one aggregation, using various graph datasets and hidden sizes.}
    \label{fig:eval-aggr-energy}
    \vspace{-4pt}
\end{figure}

Fig.~\ref{fig:eval-aggr-energy} presents the energy consumption (in Joules) for all comparison points (See~\cref{sec:methodology}) in one  \gnn aggregation using int32 data type, and various datasets and hidden sizes. In PIM executions, we use 32 PIM devices, having in total 1992 cores, and we enable the \myNameN's tuner. 
We use Intel RAPL~\cite{rapl} to measure energy in CPU execution parts, which are (i) the \CPUName  scheme, and (ii) in PIM schemes, the load, retrieve, and merge steps of aggregation. For the kernel step of aggregation, we measure the energy consumed in PIM-enabled chips using the methodology suggested by the UPMEM PIM manufacturer, which is described in a recent paper~\cite{falevoz2023energy}: the power of each PIM DIMM is 23.22W, thus the total energy of kernel time is calculated as  $kernel\_time$ $\times$ $\#PIM\_DIMMs$ $\times$ $power$.
\myName provides higher energy efficiency by on average 4.08$\times$ and 1.39$\times$ over prior PIM-based and \CPUName schemes, respectively.

\subsection{\gnn Inference Performance}\label{sec:append-endToEnd}

\begin{figure}[h]
    \centering
    \includegraphics[width=\linewidth]{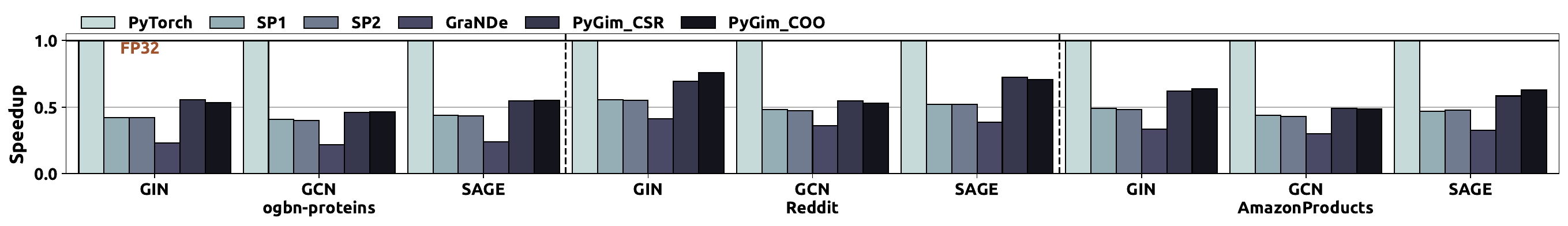}
    \caption{Performance of all comparison points in the end-to-end \gnn inference, using various graph datasets and \gnn models for fp32 data type.}
    \label{fig:eval-full-flt32}
    \vspace{-8pt}
\end{figure}

\begin{figure}[h]
    \centering
    \includegraphics[width=\linewidth]{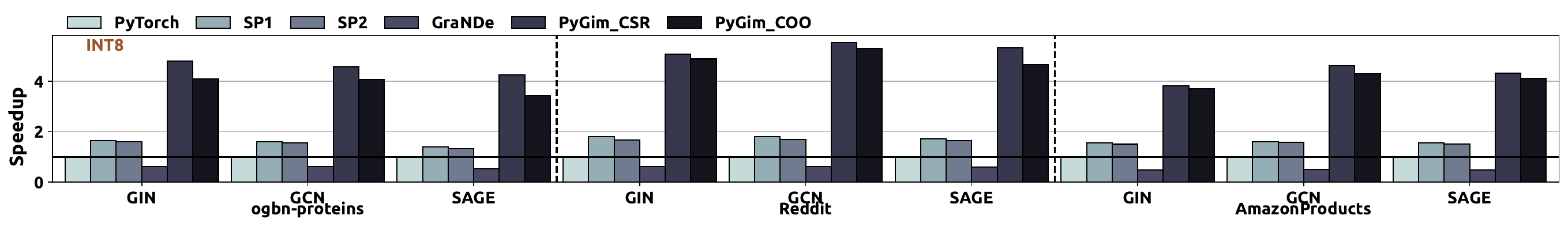}
    \caption{Performance of all comparison points in the end-to-end \gnn inference, using various graph datasets and \gnn models for int8 data type.}
    \label{fig:eval-full-int8}
    \vspace{-10pt}
\end{figure}

\begin{figure}[h]
    \centering
    \includegraphics[width=\linewidth]{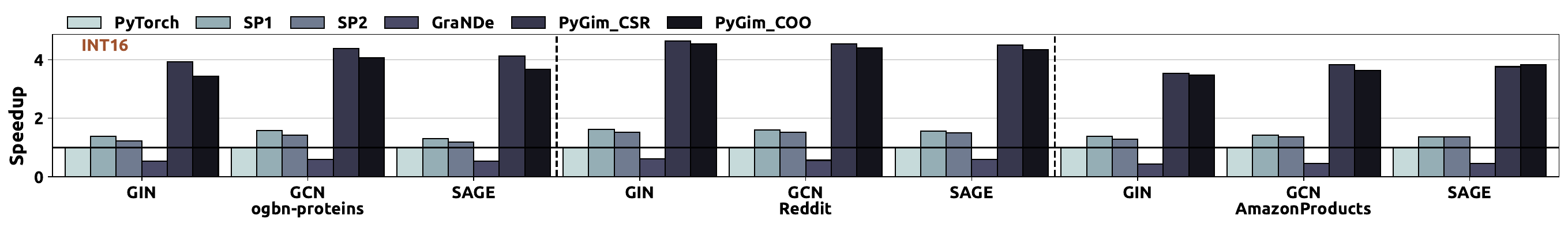}
    \caption{Performance of all comparison points in the end-to-end \gnn inference, using various graph datasets and \gnn models for int16 data type. }
    \label{fig:eval-full-int16}
    \vspace{-14pt}
\end{figure}

Figs.~\ref{fig:eval-full-flt32}, ~\ref{fig:eval-full-int8} and ~\ref{fig:eval-full-int16}  compare the performance of all comparison points (See ~\cref{sec:methodology})
in the end-to-end \gnn inference, using 32-bit float (\textbf{fp32}), 8-bit integer (\textbf{int8}) and 16-bit integer (\textbf{int16}) data types for data values, respectively. We evaluate various graph datasets and three different \gnn models, where each model has 3 layers of 256 hidden size. In PIM executions, we use 32 PIM devices, having in total $\sim$1992 cores. In \myNameN, we evaluate both CSR and COO schemes  and we enable the tuner to set the aggregation configuration.

We find that when using fp32 values, PIM \gnn execution achieves low performance, being worse than that of \CPUName  by on average 41.6\%. This is because UPMEM PIM hardware does not support floating-point operations, which are software emulated, thus they incur high performance overheads. However, we expect that ML-oriented PIM systems will be available in the market (e.g., ~\cite{Lee2021Hardware,lee20221ynm}), and will hopefully support in hardware high precision data types. Instead, when using int8 and int16 data types, \myName schemes significantly outperform  \CPUName  scheme by 4.49$\times$ (up to 5.54$\times$) and 4.03$\times$ (up to 4.63$\times$), respectively. Moreover, \myName outperforms prior state-of-the-art PIM approaches by 3.59$\times$ (up to 9.89$\times$) and 3.60$\times$ (up to 8.90$\times$) for int8 and int16 data types, respectively. When arithmetic operations are natively supported by the PIM hardware, as it happens for the int8 and int16 data type in UPMEM PIM system, \myName execution provides significant performance benefits over prior schemes. 

\subsection{\gnn Inference Accuracy}\label{sec:append-infer-acc}

Table~\ref{tab:infer-acc} shows the test accuracy achieved in \gnn inference with int32 and fp32 data types. We evaluate the accuracy of the experiments presented in Fig.~\ref{fig:eval-full-infer} and Fig.~\ref{fig:eval-full-flt32}. These models have been trained with fp32 data type, and then we run and evaluate inference using either int32 or fp32 data type.  All the comparison points compared in Fig.~\ref{fig:eval-full-infer} and Fig.~\ref{fig:eval-full-flt32}, i.e., both the CPU-based scheme (\CPUNameN) and the PIM-based schemes (SP1, SP2, GraNDe, \myNameN\_CSR, \myNameN\_COO) achieve the same accuracy,  shown in Table~\ref{tab:infer-acc} for int32 and fp32 data types. The  \gnn models are relatively simple, having only 3 layers, thus int32 and fp32 data types provide the same accuracy.

\begin{table}[H]
\begin{center}
\centering
\resizebox{0.78\linewidth}{!}{
\begin{tabular}{|l||c|c|c|c|c|c|c|c|c|}
    \hline
     \cellcolor{gray!15} &   \multicolumn{3}{c|}{\cellcolor{gray!15} \raisebox{-0.20\height}{\textbf{ ogbn-proteins}} } & \multicolumn{3}{c|}{\cellcolor{gray!15}  \textbf{Reddit}}  & \multicolumn{3}{c|}{\cellcolor{gray!15} \textbf{AmazonProducts}} \\ \hline
    \cellcolor{gray!15} &  \cellcolor{gray!15}\raisebox{-0.20\height}{\textbf{GIN}}  
    & \cellcolor{gray!15}\raisebox{-0.20\height}{\textbf{GCN}} 
    &  \cellcolor{gray!15}\raisebox{-0.20\height}{\textbf{SAGE}}
    \cellcolor{gray!15} &  \cellcolor{gray!15}\raisebox{-0.20\height}{\textbf{GIN}}  
    & \cellcolor{gray!15}\raisebox{-0.20\height}{\textbf{GCN}} 
    &  \cellcolor{gray!15}\raisebox{-0.20\height}{\textbf{SAGE}}
    \cellcolor{gray!15} &  \cellcolor{gray!15}\raisebox{-0.20\height}{\textbf{GIN}}  
    & \cellcolor{gray!15}\raisebox{-0.20\height}{\textbf{GCN}} 
    &  \cellcolor{gray!15}\raisebox{-0.20\height}{\textbf{SAGE}}\\
    \hline \hline
    INT32  &  79.89\% & 78.20\%  &   73.38\%  & 94.08\% & 91.90\%  & 94.39\% &  26.04\% & 26.04\%   &   26.04\%   \\ \hline 
    FP32  & 79.89\%  &  78.20\% &    73.38\% &  94.08\% & 91.90\%  &  94.39\% &  26.04\% & 26.04\%   &  26.04\%  \\ \hline 
\end{tabular}
}
\end{center}
\vspace{2pt}
\caption{Inference accuracy achieved by all comparison points using various graph datasets and \gnn models for int32 and fp32 data types.}
\label{tab:infer-acc}
\end{table}

\subsection{\gnn Training Performance}\label{sec:append-training}

Table~\ref{tab:gcn-train} shows the execution time of \CPUName (CPU) and \myName CSR (UPMEM PIM) schemes, when training a 2-layer GCN model for 10 epochs with 256 hidden size, int32 data type and evaluating all three datasets. Note that UPMEM PIM does not support fp32 operations in hardware. Thus, for a fair comparison, we evaluate \gnn training using int32 data type only, so that the data type used is fully supported by hardware in both evaluated systems. Table~\ref{tab:gcn-train-acc} shows the test accuracy achieved in the GCN model after the model is trained for 1000 epochs and learning rate of 0.01 with either \CPUName or \myName CSR scheme for int32 data type. Both \CPUName and \myName CSR schemes achieve the same accuracy in int32 data type. \myName improves the training performance by on average 1.25$\times$ over the \CPUName  scheme. Therefore, \myName  provides high performance benefits even for \gnn training, without degrading accuracy.

\begin{table}[H]
\begin{center}
\centering
\resizebox{0.9\linewidth}{!}{
\begin{tabular}{|l||c|c|c|}
    \hline
    \cellcolor{gray!15} &  \cellcolor{gray!15}\raisebox{-0.20\height}{\textbf{INT32 OGBN}}  
    & \cellcolor{gray!15}\raisebox{-0.20\height}{\textbf{INT32 RDT}} 
    &  \cellcolor{gray!15}\raisebox{-0.20\height}{\textbf{INT32 AMZ}} \\
    \hline \hline
    pytorch\_sparse (Intel Xeon 4215 CPU)  &  203.982 s & 339.023 s  &   440.899 s  \\ \hline 
    \myName (UPMEM PIM)  & 164.709 s  &  242.789 s &    400.998 s \\ \hline 
\end{tabular}
}
\end{center}
\vspace{2pt}
\caption{Execution time of GCN training for 10 epochs using pytorch\_sparse (CPU) and \myName (PIM cores for aggregation) libraries in aggregation step for the ogbn-proteins (\textbf{OGBN}), Reddit (\textbf{RDT}), and AmazonProducts (\textbf{AMZ}) datasets.}
\label{tab:gcn-train}
\vspace{-6pt}
\end{table}

\begin{table}[H]
\begin{center}
\centering
\resizebox{0.84\linewidth}{!}{
\begin{tabular}{|l||c|c|c|}
    \hline
    \cellcolor{gray!15} & \cellcolor{gray!15}\raisebox{-0.20\height}{\textbf{INT32 OGBN}}  
    & \cellcolor{gray!15}\raisebox{-0.20\height}{\textbf{INT32 RDT}} 
    &  \cellcolor{gray!15}\raisebox{-0.20\height}{\textbf{INT32 AMZ}}   \\
    \hline \hline
    pytorch\_sparse (Intel Xeon 4215 CPU)  & 70.34\%  &  84.25\%  &   26.04\%  \\ \hline
     \myName (UPMEM PIM)  & 70.34\%  &  84.25\%  &   26.04\% \\  \hline 
\end{tabular}
}
\end{center}
\vspace{2pt}
\caption{Test accuracy achieved after training the GCN for 1000 epochs and a learning rate of 0.01 with either \CPUName or \myName CSR scheme using int32  data type and the ogbn-proteins (\textbf{OGBN}), Reddit (\textbf{RDT}), and AmazonProducts (\textbf{AMZ}) datasets.}
\label{tab:gcn-train-acc}
\vspace{-6pt}
\end{table}

\subsection{Evaluation of \gnn Executions in GPU Systems}\label{sec:appendix-gpu-comparison}

We present the performance and energy efficiency metrics to show the readers how much performance and energy efficiency the evaluated UPMEM PIM system can achieve when it is compared over commodity GPU systems. In Table~\ref{tab:performance-energy-gpu}, we present the performance in \gnn aggregation comparing three GPU systems over the evaluated UPMEM PIM system.
Similarly, in Table~\ref{tab:performance-infer-gpu}, we present the performance in \gnn inference comparing two GPU systems over the evaluated UPMEM PIM system. 
For UPMEM PIM, we use \myName library, and for GPU systems we use pytorch\_sparse library that provides  optimized CUDA implementations.
Please note that these evaluation results are provided for completeness and \emph{not} competition purposes, since real PIM systems are still in early  manufacturing and design stages (especially compared to commercial CPU and GPU systems), and \myName  can be evaluated on other current and future real PIM systems with potentially better computation capabilities and energy efficiency than the evaluated UPMEM PIM system.

\begin{table}[h]
\vspace{1pt}
\begin{center}
\centering
\resizebox{1\linewidth}{!}{
\begin{tabular}{|l||c|c|c|c|c|c|}
    \hline
    \cellcolor{gray!15}\raisebox{-0.20\height}{\textbf{Dataset and data type }}  &  \cellcolor{gray!15}\raisebox{-0.20\height}{\textbf{OGBN INT32 }}  
    & \cellcolor{gray!15}\raisebox{-0.20\height}{\textbf{RDT INT32}} 
    &  \cellcolor{gray!15}\raisebox{-0.20\height}{\textbf{AMZ INT32}} 
    & \cellcolor{gray!15}\raisebox{-0.20\height}{\textbf{OGBN FP32}} 
    &\cellcolor{gray!15}\raisebox{-0.20\height}{\textbf{RDT FP32}} & \cellcolor{gray!15}\raisebox{-0.20\height}{\textbf{AMZ FP32}}  \\
     
    \hline \hline
    GPU GTX 1080 Ti over UPMEM PIM  & 17.7$\times$  |  8.0$\times$ &  5.3$\times$  |  3.3$\times$ & 
    6.9$\times$  |  3.8$\times$ & 112.0$\times$  |  69.9$\times$ & 35.9$\times$  |  28.1$\times$ & 44.0$\times$  |  31.7$\times$ \\ \hline 
    GPU RTX 2080 Ti over UPMEM PIM  & 15.3$\times$  |  8.0$\times$ &  7.3$\times$  |  3.8$\times$  & 8.9$\times$  |  4.2$\times$  & 102.5$\times$  |  66.4$\times$ & 49.8$\times$  |  33.4$\times$& 56.3$\times$  |  35.9$\times$\\ \hline 
    GPU RTX 3090 over UPMEM PIM   & 39.8$\times$  |  21.2$\times$ & 19.3$\times$  |  6.9$\times$ & 18.4$\times$  |  6.6$\times$  & 240.4$\times$  |  200.1$\times$ & 122.4$\times$  |  55.1$\times$ & 111.3$\times$  |  46.1$\times$ \\ \hline 
\end{tabular}
}
\end{center}
\vspace{2pt}
\caption{Performance speedup (left number in each cell) and energy efficiency improvement (right number in each cell) of the three GPU generations over the UPMEM PIM system in \gnn aggregation using INT32 and FP32 data types and the ogbn-proteins (\textbf{OGBN}), Reddit (\textbf{RDT}), and AmazonProducts (\textbf{AMZ}) datasets.}
\label{tab:performance-energy-gpu}
\end{table}

\begin{table}[h]
\vspace{6pt}
\begin{center}
\centering
\resizebox{1\linewidth}{!}{
\begin{tabular}{|l||c|c|c|c|c|c|}
    \hline
    \cellcolor{gray!15}\raisebox{-0.20\height}{\textbf{Dataset and \gnn model}}  &  \cellcolor{gray!15}\raisebox{-0.20\height}{\textbf{OGBN GIN }}  
    & \cellcolor{gray!15}\raisebox{-0.20\height}{\textbf{RDT GIN}} 
    &  \cellcolor{gray!15}\raisebox{-0.20\height}{\textbf{AMZ GIN}} 
    & \cellcolor{gray!15}\raisebox{-0.20\height}{\textbf{OGBN GCN}} 
    &\cellcolor{gray!15}\raisebox{-0.20\height}{\textbf{RDT GCN}} & \cellcolor{gray!15}\raisebox{-0.20\height}{\textbf{AMZ GCN}}  \\
     
    \hline \hline
    GPU RTX 2080 Ti over UPMEM PIM  & 17.4$\times$  |  88.1$\times$ &  9.1$\times$  |  42.5$\times$  & 11.2$\times$  |  45.3$\times$  & 17.2$\times$  |  107.3$\times$ & 8.5$\times$  |  51.8$\times$& 10.4$\times$  |  56.9$\times$\\ \hline 
    GPU RTX 3090 over UPMEM PIM   & 37.2$\times$  |  188.7$\times$ & 20.7$\times$  |  97.2$\times$ & 20.8$\times$  |  84.4$\times$  & 38.2$\times$  |  238.2$\times$ & 19.9$\times$  |  121.2$\times$ & 19.6$\times$  |  106.7$\times$ \\ \hline 
\end{tabular}
}
\end{center}
\caption{Performance speedup for int32 data type (left number in each cell) and for fp32 data type (right number in each cell) of two GPU generations over the UPMEM PIM system in the end-to-end \gnn inference using GIN and GCN models and the ogbn-proteins (\textbf{OGBN}), Reddit (\textbf{RDT}), and AmazonProducts (\textbf{AMZ}) datasets.}
\label{tab:performance-infer-gpu}
\end{table}

\subsection{Datasets}\label{sec:appendix-dataset}

\noindent\textbf{Sparse Matrices.} 
We present the characteristics of the real-world matrices that we use in our experiments to evaluate \myName when using one PIM core and when using one PIM cluster. The sparse matrices are taken from the Sparse Matrix Suite Collection~\cite{davis2011florida}. Table~\ref{tab:appendix-matrices} presents the number of rows, the number of non-zero elements (NNZs), the minimum number (min) of non-zero elements among rows, the maximum number (max) of non-zero elements among rows, the average number (avg) of non-zero elements among rows and standard deviation (std) of non-zero elements among rows.

\begin{table}[h]
\vspace{2pt}
\begin{center}
\centering
\begin{minipage}{\linewidth}
\resizebox{\linewidth}{!}{
\begin{tabular}{|l||c|r|r|r|r|r|}
    \hline
    \cellcolor{gray!15}\raisebox{-0.10\height}{\textbf{Matrix Name}} & \cellcolor{gray!15}\raisebox{-0.10\height}{\textbf{Rows}} &  \cellcolor{gray!15}\raisebox{-0.10\height}{\textbf{NNZ}} & \cellcolor{gray!15}\raisebox{-0.10\height}{\textbf{Min NNZ}} & 
    \cellcolor{gray!15}\raisebox{-0.10\height}{\textbf{Max NNZ}} & \cellcolor{gray!15}\raisebox{-0.10\height}{\textbf{Avg NNZ}} & \cellcolor{gray!15}\raisebox{-0.10\height}{\textbf{Std NNZ}}  \\
    \hline \hline
    raefsky4 & 19779 &	1328611	 & 18	& 177	& 67.17 & 15.96 \\ \hline 
    
    wing\_nodal & 10937	 &	150976  &	5	 & 28	 & 13.80	 & 2.86 \\ \hline
    
    Dubcova2 &	65025  &	1030225	 &	4  &	25  &	15.84	  & 5.76  \\ \hline
    
    mosfet2 & 	46994	 & 1499460	 &	4	 & 162	 & 31.91	 & 11.71 \\ \hline
    
    poisson3Db & 	85623  &	2374949	 &	6  &	145	  & 27.74 &	14.71\\ \hline
    
    smt & 	25710  &	3753184   &	52  &	414	 & 145.98  & 47.52 \\ \hline  
    
    av41092 & 	41092	 & 1683902 & 		2  &	2135	 & 40.98	 & 167.04 \\ \hline
    
    Zd\_Jac6  &	22835	 & 1711983	 &	1  &	1050	  & 74.97	 & 175.48  \\ \hline
    
    mycielskian15 &  24575  &	11111110  &	14  &	12287	 & 452.13	 & 664.17 \\ \hline
\end{tabular}
}
\end{minipage}%
\end{center}
\caption{Sparse matrix dataset used for one PIM core and one PIM cluster analysis.}
\label{tab:appendix-matrices}
\vspace{4pt}
\end{table}

\noindent\textbf{Graph Datasets.} 
We present the characteristics of the real-world graph datasets that we use in our large-scale experiments to evaluate \myName using multiple PIM DIMMs and devices, as well as to evaluate CPU and PIM schemes in aggregation operator and end-to-end \gnn inference. The real-world graph datasets are taken from ogbn-proteins~\cite{szklarczyk2019string}, Reddit~\cite{hamilton2017inductive} and AmazonProducts~\cite{zeng2019graphsaint}. The original  AmazonProducts dataset is too large to fit in a single machine, thus we split the dataset using cluster partition~\cite{chiang2019cluster}, and evaluate the largest subgraph in our experiments, its detailed characteristics are shown in Table~\ref{tab:appendix-graphs}. Specifically, Table~\ref{tab:appendix-graphs} presents the number of vertices, the number of edges (EDGs), the minimum number (min) of edges among vertices, the maximum number (max) of edges among vertices, the average number (avg) of edges among vertices and standard deviation (std) of edges among vertices.

\begin{table}[H]
\vspace{2pt}
\begin{center}
\centering
\begin{minipage}{\linewidth}
\resizebox{1.0\linewidth}{!}{
\begin{tabular}{|l||c|r|r|r|r|r|}
    \hline
    \cellcolor{gray!15}\raisebox{-0.10\height}{\textbf{Graph Name}} & \cellcolor{gray!15}\raisebox{-0.10\height}{\textbf{Vertices}} &  \cellcolor{gray!15}\raisebox{-0.10\height}{\textbf{EDGs}} & \cellcolor{gray!15}\raisebox{-0.10\height}{\textbf{Min EDG}} & 
    \cellcolor{gray!15}\raisebox{-0.10\height}{\textbf{Max EDG}} & \cellcolor{gray!15}\raisebox{-0.10\height}{\textbf{Avg EDG}} & \cellcolor{gray!15}\raisebox{-0.10\height}{\textbf{Std EDG}}  \\
    \hline \hline
    ogbn-proteins   & 132534    & 79122504  & 1 & 7750 & 597.00 & 621.48\\ \hline 
    
    Reddit          & 232965	& 114615892 & 1 & 21657 & 492.00 & 799.82 \\ \hline
    
    AmazonProducts  & 403598   & 156149176 & 1 & 53864 &  386.89 & 1140.91 \\ \hline
    
\end{tabular}
}
\end{minipage}%
\end{center}
\caption{Real-world graph datasets used for our large-scale experiments, when using multiple PIM DIMMs.}
\label{tab:appendix-graphs}
\vspace{4pt}
\end{table}

\end{document}